\documentclass[review]{elsarticle}

\usepackage[colorlinks,citecolor=blue,linktoc=all,linkcolor=cyan]{hyperref}
\usepackage{graphicx}

\usepackage[T1]{fontenc}
\usepackage{dsfont}               
\usepackage{mathrsfs}            
\usepackage{slashed}             
\usepackage{amsmath}
\usepackage{amssymb}
\usepackage{amsbsy}
\usepackage{amsfonts}
\usepackage{physics} 
\usepackage{diffcoeff} 
\usepackage{subcaption}
\usepackage{orcidlink}
\usepackage{pdflscape}
\usepackage{longtable} 
\usepackage{makecell} 
\numberwithin{equation}{section}
\numberwithin{table}{section}
\numberwithin{figure}{section}
\usepackage{comment}
\usepackage{multirow}

\newcommand{\MSbar}{\overline{\text{MS}}}
\newcommand{\nn}{\nonumber}

\usepackage[dvipsnames]{xcolor}
\usepackage{enumitem}

\journal{Progress in Particle and Nuclear Physics}

\usepackage{titlesec}
\usepackage{sectsty}
\titleformat{\section}{\normalfont\Large\bfseries}{\thesection}{1em}{}
\titleformat{\subsection}{\normalfont\large\bfseries}{\thesubsection}{1em}{}
\titleformat{\subsubsection}{\normalfont\normalsize\bfseries}{\thesubsubsection}{1em}{}

\begin{document}

\begin{frontmatter}

\title{Recent Advances in Transverse-Momentum-Dependent Distributions in Lattice QCD } 

\author[mymainaddress]{Huey-Wen Lin\corref{mycorrespondingauthor}}
  \cortext[mycorrespondingauthor]{Corresponding author}
  \ead{hwlin@pa.msu.edu}
  \address[mymainaddress]{Department of Physics and Astronomy, Michigan State University, East Lansing, MI 48824}

\author[mysecondaryaddress]{Wei Wang}
\ead{wei.wang@sjtu.edu.cn}
\address[mysecondaryaddress]{State Key Laboratory of Dark Matter Physics, Shanghai Key Laboratory for Particle Physics and Cosmology, Key Laboratory for Particle Astrophysics and Cosmology (MOE), Physics-AI Institute School of Physics and Astronomy, Shanghai Jiao Tong University, Shanghai 200240, China\\
Southern Center for Nuclear-Science Theory (SCNT), Institute of Modern Physics, Chinese Academy of Sciences, Huizhou 516000, Guangdong Province, China
}

\author[mythirdaddress]{Yong Zhao}
  \ead{yong.zhao@anl.gov}
  \address[mythirdaddress]{Physics Division, Argonne National Laboratory, Lemont, IL 60439, USA}

\begin{abstract}
Unraveling the three-dimensional structure of hadrons is a central objective in modern nuclear and particle physics.
transverse-momentum-dependent parton distributions (TMDs) provide crucial insights into the interplay between partonic transverse motion and spin, and are indispensable for describing a wide range of high-energy processes at facilities such as Jefferson Lab, COMPASS, the LHC, and the upcoming Electron-Ion Collider.
In recent years, significant progress has been made in accessing TMDs from lattice QCD using novel approaches. Methods based on ratios of Mellin moments, Large-Momentum Effective Theory, and Coulomb-gauge techniques have enabled first-principles investigations of the Collins–Soper kernel, soft function, TMD wave functions, and TMD parton distribution functions.
Recent lattice results for both unpolarized and spin-dependent distributions, including the Boer–Mulders function, demonstrate the feasibility of extracting transverse-momentum-–dependent information directly from QCD.
This paper reviews recent advances in lattice-QCD studies of TMDs, outlines remaining challenges such as continuum and infinite-momentum extrapolations, and discusses future directions toward precision determinations of hadron structure in three dimensions.
\end{abstract}


\end{frontmatter}

\newpage

\thispagestyle{empty}
\tableofcontents

\section{Introduction}
The study of nucleon structure has progressed from the identification of pointlike constituents inside nucleons to the modern objective of resolving the full three-dimensional momentum distribution.
Traditional high-energy deep-inelastic scattering measurements provide information encoded in collinear parton distribution functions (PDFs) that depend solely on the longitudinal momentum fraction carried by partons.
While these PDFs have generated important insights into momentum sharing inside nucleons, they do not capture any information regarding the transverse dynamics of quarks and gluons.
However, confinement inherently imposes nontrivial transverse spatial and momentum behavior on partons.
Extracting this internal motion requires observables sensitive to both a hard scattering scale and a softer transverse momentum scale.
This conceptual evolution laid the foundation for the development of transverse-momentum--dependent distributions (TMDs), which provide direct access to such intrinsic partonic motion~\cite{Collins:1981uk,Collins:1984kg,Collins:2011zzd}.

The emergence of TMDs represents a central advancement toward resolving hadrons in three dimensions.
Unlike collinear PDFs, which integrate out all transverse structure, TMDs depend simultaneously on the longitudinal momentum fraction and the intrinsic transverse momentum of partons.
This allows them to encode correlations between parton motion and the spin of both the hadron and the parton.
Theoretical development of the TMD framework required significant progress in QCD factorization, including careful treatment of gauge links and soft factors.
These advances enabled the construction of factorization theorems for two-scale processes such as semi-inclusive deep inelastic scattering (SIDIS) and Drell-Yan production at low transverse momentum.
With the upcoming Electron-Ion Collider (EIC)~\cite{AbdulKhalek:2021gbh,Accardi:2012qut,Burkert:2022hjz} and the China-based Electron-Ion Collider (EicC)~\cite{Anderle:2021wcy} currently under design, TMDs are expected to play a key role in mapping proton substructure in unprecedented detail.

TMDs describe the probability to find a parton inside a fast-moving hadron carrying a specified longitudinal momentum fraction and transverse momentum.
Their dependence on transverse momentum reflects the confined and intrinsically relativistic motion of partons inside hadrons.
Because nucleons are finite-sized bound states governed by QCD, transverse motion is an unavoidable consequence of confinement dynamics.
TMDs therefore extend the one-dimensional information provided by PDFs to reveal the rich internal momentum landscape of the proton. Their operator definitions include staple-shaped gauge links that encode the effects of color interactions between the struck parton and the hadronic remnant.
These path-dependent gauge links are responsible for process dependence, such as the predicted sign change of the Sivers function between SIDIS and Drell-Yan.
TMDs are also key to constructing three-dimensional images of the proton.
Their Fourier transforms with respect to transverse momentum yield impact-parameter space distributions that encode spatial structure.
These distributions allow the study of correlations between spin and orbital angular momentum contributions within the proton.
Spin-dependent TMDs, such as the Sivers and Boer-Mulders functions, reveal how partonic motion is influenced by nucleon polarization and the internal dynamics of QCD.
Such effects do not appear in collinear distributions and therefore represent new, independent aspects of nucleon structure.
Consequently, TMDs connect theoretical operator definitions to experimentally observable three-dimensional hadron imaging.

Extensive global analyses of SIDIS, Drell-Yan, and $e^{+}e^{-}$ hadron-pair production data have enabled phenomenological extractions of several leading  twist TMDs~\cite{Davies:1984sp,Ladinsky:1993zn,Landry:1999an,Landry:2002ix,Konychev:2005iy,Sun:2014dqm,DAlesio:2014mrz,Bacchetta:2017gcc,Scimemi:2017etj,Bertone:2019nxa,Scimemi:2019cmh,Bacchetta:2019sam,Hautmann:2020cyp,Bury:2022czx,Bacchetta:2022awv,Isaacson:2023iui,Aslan:2024nqg,Moos:2023yfa,Kang:2024dja,Bacchetta:2024qre,Moos:2025sal,Cuerpo:2025zde,Rossi:2025pwh,Barry:2025glq,Aglietti:2026bhu,Kang:2026mod,Zaccheddu:2026pyy}. 
The phenomenological extractions above draw on data from a broad set of experiments.
Semi-inclusive DIS measurements of pion and kaon production come from HERMES at DESY, COMPASS at CERN (muon beams on deuteron and proton targets), and the Jefferson Lab experiments in Halls~A, B, and C. 
These provide the low-energy, valence-dominated constraints on the flavor structure and on the nonperturbative transverse-momentum dependence. 
Drell-Yan and vector-boson production data span a wide range of scales. They include fixed-target experiments at Fermilab (E288, E605, E772), collider measurements of $Z$/$W$ transverse-momentum spectra at RHIC (STAR, PHENIX), the Tevatron (CDF, D0), and the LHC (ATLAS, CMS, LHCb), which are especially sensitive to the evolution (Collins-Soper kernel) at high scales. 
Data from $e^{+}e^{-}$ annihilation, in particular hadron-pair production from Belle, BaBar, and BESIII and event-shape and energy-energy-correlator measurements from LEP and SLD, constrain the fragmentation side and the evolution. 
The recent helicity extractions~\cite{Yang:2024drd,Bacchetta:2024yzl} rely on longitudinal double-spin asymmetries in SIDIS from HERMES, COMPASS, and Jefferson Lab, while the first gluon-TMD extraction~\cite{Anedda:2026cox} uses Higgs-boson transverse-momentum spectra from ATLAS and CMS.  
These analyses rely on QCD factorization and evolution equations~\cite{Collins:1981uk,Collins:1984kg,Collins:2011zzd} that describe how TMDs vary with momentum scales.
However, many TMDs, particularly gluon TMDs and certain spin-dependent quark TMDs, remain poorly constrained due to limited experimental sensitivity.
Moreover, global fits depend on modeling assumptions, such as parameterizations of transverse momentum dependence, which introduce systematic uncertainties.
The precision of extracted TMDs, therefore, varies widely across parton flavor, polarization, and momentum fraction.
This motivates accessing nonperturbative information from first-principles methods.

Lattice QCD offers a powerful theoretical framework for computing hadron structure from first principles~\cite{Detmold:2019ghl,Davoudi:2022bnl,Lin:2022nnj,Lin:2017snn,Constantinou:2020hdm,Constantinou:2022yye}.
However, traditional lattice calculations are performed in Euclidean spacetime and therefore cannot directly access lightcone correlations required to define TMDs.
Additionally, TMD operator definitions involve rapidity divergences and soft factors that must be treated carefully in matching between lattice observables and continuum TMDs.
To overcome these obstacles, recent developments have introduced new strategies including the use of quasi-distributions and lattice-calculable approximations that can be perturbatively matched to physical TMDs.
This motivates ongoing efforts to integrate phenomenological extractions with lattice-based nonperturbative determinations of TMDs.
Early lattice QCD studies addressed TMD-related structure using Lorentz-invariant decompositions of nonlocal Euclidean correlators~\cite{Hagler:2009mb,Musch:2010ka,Musch:2011er,Engelhardt:2015xja,Yoon:2017qzo}.
These approaches provided qualitative access to certain spin-momentum correlations but lacked a systematic connection to physically measurable TMDs.
The primary limitation was that TMDs require lightlike gauge links and lightcone correlations, which cannot be directly implemented in Euclidean space.
Additionally, treatments of rapidity divergences in higher-order corrections and the soft factors remained incomplete. 
While these early lattice studies provided valuable insights into TMDs, quantitatively precise predictions of their full kinematic dependence in 3D momentum space were still lacking. 

The development of Large-Momentum Effective Theory (LaMET) marked an important  shift in lattice-QCD studies of parton structure~\cite{Ji:2013dva,Ji:2014gla,Ji:2020ect}.
LaMET introduced the concept of quasi-distributions, computed for hadrons boosted to large but finite momenta on the lattice, which can be perturbatively matched to physical lightcone distributions.
This framework enables systematic access to parton momentum structure, including TMDs, from Euclidean lattice data.
Subsequent research extended LaMET to include soft-factor subtraction and rapidity regularization necessary for TMD observables.
These advances have opened the possibility of controlled, quantitative lattice QCD determinations of TMDs that can be directly compared with experiment.
LaMET provides a formal bridge connecting Euclidean lattice matrix elements to physical lightcone TMDs.
In this approach, quasi-TMDs computed with highly boosted hadrons are perturbatively matched to continuum TMDs, accounting for renormalization differences and soft-factor structures.
Successful application of LaMET requires careful control over hadron momenta, lattice spacing, and finite-volume effects to ensure reliable extrapolations.
Current calculations have demonstrated feasibility for both unpolarized and polarized TMDs, establishing LaMET as a central tool in modern hadron structure research.
As computational resources expand, LaMET-based lattice determinations are expected to play an increasingly important role in global TMD analyses.

The structure of this review is as follows.
In Sec.~\ref{sec:Methodology}, we start with the definition of TMDs, followed by an introduction to the LaMET methodology and its application to various TMD-related quantities, including TMD parton distribution functions (TMDPDFs).
In Sec.~\ref{sec:Results}, we review recent lattice-QCD results obtained within the LaMET framework, focusing on the extraction of the Collins-Soper  (CS) kernel, soft function, TMD wave functions (TMDWFs) and TMDPDFs.
In Sec.~\ref{sec:Developments}, we discuss ongoing efforts and future directions, including the impact on phenomenology, the exploration of additional TMD observables, and developments toward improved precision and methodology.
Finally, we conclude and provide an outlook on near-term progress in lattice-QCD studies of TMDs and their role in advancing three-dimensional hadron structure.

\section{Methodology}\label{sec:Methodology}
In this section, we present the theoretical foundations and lattice QCD formulations of TMDPDFs.
We begin by defining TMDPDFs through a representative factorization theorem and introducing their operator definitions in QCD, emphasizing the Collins scheme and its treatment of rapidity divergences, beam and soft functions, and evolution equations.
We then review the Lorentz-invariant approach, in which TMD observables are accessed via a general off-lightcone correlator and decomposed into invariant amplitudes suitable for lattice calculations, enabling the extraction of ratios of moments.
Next, we describe the LaMET framework, where quasi-TMDPDFs, defined by equal-time correlators in boosted hadron states, are matched perturbatively to their lightcone counterparts, providing a systematic path toward computing the full $x$-dependence.
Finally, we discuss practical challenges in lattice implementations, including Wilson-line artifacts and power divergences, and introduce alternative formulations such as the Coulomb-gauge approach within the universality class of quasi-observables, which aim to improve numerical efficiency while preserving the infrared (IR) physics relevant for TMD factorization.

\subsection{Definition of TMDs} 
\label{subsec:def}

The TMDs are nonperturbative inputs for hard processes in which the measured observable is sensitive to the transverse momentum of the probed parton.
As an example, consider the hadron-hadron scattering process
\begin{align}
\label{eq:DY}
    h_1(P_1) + h_2(P_2) \longrightarrow L(q) + X\,,
\end{align}
where $h_{1,2}$ denote the colliding hadrons with momenta $P_{1,2}$, $L$ is a colorless final state with momentum $q^\mu$, and $X$ represents all unmeasured final-state particles.
Let $Q$, $Y$, and $\vec{q}_T$ denote the invariant mass, rapidity, and transverse momentum of $L$, respectively.
Then, in the limit $q_T = |\vec{q}_T| \ll Q$, the differential cross section factorizes as~\cite{Collins:1981uk,Collins:1981va,Collins:1984kg,Collins:2011zzd}
\begin{align}\label{eq:fact_thm_generic}
 \frac{d\sigma}{dQ^2\, dY\, d^2\vec{q}_T}
 = \sigma_0 \sum_{i,j} H_{ij}(Q, \dots)
 \int\!\frac{d^2\vec{b}_\perp}{(2\pi)^2}
 e^{i \vec{q}_T \cdot \vec{b}_\perp}
 f_{i/h_1}\bigl(x_1, \vec{b}_\perp, \dots\bigr)
 f_{j/h_2}\bigl(x_2, \vec{b}_\perp, \dots\bigr)
\,,
\end{align}
where $\sigma_0$ is the Born cross section, $H_{ij}$ is the hard coefficient function, and $f_{i/h_{1,2}}$ are the TMDPDFs associated with the two hadrons.
The indices $i,j$ run over the flavors of all contributing partons.
The struck partons carry longitudinal momentum fractions
$ x_{1,2} = \frac{Q e^{\pm Y}}{\sqrt{S}}$
where $S = (P_1 + P_2)^2$ is the hadronic center-of-mass energy squared.

In QCD, the TMD factorization theorem has been formulated in a variety of schemes, which mainly differ in how the so-called rapidity divergences are regulated~\cite{Collins:1981uk,Collins:1984kg,Collins:2011zzd,Beneke:2003pa,Ji:2004wu,Chiu:2007yn,Becher:2011dz,Chiu:2011qc,Chiu:2012ir,Chiu:2009yx,Echevarria:2011epo,Li:2016axz,Ebert:2018gsn,Ebert:2019okf}.
See Refs.~\cite{Ebert:2019okf,Collins:2017oxh,Boussarie:2023izj} for an overview.
After taking the lightcone (or infinite-rapidity) limit, most of these schemes become equivalent to the Collins scheme~\cite{Collins:2011zzd}, which is also found to be most closely connected to the LaMET approach~\cite{Ji:2019sxk,Ji:2019ewn,Ebert:2022fmh}.
Therefore, we adopt the TMDPDF definition in the Collins scheme~\cite{Ebert:2022fmh} as
\begin{align}
\label{eq:tmdpdf_Collins}
 f_{i/h}(x, \vec{b}_\perp, \mu, \zeta)
 &= \lim_{\epsilon\to0} Z_\text{uv}^R(\epsilon, \mu, \zeta)
   \lim_{y_B \to -\infty}
   \frac{B_{i/h}(x, \vec{b}_\perp, \epsilon, y_P - y_B)}
        {\sqrt{S^R(b_\perp, \epsilon, 2y_n, 2y_B)}}
\,,
\end{align}
where $B_{i/h}$ and $S^R$ denote the beam and soft functions, respectively. They encode the partonic structure of the incoming hadron and the associated soft-gluon radiation.
Here $y_P = \frac{1}{2}\ln(P^+/P^-)$, with $P^\pm = (P^t \pm P^z)/\sqrt{2}$, is the rapidity of the hadron, while $y_B$ is a regulator parameter for rapidity divergences.
The ultraviolet (UV) renormalization factor $Z_\text{uv}^R$ absorbs the $\epsilon$-poles arising from dimensional regularization in $d=4-2\epsilon$.
Both $Z_\text{uv}^R$ and $S^R$ depend on the color representation $R$ of the parton $i$ (with $R=q$ for quarks in the fundamental representation and $R=g$ for gluons in the adjoint representation), but are independent of the parton flavor.
The rapidity divergences (i.e., the $y_B$ dependence) cancel between the beam and soft functions, leading to a dependence on the Collins-Soper (CS) scale $\zeta = 2(xP^+)^2 e^{-2y_n}$, where $y_n$ is a finite parameter.
Consequently, in addition to the UV renormalization group equation, the TMDPDFs satisfy the rapidity evolution equation~\cite{Collins:1981uk,Collins:1984kg},
\begin{align}
    \frac{d}{d\ln \zeta} \ln f_{i/h}(x, \vec{b}_\perp, \mu, \zeta)
    &= -\frac{1}{2} \frac{d}{d y_n} \ln f_{i/h}(x, \vec{b}_\perp, \mu, \zeta)
    = \frac{1}{2} K^R(b_\perp,\mu)\,,
\end{align}
where the anomalous dimension $K^R(b_\perp,\mu)$ is known as the  kernel.
Owing to its dependence on \(b_\perp\), the  kernel becomes nonperturbative for \(b_\perp \gtrsim \Lambda_\text{QCD}^{-1}\), where \(\Lambda_\text{QCD}\) characterizes the QCD confinement scale, of order a few hundred MeV, below which perturbation theory in \(\alpha_s\) breaks down. 
The Collins beam and soft functions for quarks and gluons are defined as
\begin{align} \label{eq:def_Collins}
 B^\Gamma_{q/h}(x, \vec{b}_\perp, \epsilon, y_P - y_B)
 &= \int\!\frac{d b^-}{2\pi} \, e^{-i b^- (x P^+)}
   \Bigl\langle h(P) \Big| \bar q\Bigl(\tfrac{b}{2}\Bigr) \frac{\Gamma}{2}
   W^q\Big[\tfrac{b}{2}, -\tfrac{b}{2}; \mp \infty n_B(y_B)\Big]
   q\Bigl(-\tfrac{b}{2}\Bigr) \Big| h(P) \Bigr\rangle
\,, \nonumber\\
 B_{g/h}^{\rho\sigma}(x, \vec{b}_\perp, \epsilon, y_P - y_B)
 &= \int\!\frac{d b^-}{2\pi} \, \frac{e^{-i b^- (x P^+)}}{x P^+}
   \Bigl\langle h(P) \Big| F^{+\rho}\Bigl(\tfrac{b}{2}\Bigr)
   W^g\Big[\tfrac{b}{2}, -\tfrac{b}{2}; \mp \infty n_B(y_B)\Big]
   F^{+\sigma}\Bigl(-\tfrac{b}{2}\Bigr) \Big| h(P) \Bigr\rangle
\,, \nonumber\\
 S^R(b_\perp,\epsilon, y_A, y_B)
 &= \frac{1}{d_R}
    \Bigl\langle 0 \Big|
    \Tr \Bigl[
        W^{R \dagger}[b_\perp,0; -\infty n_A(y_A)]
        W^{R}[b_\perp,0; -\infty n_B(y_B)]
    \Bigr]
    \Big| 0 \Bigr\rangle
\,,
\end{align}
where the trace is taken over color indices, and the normalization factor $d_R$ is given by $d_q = N_c$ for quarks and $d_g = N_c^2 - 1$ for gluons.
The Wilson lines $W^{q,g}[a,b; \mp \infty n]$ represent a pair of semi-infinite gauge links extending along the direction $\mp n$, with color flowing from $b$ to $a$.
In lightcone coordinates $n^\mu = (n^+, n^-, n_\perp)$, the spacelike directions
\begin{align}\label{eq:nb}
 n_A^\mu(y_A) &\equiv \bigl(1, -e^{-2 y_A}, 0_\perp\bigr)
\,, \qquad
 n_B^\mu(y_B) \equiv \bigl(-e^{2 y_B}, 1, 0_\perp\bigr)
\,,
\end{align}
are parametrized by the rapidities $y_A > 0$ and $y_B < 0$.

The Dirac structure $\Gamma$ and the Lorentz indices $\rho,\sigma$, together with the polarization of the external hadron, define the eight leading-power (twist-2) quark and gluon TMDPDFs~\cite{Ralston:1979ys,Tangerman:1994eh,Boer:1997nt,Mulders:1995dh}.
In the quark case,
\begin{align} \label{eq:Gamma_LP}
 \Gamma \in \{\, \gamma^+ \,,\, \gamma^+\gamma_5 \,,\, i\sigma^{\alpha+}\gamma_5 \,\}
\,,
\end{align}
where $\sigma^{\mu\nu} = \frac{i}{2} [\gamma^\mu, \gamma^\nu]$.
Following the notation of Ref.~\cite{Bacchetta:2006tn}, the leading-power decomposition in momentum space reads
\begin{align} \label{eq:tmd_decomposition}
 f_{q/h_S}^{[\gamma^+]}(x,\vec{k}_\perp)
 &= f_1(x,k_\perp)
 - \frac{\epsilon_{\rho\sigma} k_\perp^\rho S_\perp^\sigma}{M} \, f_{1T}^\perp(x,k_\perp)
\,, \nonumber\\
 f_{q/h_S}^{[\gamma^+\gamma_5]}(x,\vec{k}_\perp)
 &= S_L \, g_{1L}(x,k_\perp)
 - \frac{k_\perp \cdot S_\perp}{M} \, g_{1T}(x,k_\perp)
\,, \nonumber\\
 f_{q/h_S}^{[i \sigma^{\alpha+}\gamma_5]}(x,\vec{k}_\perp)
 &= S_\perp^\alpha \, h_1(x,k_\perp)
 + \frac{S_L \, k_\perp^\alpha}{M} \, h_{1L}^\perp(x,k_\perp)
 \nonumber\\
 &\quad
 + \frac{k_\perp^2}{M^2}
   \biggl(\frac{1}{2} g_\perp^{\alpha\rho}
   - \frac{k_\perp^\alpha k_\perp^\rho}{k_\perp^2}\biggr)
   S_{\perp\,\rho} \, h_{1T}^\perp(x,k_\perp)
 - \frac{\epsilon^{\alpha\rho} k_{\perp\rho}}{M} \, h_1^\perp(x,k_\perp)
\,,
\end{align}
where $\epsilon^{\rho\sigma} = \epsilon^{+-\rho\sigma}$ with $\epsilon^{12} = -\epsilon^{21} = 1$, and $M$ denotes the hadron mass. The longitudinal and transverse spin vectors are given by
\begin{align}
 S_L^\mu &= (P^-, P^+, 0_\perp)/M
\,, \qquad
 S_\perp^\mu = (0, 0, \vec{S}_\perp)
\,,
\end{align}
with $S_L^2 = S_\perp^2 = -1$. Each function in Eq.~\eqref{eq:tmd_decomposition} corresponds to a specific choice of quark and hadron polarizations, as summarized in Table~\ref{tbl:tmds}.

Six of these distributions are time-reversal even ($\mathcal{T}$-even), namely the unpolarized $f_1$, helicity $g_{1L}$, transversity $h_1$, pretzelosity $h_{1T}^\perp$, and the worm-gear functions $g_{1T}$ and $h_{1L}^\perp$. The remaining two, the Sivers $f_{1T}^\perp$~\cite{Sivers:1989cc} and Boer-Mulders $h_1^\perp$~\cite{Boer:1997nt} functions, are $\mathcal{T}$-odd.
The spin-dependent gluon TMDPDFs admit a similar decomposition as in Eq.~\eqref{eq:tmd_decomposition} (see, e.g., Ref.~\cite{Boussarie:2023izj}), and a complete classification at twist-3 and twist-4 can be found in Refs.~\cite{Bacchetta:2004zf,Goeke:2005hb,Bacchetta:2006tn}.
A central goal of experimental programs is to extract the TMDPDFs, together with the  kernel, in the nonperturbative region $b_\perp \gtrsim \Lambda_\text{QCD}^{-1}$, thereby enabling a three-dimensional tomography of the nucleon in momentum space.
First-principles lattice QCD calculations of these quantities would provide essential benchmarks and guidance for phenomenological studies.
However, due to the real-time dependence of the TMD correlators in Eq.~\eqref{eq:def_Collins}, they cannot be directly simulated in Euclidean lattice QCD.
Although the Collins beam function at finite $y_B$ can be related to equal-time correlators (quasi-TMDs) in a large-momentum hadron state~\cite{Ebert:2022fmh}, taking the $y_B \to -\infty$ limit prior to UV renormalization (i.e., before the $a \to 0$ limit) remains prohibited in lattice calculations.
Furthermore, the soft function involves two time-like directions and cannot be mapped onto any equal-time matrix element accessible on the lattice.

\begin{table}[t]
 \centering
 \begin{tabular}{|*{5}{c|}}
 \cline{3-5}
 \multicolumn{2}{c|}{} & \multicolumn{3}{c|}{\textbf{Quark}}
 \\ \cline{3-5}
 \multicolumn{2}{c|}{} & \hspace{0.5cm}U\hspace{0.5cm} & \hspace{0.5cm}L\hspace{0.5cm} & \hspace{0.5cm}T\hspace{0.5cm}
 \\ \hline
 \multirow{3}{*}{\parbox{1.5ex}{\vspace{0.5cm}\rotatebox[origin=c]{90}{\textbf{Hadron}}}}
 & \parbox{1.5ex}{U} & $f_1$ (unpolarized) & & $h_1^\perp$ (Boer-Mulders)
 \\ \cline{2-5}
 & \parbox{1.5ex}{L} & & $g_{1L}$ (helicity) & $h_{1L}^\perp$ (worm-gear)
 \\ \cline{2-5}
 & \parbox{1.5ex}{\vspace{3ex}T} & $f_{1T}^\perp$ (Sivers) & $g_{1T}$ (worm-gear) & $h_{1}$ (transversity)
 \\ & &  &  & $h_{1T}^\perp$ (pretzelosity)
 \\ \hline
 \end{tabular}
 \caption{Leading-power quark TMDPDFs of a spin-$\frac12$ hadron in Eq.~\eqref{eq:tmd_decomposition}, with U, L, T standing for unpolarized, longitudinally polarized and transversely polarized, respectively.}
 \label{tbl:tmds}
\end{table}

\subsection{Moments from the Lorentz-Invariant Method} 

\begin{figure*}
  \centering
  \includegraphics[width=0.4\textwidth]{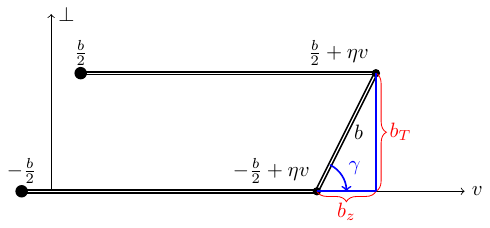}
  \caption{The staple-shaped Wilson line in the Lorentz-invariant approach~\cite{Musch:2010ka}. The black double lines extend along $\eta v$, while the blue segment along $b$ closes the staple. Image credit: Ref.~\cite{Ebert:2022fmh}.
}
  \label{fig:MHENS_staple}
\end{figure*}

The first lattice QCD studies of TMDs employed a Lorentz-invariant approach, as developed in Refs.~\cite{Hagler:2009mb,Musch:2010ka,Musch:2011er,Engelhardt:2015xja,Yoon:2015ocs,Yoon:2017qzo}.
In this framework, one starts from a general TMD correlator with staple-shaped Wilson lines,
\begin{align}\label{eq:beam_MHENS}
 \tilde\Phi^{[\Gamma]}(b, P, a, \eta v)
 &= \Bigl\langle h(P) \Big| \bar q\Bigl(\tfrac{b}{2}\Bigr) \frac{\Gamma}{2}
   W^q\Big[\tfrac{b}{2}, -\tfrac{b}{2}; \mp \eta v\Big]
   q\Bigl(-\tfrac{b}{2}\Bigr) \Big| h(P) \Bigr\rangle
\,,
\end{align}
where the Wilson lines extend along the $\mp v$ direction with finite length $\eta$, and are closed by an additional Wilson line connecting $-\tfrac{b}{2} \mp \eta v$ to $\tfrac{b}{2} \mp \eta v$, as illustrated in Fig.~\ref{fig:MHENS_staple}. 

The matrix element of this correlator is uniquely determined by the Lorentz vectors, $b^\mu$, $P^\mu$, and $\eta v^\mu$. The limits $v^2\to\infty$ and $\eta\to\infty$ correspond to the standard beam function definition. To make connections to the physical TMDs, Ref.~\cite{Musch:2011er} decomposes it
for generic choices of $b, P$ and $v$ as
\begin{align} \label{eq:MHENS_decomposition}
 \frac12 \tilde\Phi^{[\gamma^\mu]}(b, P, a, \eta v) &
 = P^\mu \tilde A_2\Bigl(b^2, b \cdot P, \frac{v \cdot b}{v \cdot P}, \frac{v^2}{(v \cdot P)^2}, \eta v \cdot P\Bigr)
 \nn\\&\quad
 + \frac{P^2 v^\mu}{v \cdot P} \tilde B_1 \Bigl(b^2, b \cdot P, \frac{v \cdot b}{v \cdot P}, \frac{v^2}{(v \cdot P)^2}, \eta v \cdot P\Bigr) + \ldots
\,,\end{align}
where ``$\ldots$'' elides other spin-dependent and higher-twist structures.
The full parametrization and similar decompositions for all $\Gamma$ can be found in Ref.~\cite{Musch:2011er}.

The scalar amplitudes $\tilde A_2$ and $\tilde B_1$ depend on all six Lorentz invariants formed by $b^\mu$, $P^\mu$ and $\eta v^\mu$, as well as the hadron mass $P^2 = m_h^2$ and nonperturbative QCD scale $\Lambda_\text{QCD}$. In the lightcone beam function, $b^+ = 0$ and $v_\perp = P_\perp = 0$, so the third argument of $\tilde A_2$ and $\tilde B_1$ in Eq.~(\ref{eq:MHENS_decomposition}) is not independent and can be expressed in a Lorentz-invariant fashion~\cite{Musch:2011er},
\begin{align} \label{eq:MHENS_constraint}
 \frac{v \cdot b}{v \cdot P}
 = \frac{b \cdot P}{P^2} R\bigl(\hat \zeta^2\bigr)
\,,\quad
 R\bigl(\hat \zeta^2\bigr) = 1 - \sqrt{1 + \hat \zeta^{-2}}
\,.\end{align}
As a result, on the lattice one must choose $b$ and $v$
such that Eq.~(\ref{eq:MHENS_constraint}) is fulfilled.
The --like parameter $\hat\zeta$ entering Eq.~(\ref{eq:MHENS_constraint}) is defined as
\begin{align} \label{eq:hatzeta}
 \hat \zeta =  \frac{v \cdot P}{\sqrt{|v^2| P^2}}
\,.\end{align}
If $P$ and $v$ are parameterized by their rapidities, i.e.,
\begin{align}
 P^\mu = \frac{m_h}{\sqrt 2} (e^{y_P}, e^{-y_P}, 0_\perp)
\,,\quad
v^\mu \propto (e^{y_v}, -e^{- y_v}, 0_\perp)
\,.\end{align}
then
\begin{align}
 \hat\zeta = \sinh(y_v - y_P)
\end{align}
essentially is a rapidity difference, even though $v^2<0$. 
The physical TMDs correspond to the large rapidity limit $\hat\zeta\to \infty$, equivalent to $v\cdot P\to \infty$ with $P^2$ fixed.

For the unpolarized TMD,  $\Gamma = \gamma^+$, leading to the beam function
\begin{align}\label{eq:beam_MHENS_2}
 \frac12 B_{q/h}^{[\gamma^+]}(x, \vec{b}_\perp, P, a, \eta v) &
 = \int\frac{d (b {\cdot} P)}{2 \pi} e^{-i x (P {\cdot} b)}
   \tilde A_{2B}\Bigl(b^2, b {\cdot} P, \frac{b {\cdot} P}{P^2} R\bigl(\hat \zeta^2\bigr), \frac{v^2}{(v {\cdot} P)^2}, \eta v {\cdot} P\Bigr)\,,
\end{align}
where $\tilde A_{2B} = \tilde A_{2} + R(\hat\zeta^2) \tilde B_1$.
The other leading-twist spin-dependent beam functions can be obtained in a similar fashion. Thanks to Lorentz invariance, $\tilde A_{2B}$ can be evaluated in
any frame so long as Eq.~\eqref{eq:MHENS_constraint} is satisfied. Therefore, one can pick a time-independent frame where $b^0 = v^0 = 0$ so that $\tilde A_{2B}$ is directly calculable on the lattice.

A soft function subtraction is necessary to relate the above beam function to the lightcone TMDPDF. In addition, since the lattice beam function depends dynamically on the scale $\sqrt{|{(v {\cdot} P)^2}/{v^2}|}$, the $\hat{\zeta}\to\infty$ or $v\cdot P\to \infty$ limit cannot be analytically taken. Instead, a $v\cdot P$-dependent perturbative matching is expected to compensate for the difference between finite $v\cdot P$ and the $v\cdot P\to\infty$ limit.

Nevertheless, at the lowest order in perturbation theory, the $\langle x^n\rangle$ moments of lightcone TMDs can be related to the $b\cdot P=0$ matrix elements and their derivatives with respect to $b\cdot P$, up to the soft function subtraction. Then by forming ratios of different spin-dependent matrix elements, one can cancel the soft function and directly calculate the $b_\perp$-dependence of their lightcone counterparts on the lattice. This approach has been used to calculate the ratios of the moments of spin-dependent TMDs to those of the unpolarized TMD~\cite{Musch:2010ka,Musch:2011er,Engelhardt:2015xja,Yoon:2015ocs,Yoon:2017qzo}. It has also been extended to the off-forward matrix elements of the TMD correlators to extract the quark canonical orbital angular momentum on the lattice~\cite{Engelhardt:2017miy,Engelhardt:2020qtg}. See Ref.~\cite{Boussarie:2023izj} for a review of lattice results from the Lorentz-invariant approach.

\subsection{Large-Momentum Effective Theory} 

While the Lorentz-invariant approach provides valuable qualitative insights into TMD structure through ratios of moments, it does not offer a systematic pathway to reconstruct the full kinematic dependence of TMDs in the $(x,k_\perp)$ or $(x,b_\perp)$ space.
This limitation arises from the fact that the underlying lattice observables are defined in Euclidean spacetime and are not directly connected to the lightcone correlations that define physical parton distributions, and that a soft-factor subtraction is necessary for reproducing the correct IR physics in physical TMDs.
A key challenge in accessing TMDPDFs, therefore, is to develop a framework that preserves the IR physics of QCD while reformulating the problem in terms of equal-time correlators that are computable on the lattice.
This requirement motivates the introduction of LaMET~\cite{Ji:2013dva,Ji:2014gla,Ji:2020ect}, which provides a rigorous approach to relate Euclidean lattice matrix elements to lightcone observables through a systematic expansion in the inverse hadron momentum.
The essential idea of LaMET is to compute quasi-distributions defined from spatial correlations in a hadron boosted to large but finite momentum and then match them perturbatively to the corresponding lightcone distributions.
Because quasi-observables share the same IR structure as their physical counterparts, the differences between them are purely ultraviolet and can be accounted for through factorization and matching procedures.
This framework has significantly advanced the lattice-QCD program by enabling direct calculation of the $x$-dependence of parton distributions, including TMDPDFs, and by providing a unified treatment of renormalization, rapidity evolution, and soft-factor subtraction.
In the following subsections, we review the formulation of LaMET for collinear PDFs and its extension to TMD observables, highlighting the key theoretical ingredients and lattice implementation strategies that enable first-principles calculations of transverse-momentum--dependent quantities.

\subsubsection{LaMET for PDFs}

LaMET is a framework developed to calculate parton physics from lattice QCD.
In QCD, parton distribution functions  are defined from lightcone correlations, which are inaccessible in Euclidean lattice calculations.
LaMET circumvents this by relating equal-time spatial correlations in a boosted hadron to lightcone PDFs via effective field theory matching and power corrections.
It builds on the observation that the structure of a hadron at large momentum becomes approximately independent of its momentum, allowing for a systematic expansion in powers of $\Lambda_{\text{QCD}}/P$~\cite{Ji:2020byp,Ji:2022ezo,Ji:2024oka}.
This leads to the concept of quasi-PDFs, which are calculable on the lattice and matched to standard PDFs using perturbative techniques.
The framework has enabled significant progress in computing PDFs, lightcone distribution amplitudes,  generalized PDFs (GPDs), TMDs, and other partonic observables directly from first principles.
This review summarizes the theoretical foundations, key developments, and applications of LaMET.

The parton model, introduced by Feynman, describes hadrons as composed of noninteracting partons in the infinite-momentum frame (IMF).
PDFs are defined via lightcone correlations, such as for the unpolarized quark case, 
\begin{equation}\label{eq:def_quasi}
q(x) = \frac{1}{2} \int \frac{d\xi^-}{2\pi} e^{-ixP^+\xi^-} \langle P | \bar{\psi}(\xi^-) \gamma^+ W( \xi^-,0) \psi(0) | P \rangle\,,
\end{equation}
where $W$ is a Wilson line along the lightcone direction $n^\mu=({1\over \sqrt{2}},0,0,-{1\over \sqrt{2}})$. 
However, lattice QCD operates in Euclidean space, making direct access to lightcone dynamics impossible.
This has historically limited lattice calculations to moments of PDFs~\cite{Kronfeld:1984zv,Martinelli:1987si}, which are insufficient to reconstruct the full $x$-dependence. 

LaMET offers a new path by exploiting the approximate momentum-independence of hadron structure at large $P^z$. To start with, one can construct a lattice-calculable quasi-PDF from a universality class~\cite{Ji:2013fga,Ji:2020ect} of candidates which have the same infinite-momentum limit as the lightcone PDF. For example, the first and most studied definition is given for the unpolarized quark case by~\cite{Ji:2013dva}
\begin{equation}
\tilde{q}(y, P^z) = N_\Gamma \int \frac{dz}{4\pi} e^{iyP^z z} \langle P | \bar{\psi}(z) \Gamma W(z,0) \psi(0) | P \rangle\,,
\end{equation}
where $W(z,0)$ is a spatial Wilson line along the $z$-direction, and $\Gamma=\gamma^t$ or $\gamma^z$ with $N_{\gamma^t}=P^z/P^t$ and $N_{\gamma^z}=1$. Other choices include quasi-PDFs defined from Coulomb-gauge correlators~\cite{Gao:2023lny,Zhao:2023ptv} and current-current correlators~\cite{Zhang:2026lle}. Unlike the lightcone PDF, the quasi-PDF has support over the entire range $-\infty<y<\infty$, although contributions outside the physical region $-1<y<1$ are typically suppressed and fall off rapidly.

At $P^z\gg\Lambda_\text{QCD}$, the lightcone PDF can be calculated from the quasi-PDF through a large-$P^z$ expansion and perturbative matching at leading power as~\cite{Ji:2020ect,Ji:2020byp,Ji:2022ezo,Ji:2024oka}
\begin{align}\label{eq:lamet-fact}
  q(x, \mu) &= \int_{-\infty}^{\infty} \frac{dy}{|y|}  C\left(\frac{x}{y}, \frac{\mu}{xP^z}, {\tilde\mu \over \mu}\right)  \tilde q(y,P^z,\tilde\mu)  + {\cal O}\bigg(\frac{\Lambda_{\text{QCD}}^2}{(xP^z)^2},
  \frac{\Lambda_{\text{QCD}}^2}{((1-x)P^z)^2}
  \bigg)\,,
\end{align}
where $\tilde\mu$ is the renormalization scale of the quasi-PDF, and $C$ is the matching coefficient which has been calculated up to three-loop order for the nonsinglet quark case~\cite{Xiong:2013bka,Izubuchi:2018srq,Li:2020xml,Chen:2020ody,Cheng:2024wyu} and one-loop order for the singlet-quark and gluon channels~\cite{Wang:2019tgg,Balitsky:2019krf,Balitsky:2021qsr,Balitsky:2021cwr,Yao:2022vtp}. 
The matching coefficient $C$ contains logarithms of the ratio of the $\overline{\text{MS}}$ scale $\mu$ to the hard parton momentum $xP^z$; these logarithms become large as $x$ approaches zero.
Here, $\overline{\text{MS}}$ denotes the modified minimal-subtraction renormalization scheme commonly used in perturbative QCD.
The power corrections are suppressed by the hard parton and spectator momenta $xP^z$ and $(1-x)P^z$, respectively, so LaMET can make reliable predictions for PDFs with a moderate range of $x$ at a given finite $P^z$.

The matching/factorization relation between the quasi- and lightcone PDFs at   all orders in perturbation theory is proven using leading-region analysis of Feynman diagrams~\cite{Libby:1978qf,Curci:1980uw,Collins:2011zzd} in Refs.~\cite{Ma:2014jla,Ji:2020ect}, where it is shown that all the collinear divergences in the quasi-PDF factorize and can be absorbed into the lightcone PDF. This relation was also derived using the operator product expansion (OPE) of the Wilson-line operator in coordinate space~\cite{Izubuchi:2018srq,Ma:2017pxb,Wang:2019tgg}. Recently, this factorization relation was derived with Soft-Collinear Effective Theory (SCET)~\cite{Holligan:2025baj}.

When taking into account the mixing between gluon and singlet-quark channels, the full LaMET matching formulas for the flavor-separated PDFs are~\cite{Wang:2019tgg}
\begin{align}
q_i(x,\mu) &= \int_{-\infty}^{\infty} \frac{dy}{y} \Bigg[ \sum_j \tilde C_{q_iq_j}\!\left(\frac{x}{y},\frac{\mu}{xP^z}\right)\tilde q_j(y,P^z,\mu)
+ \tilde C_{qg}\!\left(\frac{x}{y},\frac{\mu}{xP^z}\right)\tilde g(x,P^z,\mu)\Bigg] + \cdots,
\label{eq:factorization1}\\
g(x,\mu) &= \int_{-\infty}^{\infty} \frac{dy}{y} \Bigg[ \sum_j \tilde C_{gq}\!\left(\frac{x}{y},\frac{\mu}{xP^z}\right)\tilde q_j(y,P^z,\mu)
+ \tilde C_{gg}\!\left(\frac{x}{y},\frac{\mu}{xP^z}\right)\tilde g(y,P^z,\mu)\Bigg] + \cdots,
\label{eq:factorization2}
\end{align}
where $i,j$ run over quark flavors. The ellipsis denotes mass corrections of all orders in $M^2/(P^z)^2$~\cite{Chen:2016utp} and dynamical power corrections of $\mathcal{O}(\Lambda_{\text{QCD}}^2/(xP^z)^2,\Lambda_{\text{QCD}}^2/((1-x)P^z)^2)$. 

Apart from LaMET, several other approaches have been proposed to compute PDFs on the lattice, including pseudo-PDFs~\cite{Radyushkin:2017cyf}, current-current correlators~\cite{Braun:2007wv,Ma:2017pxb}, the operator product expansion  of Compton amplitudes of light~\cite{Chambers:2017dov} or heavy-light quark currents~\cite{Detmold:2005gg,Detmold:2021uru}, and the hadronic tensor approach~\cite{Liu:1993cv}.
There have also been proposals to extend the direct computation of Mellin moments beyond the lowest orders using smearing~\cite{Davoudi:2012ya} and gradient-flow~\cite{Shindler:2023xpd} techniques.
Among these, the pseudo-PDF approach is closely related to LaMET, as it is based on the same equal-time Wilson-line operator in Eq.~\eqref{eq:def_quasi} and performs a short-distance expansion in coordinate space at small $|z|$~\cite{Ji:2017rah,Radyushkin:2017lvu,Balitsky:2019krf}.
This allows one to reconstruct the lightcone correlation function over a finite range characterized by $\lambda = z_{\max} P^z_{\max}$, with $z_{\max} \ll \Lambda_\text{QCD}^{-1}$.
In practical applications, this finite segment of the correlation function can be used either to extract the lowest Mellin moments via an OPE~\cite{Izubuchi:2018srq} or to fit the $x$-dependence of the PDFs.
Since LaMET provides reliable access at intermediate values of $x$, it was proposed in Ref.~\cite{Ji:2022ezo} that it be complemented by short-distance expansions to better constrain the endpoint behavior of PDFs, thereby enabling coverage of the full kinematic range.

Based on the expansion formula in Eq.~\eqref{eq:factorization2}, the LaMET analysis of PDFs can be organized into the following steps:
\begin{itemize}
\item {\bf Renormalization.} The nonlocal Wilson-line operator in Eq.~\eqref{eq:def_quasi} is multiplicatively renormalizable in coordinate space but contains a linear power divergence of the form $e^{-\delta m(a)|z|}$, with $\delta m(a)\sim 1/a$ on a lattice with spacing $a$~\cite{Ji:2017oey,Ishikawa:2017faj,Green:2017xeu,Zhang:2018diq,Li:2018tpe}.
To ensure a well-defined continuum limit, this linear divergence must be subtracted to all orders in perturbation theory.
Early implementations employed regularization-independent momentum-subtraction (RI/MOM) schemes~\cite{Chen:2017mzz,Constantinou:2017sej,Stewart:2017tvs,Alexandrou:2017huk}, but they introduce uncontrolled nonperturbative contributions at large $|z|$.
In modern LaMET analyses, a hybrid renormalization scheme is typically adopted~\cite{Ji:2020brr}, which combines a ratio scheme~\cite{Orginos:2017kos,Braun:2018brg,Li:2020xml,Fan:2020nzz} or RI/MOM renormalization~\cite{Chou:2022drv} at short distances with a Wilson-line mass subtraction at long distances.
The Wilson-line mass can be determined either from the static quark potential~\cite{Gao:2021dbh} or via a self-renormalization procedure~\cite{LatticePartonLPC:2021gpi}.
However, it suffers from a linear renormalon ambiguity associated with the asymptotic behavior of the perturbative series in the short-distance $\overline{\text{MS}}$ Wilson coefficients.
This ${\cal O}(\Lambda_\text{QCD})$ ambiguity must be properly regulated and subtracted in order to match lattice-renormalized matrix elements to continuum schemes.
To address this issue, the leading renormalon resummation (LRR) approach was developed~\cite{Holligan:2023rex,Zhang:2023bxs}. This method improves the convergence of perturbation theory and enhances the power accuracy from $O(\Lambda_\text{QCD}/P^z)$ to $O(\Lambda_\text{QCD}^2/(P^z)^2)$.
\item {\bf Fourier transform.} Since lattice calculations are performed in finite volumes, and the noise-to-signal ratio of renormalized matrix elements in Eq.~\eqref{eq:def_quasi} grows exponentially with $z$, the Fourier integral is typically truncated at $z_L \sim 1.0\text{ fm}$, which requires an extrapolation $|z|\to\infty$.
This issue has been the subject of extensive recent discussions, particularly regarding how to perform the Fourier transform and quantify the associated uncertainties~\cite{Dutrieux:2025jed,Chen:2025cxr,Xiong:2025obq,Chu:2025jsi,Dutrieux:2025axb,Ling:2025olz}. The analyticity of Green functions in quantum field theory implies that the correlation functions are smooth, which ensures that the Fourier integral can be computed with controlled precision with a finite-element approach. Then, due to color confinement, spatial correlation functions decay exponentially at large $|z|$, a behavior that can be rigorously derived in quantum field theory~\cite{Gao:2021dbh,Ji:2026vir}. This property enables a physics-motivated extrapolation $|z|\to\infty$ using the asymptotic form of the correlation functions~\cite{Chen:2025cxr}.
The associated systematic uncertainties can then be estimated from the truncation error of the asymptotic expansion~\cite{Ji:2026vir,Gao:2026wlz}. This approach has also been cross-validated by a direct lattice simulation of the quasi-distribution in momentum space~\cite{Grebe:2026qmt}.
\item {\bf Perturbative matching.} In addition to higher-order corrections~\cite{Li:2020xml,Chen:2020ody,Cheng:2024wyu,Ji:2025mvk}, significant progress in the matching has been achieved through the development of resummation techniques~\cite{Gao:2021hxl,Su:2022fiu,Ji:2023pba,Ji:2024hit}.
As discussed above, the matching kernel in Eq.~\eqref{eq:factorization1} contains large logarithms of $xP^z/\mu$ in the limit $x\to 1$. These logarithms obey the DGLAP evolution equation and can be resummed using standard methods~\cite{Su:2022fiu}.
On the other hand, in the threshold region $x\to 1$, there also appear enhanced threshold logarithms, whose resummation has been developed both in coordinate space~\cite{Gao:2021hxl} and in momentum space~\cite{Ji:2023pba,Liu:2023onm,Ji:2024hit}.
After resumming both small- and large-$x$ logarithms, as well as applying the LRR, the perturbative series exhibits improved convergence from next-to-leading order (NLO) to next-to-next-to-leading order (NNLO), together with a substantial reduction in renormalization- and factorization-scale uncertainties~\cite{Ji:2024hit}.
Moreover, the breakdown of perturbation theory becomes increasingly evident as $x$ approaches the endpoint regions, providing a quantitative criterion for the range of $x$ in which LaMET remains under systematic control. This region is typically given by $x\in[x_0,1-x_0]$, where $x_0\sim\Lambda_\text{QCD}/P^z$.
\item \textbf{Infinite-momentum and continuum extrapolations.}
After perturbative matching, residual power corrections can be estimated via an infinite-momentum extrapolation.
Convergence of the final results in the moderate-$x$ region provides a clear indication that the LaMET predictions are under systematic control.
However, since lattice discretization effects may also exhibit polynomial dependence on $P^z$, the limit $P^z \to \infty$ should be performed together with a continuum extrapolation.
\end{itemize}

Apart from the analyses discussed above, residual lattice systematics, including excited-state contamination, unphysical quark masses, and finite-volume effects, can also impact the precision of LaMET calculations.
While the latter two sources of uncertainty can be systematically controlled through calculations at physical quark masses and/or chiral extrapolations, together with extrapolations to the infinite-volume limit, excited-state contamination remains a more challenging issue.
In particular, it has been observed to be sizable in the imaginary parts of matrix elements~\cite{Alexandrou:2020qtt,Gao:2026hix} and thus warrants further investigation~\cite{Barca:2024hrl,Barca:2025det}.

\begin{figure}[thbp]
\includegraphics[width=.33\textwidth]{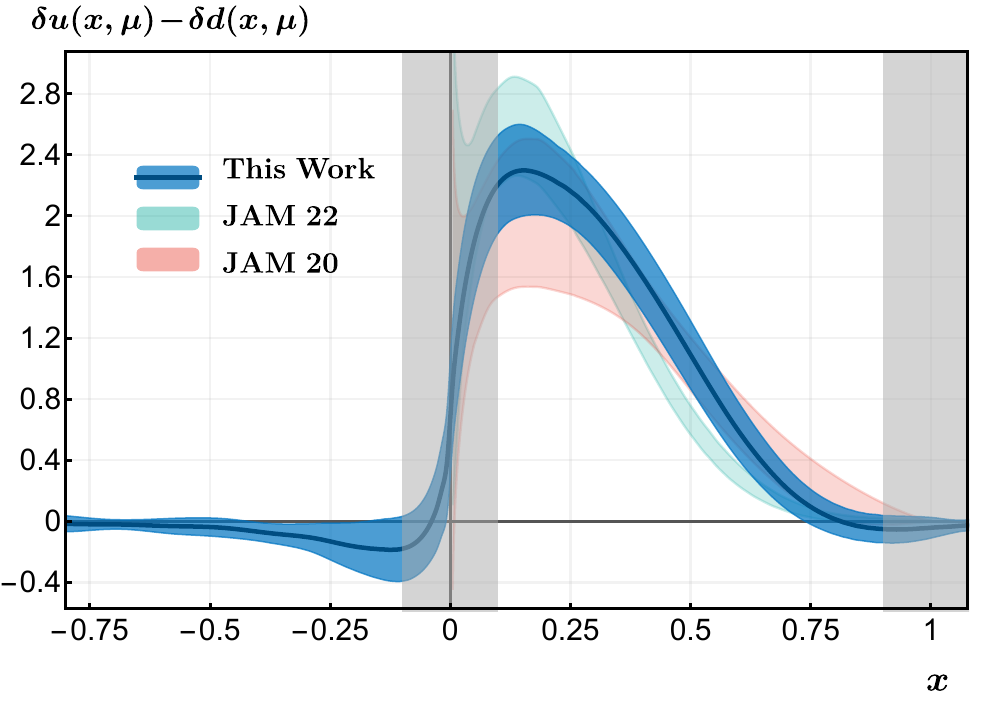}
\includegraphics[width=.32\textwidth]{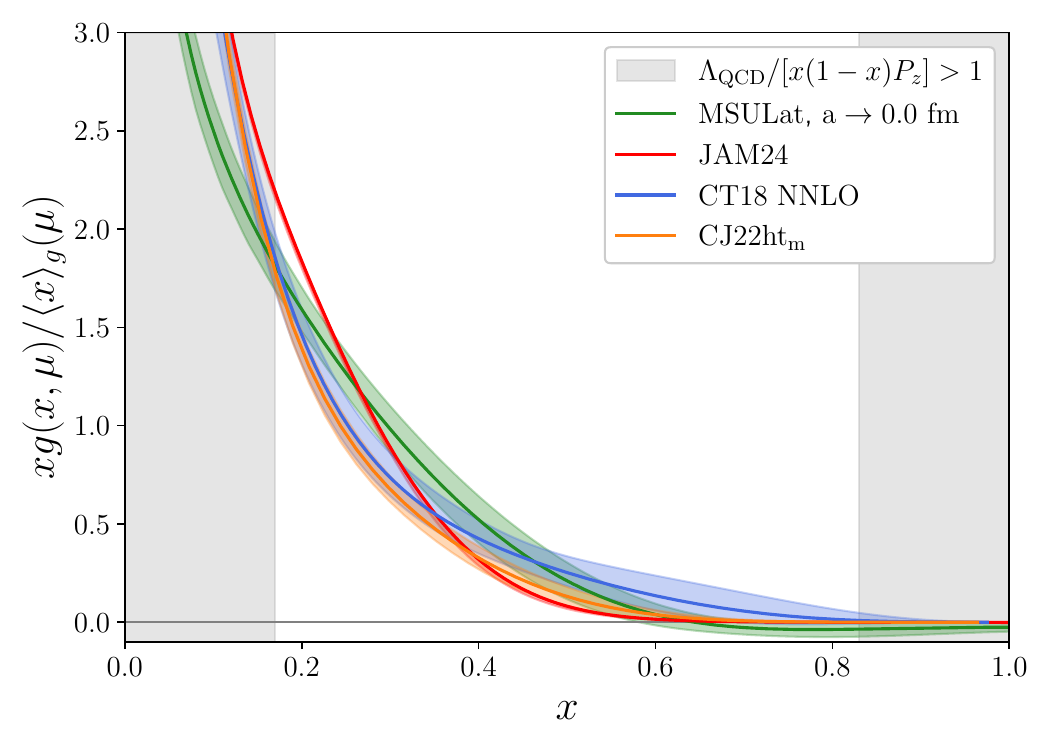}
\includegraphics[width=.33\textwidth]{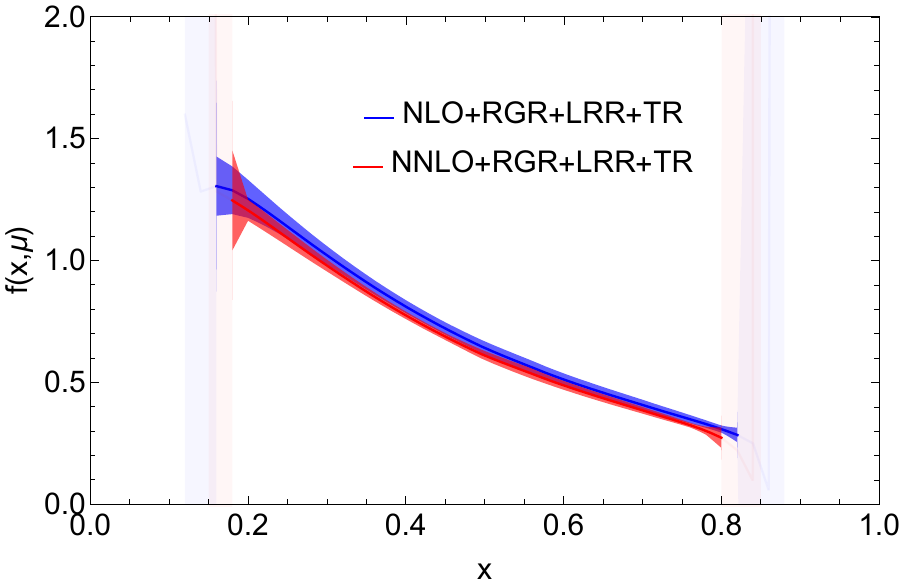}
\caption{\label{fig:SelectedPDFs}
(left) LPC's proton isovector transversity PDF, normalized by the nucleon isovector tensor charge $g_T$~\cite{LatticeParton:2022xsd}, extrapolated to the continuum, physical pion mass, and infinite-momentum limits ($a\to0$, $m_\pi\to m_{\pi,\text{phys}}$, $P_z\to\infty$), compared with the corresponding results from the JAM20~\cite{Cammarota:2020qcw} and JAM22~\cite{Gamberg:2022kdb} global analyses.
(middle) The MSULat unpolarized gluon PDF~\cite{NieMiera:2025mwj} at $P_z = 2.12$~GeV, compared with selected phenomenological determinations.
(right) The state-of-the-art pion valence-quark PDF at $P^z = 1.94\text{ GeV}$, including all perturbative improvements (LRR, RGR, and TR) up to NNLO~\cite{Ji:2024hit}.
All results are presented at the renormalization scale $\mu = 2$~GeV.
The vertical light-gray bands indicate the endpoint regions where LaMET predictions are not reliable.
}
\end{figure}

LaMET has been successfully and widely applied to compute flavor nonsinglet quark distributions (e.g., the $u-d$ distribution of the proton and the valence-quark distribution of the pion), as well as helicity and transversity PDFs and GPDs. Gluon PDFs, previously inaccessible, are now also within reach using LaMET. Figure~\ref{fig:SelectedPDFs} shows several representative state-of-the-art PDF results obtained from LaMET calculations.
LPC's nucleon isovector transversity PDF~\cite{LatticeParton:2022xsd} is computed using four lattice spacings $a \in [0.049, 0.098] \text{ fm}$ and pion masses $m_\pi \in [220, 350]\text{ MeV}$, with proton momenta up to $2.8\text{ GeV}$.
The calculation employs nonperturbative renormalization in the hybrid scheme with self-renormalization and performs controlled extrapolations to the continuum, physical pion mass, and infinite-momentum limits.
The resulting PDF is compared with recent global analyses from the JAM collaboration and exhibits smaller uncertainties, providing valuable guidance for future measurements at JLab and the EIC.
The middle panel of Fig.~\ref{fig:SelectedPDFs} shows the first lattice-QCD determination of the nucleon gluon PDF within the LaMET framework, obtained using the hybrid scheme with self-renormalization in the continuum limit.
High-statistics calculations are performed with boost momenta in the range $2.0$--$2.2\text{ GeV}$ on three lattice spacings $a\in[0.09,0.15]\text{ fm}$, at a pion mass of $310\,\mathrm{MeV}$ with $N_f=2+1+1$ quark flavors.
The renormalized gluon matrix elements and resulting PDFs show consistent behavior across lattice spacings, with the finest lattice ($a=0.09\text{ fm}$) statistically compatible with the continuum extrapolation. The extracted gluon PDF agrees reasonably well with selected global-fit determinations, particularly favoring analyses with near-zero gluon density in the large-$x$ region ($x>0.6$). Independent LPC calculations~\cite{ChenChen:2025amm}, using different ensembles and lattice spacings, report similarly encouraging results.
The right panel of Fig.~\ref{fig:SelectedPDFs} presents a determination of the pion valence-quark PDF on a superfine lattice ensemble ($a=0.04$~fm, $m_\pi=300$~MeV)~\cite{Gao:2020ito} with state-of-the-art perturbative improvements~\cite{Ji:2024hit}, including small- and large-$x$ resummation, LRR, and NNLO matching. Within the range of LaMET control, the NNLO results show good perturbative convergence and significantly reduced scale-variation uncertainties.

These examples illustrate the current generation of LaMET calculations, which typically employ multiple lattice spacings to enable reliable continuum extrapolations and/or systematic improvements in perturbative matching.
The field continues to evolve rapidly, with ongoing efforts aimed at further reducing both statistical and systematic uncertainties.
Since the predictive power of LaMET is rooted in the convergence of the large-momentum expansion and perturbative matching, both of which are independent of global fitting procedures, it can be directly tested in lattice QCD calculations and benchmarked against phenomenology.
Future directions will focus on quantifying the remaining lattice systematics using multiple ensembles, improving the extraction of hadronic matrix elements, and incorporating state-of-the-art LaMET analyses.

\subsubsection{LaMET for TMDPDFs}
\label{subsec:lamet-tmd}

Transverse-momentum-dependent parton distributions generalize collinear PDFs by incorporating both longitudinal momentum fraction $x$ and transverse momentum $\vec{k}_\perp$ of partons~\cite{Collins:1981uk,Collins:1984kg}.
They describe the probability of finding a parton of a given flavor, momentum, and polarization inside a hadron. TMDPDFs are essential for processes where small transverse momenta are measured, such as Drell-Yan, SIDIS, and electroweak boson production.
In these cases, the differential cross section $d\sigma/dQ_T^2$ peaks at low $Q_T$, where nonperturbative effects dominate~\cite{Collins:1984kg}.
A precise understanding of TMDPDFs is therefore critical for QCD factorization and precision phenomenology at current and future colliders.

Historically, TMDPDFs were introduced to explain transverse-momentum spectra in DY and $e^+e^-$ annihilation~\cite{Collins:1981va,Collins:1982wa}.
Later, factorization was formalized in Soft-Collinear Effective Theory~\cite{Bauer:2000yr,Becher:2010tm,Echevarria:2012js}, making explicit the role of collinear and soft modes. Modern formulations define scheme-independent TMDPDFs that separate collinear and soft contributions~\cite{Collins:2012uy}.
These distributions are sensitive to spin-dependent effects, such as the Sivers and Boer-Mulders functions, which encode correlations between parton transverse momentum and hadron spin~\cite{Boer:1997nt,Sivers:1989cc}.
Such functions exhibit process-dependent sign changes between DY and SIDIS~\cite{Collins:2011zzd}, providing stringent tests of QCD dynamics.

Beyond phenomenology, TMDPDFs enable a three-dimensional tomography of hadrons, combining longitudinal and transverse structure~\cite{Burkardt:2000za,Boer:2011fh}.
They also play a role in small-$x$ physics and gluon saturation studies~\cite{Kuraev:1977fs,Kovchegov:2012mbw}.
Current knowledge relies heavily on global fits to limited experimental data~\cite{Landry:1999an,Konychev:2005iy,Bacchetta:2017gcc}, motivating first-principles approaches for their determination. Lattice QCD combined with LaMET offers a systematic framework to compute TMDPDFs nonperturbatively~\cite{Ji:2014hxa,Ji:2018hvs,Ebert:2018gzl,Ebert:2019okf,Ebert:2019tvc,Ji:2019sxk,Ji:2019ewn,Vladimirov:2020ofp,Ebert:2020gxr,Ji:2020jeb,Ji:2021znw,Ebert:2022fmh,Schindler:2022eva,Deng:2022gzi,Zhu:2022bja,delRio:2023pse,Ji:2023pba,Ji:2024hit,Liu:2023onm,Xie:2025rrw}. 
Recent progress includes establishing the LaMET matching relation between the quasi- and physical TMDPDFs~\cite{Ebert:2019okf,Ebert:2022fmh,Ji:2019sxk,Ji:2019ewn}, and developing methods to calculate the  kernel~\cite{Ji:2014hxa,Ebert:2018gzl} and the soft factor~\cite{Ji:2019sxk}.
These developments pave the way for lattice-based determinations of TMD observables relevant to the Electron-Ion Collider programs.

TMDPDFs relevant for high-energy processes are defined using light-like Wilson lines to ensure gauge invariance~\cite{Belitsky:2002sm}.
These operators arise naturally in effective field theory descriptions such as SCET, where the infinite-momentum limit introduces additional divergences requiring special treatment.
For illustration, consider the unpolarized quark TMDPDF:
\begin{align}\label{eq:beam}
f(x,\vec{k}_\perp) &= \frac{1}{2}\int\frac{d\xi^-}{2\pi}\frac{d^2\vec{b}_\perp}{(2\pi)^2}
e^{-i x P^+\xi^-+i\vec{k}_\perp\cdot\vec{b}_\perp} 
\left \langle P|\bar\psi \left({\xi^-\over2},\vec{b}_\perp \right)\gamma^+{\cal W}_n\left({\xi^-\over2},\vec{b}_\perp;-{\xi^-\over2},0\right)\psi\left(-{\xi^-\over2},0\right)|P\right \rangle,
\end{align}
where ${\cal W}_n$ denotes a staple-shaped gauge link along the lightcone direction:
\begin{align}\label{eq:staplen}
{\cal W}_n(\xi;\xi')=W^\dagger_n(\xi)W_\perp W_n(\xi'), \qquad
W_n(\xi)= {\cal P}\exp\Big[-ig\int_0^{-\infty} d\lambda\, n\cdot A(\xi+\lambda n)\Big].
\end{align}

The presence of infinitely long light-like gauge links introduces rapidity divergences that cannot be regulated by standard UV methods.
These divergences originate from gluon radiation collinear to the gauge link and manifest as unregulated integrals over rapidity variables.
For example, in dimensional regularization one encounters:
\begin{equation}\label{eq:RD}
I = \int dk^+dk^- \frac{f(k^+k^-)}{(k^+k^-)^{1+\epsilon}} = \frac{1}{2}\int \frac{dy}{y}\int dm^2 \frac{f(m^2)}{m^{2+2\epsilon}},
\end{equation}
where $y=k^+/k^-$ is the rapidity parameter. Divergences at large and small $y$ require dedicated regulators.

Several schemes have been developed to handle rapidity divergences~\cite{Ebert:2019okf}. On-lightcone regulators, such as the $\delta$ regulator~\cite{Echevarria:2015usa}, modify Wilson lines by introducing exponential damping:
\begin{align}
W_n(\xi)\rightarrow W_{n,\delta^-}(\xi)= {\cal P}\exp\Big[-ig\int_0^{-\infty} d\lambda\, A^+(\xi+\lambda n)e^{-\frac{\delta^-}{2p^+}\lambda}\Big].
\end{align}
Off-lightcone regulators deform the gauge link direction into space-like or time-like regions~\cite{Collins:2011zzd}, introducing a parameter $Y$ that acts as a rapidity cutoff:
\begin{align}
n\rightarrow n_Y=n-e^{-2Y}\frac{p}{(p^+)^2}.
\end{align}
These methods preserve boost invariance and allow consistent definitions of TMDPDFs for lattice calculations.
In practice, rapidity regularization is combined with soft function subtraction to obtain scheme-independent physical TMDPDFs, as discussed in the next subsection.

Before LaMET, lattice studies of TMD physics focused on ratios of $x$-moments of TMDPDFs~\cite{Hagler:2009mb,Musch:2010ka}, which avoid soft-function complications but do not provide full $x$ and $\vec{k}_\perp$ dependence.
LaMET enables direct access to TMDPDFs by computing quasi-TMDPDFs at large hadron momentum and matching them to physical distributions~\cite{Ebert:2019okf,Ebert:2022fmh,Ji:2019sxk,Ji:2019ewn}.
A proper treatment of soft-function subtraction and perturbative matching is essential for consistency.
Early proposals using bent or rectangular Wilson loops captured correct IR behavior only at one loop~\cite{Ji:2018hvs,Ebert:2019okf}, but failed at higher orders. Recent work established that quasi-TMDPDFs combined with a reduced soft function reproduce the correct IR physics to all orders~\cite{Ji:2019sxk,Ji:2019ewn}.

The quasi-TMDPDF is defined with a staple-shaped gauge link along the spatial $z$ direction:
\begin{align}\label{eq:quasi_TMD}
\tilde f(z,b_\perp,\mu,\zeta_z)=\lim_{L\to\infty}\frac{\langle P|\bar\psi(\tfrac{z}{2},\vec{b}_\perp)\gamma^z{\cal W}_z(\tfrac{z}{2},\vec{b}_\perp;-\tfrac{z}{2},0;L)\psi(-\tfrac{z}{2})|P\rangle}{\sqrt{Z_E(b_\perp,\mu,2L)}},
\end{align}
where ${\cal W}_z$ is the spatial staple with length $L$ and $Z_E$ subtracts linear divergences:
\begin{align}\label{eq:Z_E}
Z_E(b_\perp,\mu,2L)=\frac{1}{N_c}\Tr\langle 0|W_\perp{\cal W}_z(0,\vec{b}_\perp;0,0;2L)|0\rangle.
\end{align}
The  scale associated with the quasi-TMDPDF is given by
$\zeta_z=(2xP^z)^2$.
After the subtraction procedure, logarithmic UV divergences remain and must be removed through nonperturbative renormalization, typically performed in the RI/MOM scheme and subsequently matched to the $\MSbar$ scheme~\cite{Shanahan:2019zcq,Ebert:2019tvc}.
More recently, a hybrid renormalization scheme for quasi-TMDs was introduced in Ref.~\cite{Zhang:2022xuw}.
An RI-xMOM renormalization scheme was proposed in Refs.~\cite{Green:2017xeu,Green:2020xco} and subsequently implemented in Refs.~\cite{Avkhadiev:2023poz,Avkhadiev:2024mgd}.

In Ref.~\cite{Ebert:2022fmh}, it was shown that the quasi-TMDPDF is equivalent, under a Lorentz boost, to the beam function defined in the Collins scheme~\cite{Collins:2011zzd}.
In the Collins scheme, TMD beam functions are constructed using spacelike Wilson lines that are nearly lightlike. By identifying the hadron rapidity,
$y_P$,
appearing in the quasi-TMDPDF with the rapidity difference between the Wilson line and the hadron in the Collins-scheme beam function, one can establish an exact equivalence between the two. Consequently, the quasi soft factor associated with the quasi-TMDPDF can be identified as~\cite{Ebert:2022fmh}
\begin{align}\label{eq:etalimit}
\lim_{y_B\to-\infty }\lim_{L\to\infty}\frac{\tilde Z_E(b_\perp, \mu,2L)}{\tilde S(b_\perp, \mu, 2y_n, 2y_B,2L)} = S_r(b_\perp,\mu) e^{2(y_n-y_B)K(b_\perp,\mu)}\,,
\end{align}
where $\tilde S(b_\perp,\mu,2y_n,2y_B,2L)$ denotes the Collins-scheme soft function with rapidities $y_n$ and $y_B \sim -y_P$, and finite Wilson-line length $L$.

Despite the above relation, the quasi- and physical TMDPDFs differ by the order of the $y_B\to-\infty$ (or $P_z\to\infty$) limit and UV renormalization after soft function subtraction~\cite{Ebert:2022fmh}. In the physical TMDPDF, the $y_B\to-\infty$ limit must be taken first before UV renormalization, which is not just limited to the Collins scheme. However, the quasi-TMDPDF is always renormalized first on the lattice at a finite large $P_z$, so it corresponds to the opposite order. Thanks to the asymptotic freedom of QCD, the difference between the two orders can be matched perturbatively in LaMET, thus allowing for the factorization relation below:
\begin{align}\label{eq:quasiTMDmatch}
f^\text{TMD}(x,b_\perp,\mu,\zeta)=H\!\left(\frac{\zeta_z}{\mu^2}\right)e^{-\ln(\zeta_z/\zeta)K(b_\perp,\mu)}\tilde f(x,b_\perp,\mu,\zeta_z)S_r^{1/2}(b_\perp,\mu)+\dots,
\end{align}
where $S_r(b_\perp,\mu)$ removes rapidity-scheme dependence~\cite{Ji:2019ewn}.
The kernel $H$ resums large logarithms of $\zeta_z/\mu^2$ and satisfies an RG equation involving cusp and soft anomalous dimensions. It has been worked out up to NNLO~\cite{Ji:2014hxa,Ebert:2019okf,Ji:2019ewn,Ji:2023pba,delRio:2023pse}. 
The above factorization formula has also been derived by using the region analysis of Feynman diagrams in perturbative QCD~\cite{Liu:2022nnk}.
This framework ensures that lattice-computed quasi-TMDPDFs can be consistently matched to physical TMDPDFs, enabling first-principles predictions for transverse-momentum--sensitive observables.

The soft function plays a central role in TMD factorization by removing rapidity divergences and capturing nonperturbative effects from soft-gluon radiation~\cite{Collins:1984kg,Collins:2011ca}.
For processes such as DY and SIDIS, where small transverse momentum is measured, incomplete cancellation of soft radiation requires introducing a dedicated soft function.
In the lightcone limit, the soft function is defined through Wilson loops involving staple-shaped gauge links, and its divergences are multiplicative in position space~\cite{Vladimirov:2017ksc}.
To compute this quantity nonperturbatively, we need a Euclidean formulation suitable for lattice QCD.
This motivates expressing the soft function as a lattice-calculable observable using heavy-quark effective theory (HQET)~\cite{Ji:2019sxk}.

We start from the Minkowski definition of the soft function as a ratio of Wilson loops:
\begin{align}\label{eq:S_t}
S(b_\perp,\mu,Y,Y')=\lim_{\substack{t\to\infty\\ t'\to\infty}}\frac{W(t,t',b_\perp,\mu,Y,Y')}{\sqrt{Z(2t,b_\perp,\mu,Y)Z(2t',b_\perp,\mu,Y')}},
\end{align}
where $W$ denotes a Wilson loop with opposite time-like staple directions $v$ and $v'$ approaching the lightcone, and $Z$ subtracts pinch-pole singularities.
After subtraction, the remaining UV divergences are associated with the cusp angle $Y+Y'$, and the soft function satisfies an RGE involving the cusp anomalous dimension:
\begin{align}
\mu^2\frac{d}{d\mu^2}\ln S(b_\perp,\mu,Y,Y')=-(Y+Y')\Gamma_\text{cusp}(\alpha_s)-\Gamma_S(\alpha_s).
\end{align}
{At large rapidity, the soft function factorizes as 
\begin{equation}
S (b_\perp,\mu,Y,Y') =  e^{(Y+Y')K(b_\perp,\mu)+\mathcal{D}(b_\perp,\mu)}, 
\end{equation}
where $K$ is the  kernel and $\mathcal{D}$ collects the contributions that are independent of the rapidity regulators $Y$ and $Y'$. The rapidity
dependence enters only through the combination $Y+Y'$, whose coefficient is fixed by
the cusp anomalous dimension that also controls the RGE above,
$\mu^2\,dK/d\mu^2=-\Gamma_\text{cusp}(\alpha_s)$; the rapidity divergence is therefore
multiplicative and exponential, and it is removed by absorbing the factor
$e^{(Y+Y')K}$ into the evolution of the TMD distributions. The remainder is collected in  reduced soft function, which has also been referred to as the intrinsic soft function in the literature:
\begin{equation}
S_r(b_\perp, \mu) = e^{-\mathcal{D}(b_\perp,\mu)}. 
\end{equation}
It should be noted that while the regularization of the rapidity divergence is scheme-dependent, the reduced soft functions and  kernel are independent of the rapidity regularization scheme. In lattice determinations, these quantities are converted to the $\overline{\text{MS}}$ scheme and are usually calculated at $\mu=2$~GeV. 
}

The definition in Eq.~\eqref{eq:S_t} is Minkowski; the two Wilson-line directions
approach the lightcone and the loops extend to large times. Due to their lightlike nature, the
soft function cannot be evaluated in this form. 
Nevertheless, the timelike Wilson lines in $W$ and $Z$ are the propagators of infinitely heavy moving quarks. The latter can be analytically continued to the Euclidean space and constructed via the moving Heavy Quark Effective Theory (HQET) formulation~\cite{Aglietti:1993hf,Hashimoto:1995in,Horgan:2009ti}:
\begin{align}\label{eq:L_HQ}
\mathcal{L}_\text{HQET}^E=Q_v^\dagger(i\tilde v\cdot D_E)Q_v+\bar Q_v^\dagger(i\tilde v\cdot D_E)\bar Q_v,
\end{align}
with $\tilde v^\mu=\gamma(-i,-\beta,\vec{0}_\perp)$.
This formulation enables Monte Carlo evaluation of $S(b_\perp,\mu,Y,Y')$ through
\begin{align}\label{eq:S_lattice}
S(b_\perp,\mu,Y,Y')=\lim_{\substack{T\to\infty\\ T'\to\infty}}\frac{W_E(T,T',b_\perp,\mu,Y,Y')}{\sqrt{Z_E(2T,b_\perp,\mu,Y)Z_E(2T',b_\perp,\mu,Y')}},
\end{align}
where $T$ is the Euclidean time. Up to lattice renormalization and matching, the above is identical to the Minkowski soft function through analytical continuation~\cite{Liu:2022nnk}. Therefore, the reduced soft function $S_r(b_\perp,\mu)$ and the  kernel can be extracted from $S(b_\perp,\mu,Y,Y')$ in the large rapidity limit.

A second route to the same reduced soft function avoids the Wilson-loop correlator and is the one used in the lattice calculations. It extracts the soft function from the large-momentum factorization of a meson form factor. This also makes use of the light-front wave functions (LFWFs). 
Light-front quantization (LFQ) provides a natural framework for describing partonic structure in terms of wave-function amplitudes~\cite{Brodsky:1997de}.
In principle, LFWFs encode complete nonperturbative information about hadrons and allow calculation of parton densities and correlations.
However, direct diagonalization of the QCD Hamiltonian is impractical due to the complexity of the vacuum and Lorentz symmetry requirements.
LaMET offers an alternative by relating equal-time correlators at large momentum to light-front quantities~\cite{Ji:2020baz}.
This approach enables lattice QCD to access LFWFs and related observables through quasi-distributions combined with perturbative matching.

The form-factor factorization then provides this second route. A form factor in
which the two mesons move in opposite directions involves two collinear sectors, one
for each meson, together with the soft radiation exchanged between them, so the same
soft function as in Eq.~\eqref{eq:S_t} enters its leading-power expression. Since
the form factor and the quasi-TMDWFs that approximate the two collinear amplitudes
are computed directly on the lattice, this relation determines the soft function
once the hard kernel is supplied perturbatively. To be explicit, one can investigate the pseudoscalar meson
form factor:
\begin{align}\label{eq:meson_FF}
F(b_\perp,P,P',\mu)=\langle P'|\overline{\eta}(\vec b_\perp)\Gamma'\eta(\vec b_\perp)\,\overline{\psi}(0)\Gamma\psi(0)|P\rangle,
\end{align}
where $\psi$ and $\eta$ are light-quark fields and $P^\mu=(P^t,0,0,P^z)$, $P'^\mu=(P^t,0,0,-P^z)$ denote large momenta approaching opposite lightcone directions.
At leading power, the form factor factorizes into hard, collinear, and soft components:
\begin{align}\label{eq:form_fac_bare}
F(b_\perp,P,P',\mu)=\int dx\,dx'\,H_F(Q^2,\bar Q^2,\mu^2)\,\psi^\dagger(x',b_\perp)\psi(x,b_\perp)\,S(b_\perp,\mu,\delta^+,\delta^-),
\end{align}
where $H_F$ is the hard kernel~\cite{Ji:2019sxk,Deng:2022gzi,Ji:2021znw}, $\psi(x,b_\perp)$ denotes collinear LFWF amplitudes, and $S$ is the soft function. Rapidity divergences cancel among these components, leaving a scheme-independent result.

\begin{figure}[http]
\centering
\includegraphics[width=0.4\textwidth]{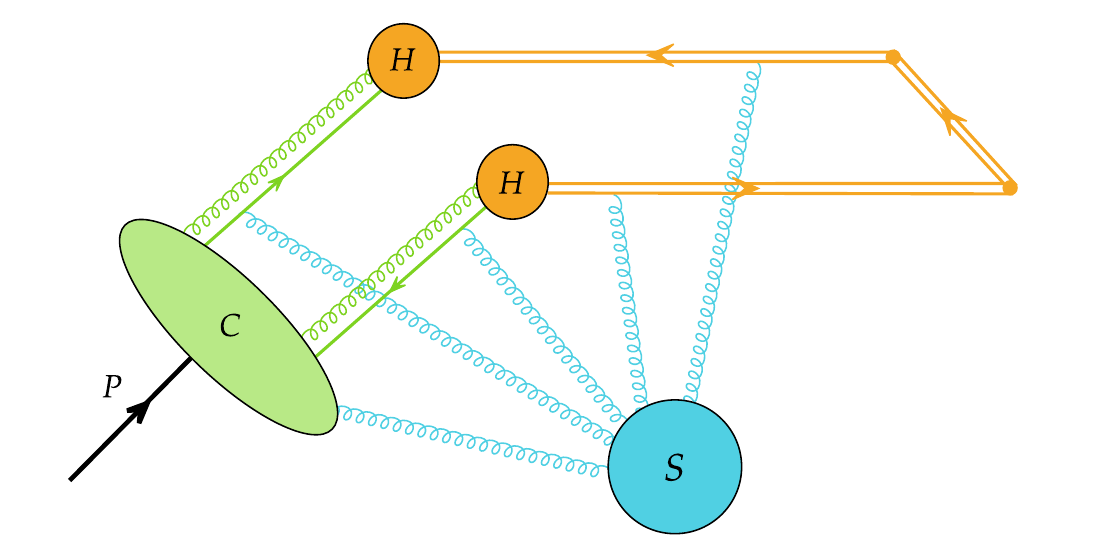}\hspace{1cm}
\includegraphics[width=0.4\textwidth]{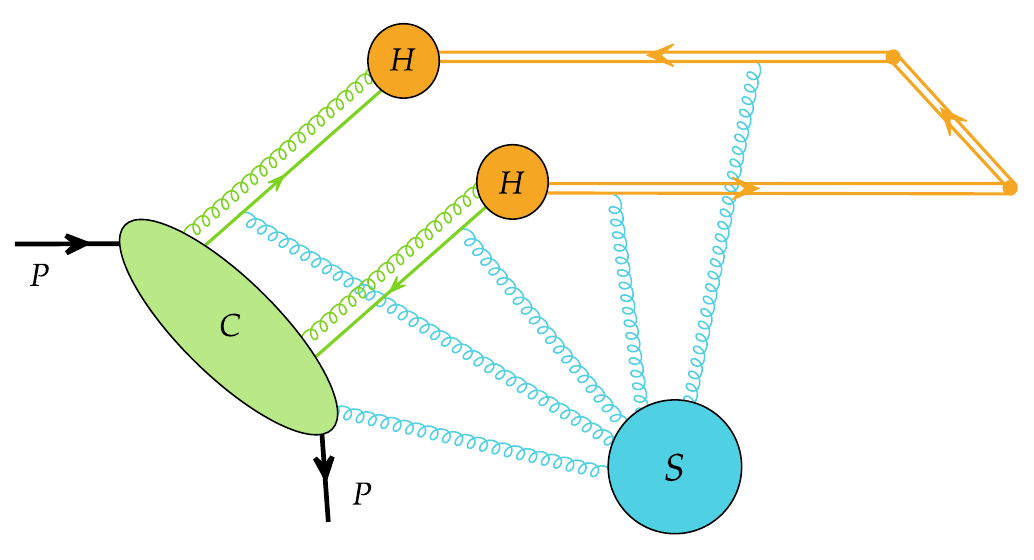}
\caption{
The leading-power reduced graphs for pseudoscalar-meson quasi-TMDWFs (left) and quasi-TMDPDFs (right) are shown. Hard, collinear, and soft modes are denoted by ``$H$'', ``$C$'', and ``$S$'', respectively.
Hard interactions between the two collinear sectors are power suppressed.
 }
\label{fig:factorization_quasi_TMD}
\end{figure}

On the lattice, quasi-LFWF amplitudes are defined analogously to quasi-TMDPDFs:
\begin{align}
\tilde\psi_{\bar qq}(x,b_\perp,\mu,\zeta_z)
= \int \frac{d\lambda}{4\pi} \, e^{i x \lambda}
\frac{\langle 0 | \bar\psi\bigl(\tfrac{\lambda n_z}{2} + \vec b_\perp\bigr)\gamma^z
{\cal W}_z\bigl(\tfrac{\lambda n_z}{2} + \vec b_\perp; L\bigr)
\psi\bigl(-\tfrac{\lambda n_z}{2}\bigr) | P \rangle}
{Z_E(b_\perp,\mu,2L)}
\,.
\end{align}
Figure~\ref{fig:factorization_quasi_TMD} illustrates the reduced diagrams for quasi-TMD wave functions and quasi-TMDPDFs.
By combining the quasi-amplitudes with the reduced soft function $S_r(b_\perp,\mu)$, one obtains a Euclidean factorization formula,
\begin{align}\label{eq:form_qfac}
F(b_\perp,P,P',\mu)
= \int dx\, dx'\,
H(x,x') \,
\tilde\psi^\dagger(x',b_\perp)\,
\tilde\psi(x,b_\perp)\,
S_r(b_\perp,\mu)
\,,
\end{align}
where $H(x,x')$ is perturbatively calculable.
This framework enables a lattice determination of the reduced soft function $S_r$ and the matching kernels, thereby providing a pathway to compute LFWFs and related observables from first principles.
Future work will further refine this approach, particularly in controlling endpoint behavior and improving the precision needed for phenomenological applications.
From the factorization of the form factor in Eq.~\eqref{eq:form_qfac}, one can note that the reduced soft function can be determined through:
\begin{eqnarray}
\label{eq:softfunction}
S_r(b_\perp,\mu) = \frac{F(b_\perp,P,P',\mu)}{\int dx\, dx'\,
H(x,x') \,
\tilde\psi^\dagger(x',b_\perp)\,
\tilde\psi(x,b_\perp)}.
\end{eqnarray}
Both the form factor and the quasi-TMDWFs can be directly calculated on the lattice.

The form factor therefore provides a complete, lattice-accessible route to the
reduced soft function: the form factor and the quasi-TMDWFs are measured on the
lattice, the hard kernel of Eq.~\eqref{eq:form_fac_bare} is supplied perturbatively,
and $S_r$ follows from the ratio above. Its accuracy is controlled by several effects
that have not yet been eliminated. The factorization holds at leading power in
$1/P^z$, so the power corrections present at the finite momenta used on the lattice
are inherited by $S_r$; these are known to be sizable in the transition region
$xP^z b_\perp\sim 1$, where neither collinear nor TMD factorization alone applies. The
hard kernel is currently known only at one loop~\cite{Ji:2019sxk,Ji:2021znw,Deng:2022gzi}, and the matching between the
quasi-TMDWFs and the collinear amplitudes is perturbative as well, with the imaginary
part of the quasi-TMDWF kernel containing a linear renormalon~\cite{Liu:2023onm}. The renormalization of
the quasi-TMDWFs (the subtraction of the linear divergence and the control of
operator mixing) remains a dominant systematic, and the conversion of the extracted
$S_r$ to the $\overline{\text{MS}}$ scheme at $\mu=2$~GeV is itself carried out
perturbatively. 

It is interesting that, 
unlike the quasi-PDF case, the factorization of quasi-TMDs takes a multiplicative form, as shown in Eq.~\eqref{eq:quasiTMDmatch}.
An intuitive understanding can be obtained from a region expansion analysis, which shows that hard contributions are absent in exchange diagrams~\cite{Deng:2022gzi}.
When the exchanged gluon carries hard momentum $q \sim Q$ in all components, the amplitude contains an oscillatory factor
$e^{i q_\perp \cdot b_\perp}$ arising from the transverse separation of the fields. In the hierarchy $\Lambda_{\text{QCD}} \ll 1/b_\perp \ll P^z$, this exponential varies rapidly, and after integration over the loop momentum the contribution is power suppressed, scaling as $\sim 1/(b_\perp P^z)$.
Physically, a hard gluon probes distance scales much smaller than $b_\perp$ and therefore cannot coherently connect the spatially separated operator insertions. As a result, hard contributions are purely local.
Several important features follow:
\begin{itemize}
\item Ratios of quasi-TMD distributions at different hadron momenta are sensitive only to the  kernel. Consequently, lattice calculations of quasi-TMDs can be used to determine the  kernel~\cite{Ji:2014hxa,Ebert:2018gzl}.
\item Since exchange diagrams do not contribute to the hard kernel at leading power, the perturbative matching kernel arises entirely from local contributions.
It is therefore independent of the Lorentz structure and insensitive to spin~\cite{Ebert:2020gxr}.
In ratios of TMD distributions with the same hadron momentum, both the  kernel and soft functions cancel, and the hard kernels also drop out, leaving the ratios of TMDPDFs equal to those of quasi-TMDs.
\item Quark-gluon mixing can arise only through exchange-type diagrams.
The absence of hard contributions at leading power therefore forbids mixing between quark and gluon TMDPDFs~\cite{Ebert:2022fmh}.
\end{itemize}

The results discussed above are derived at leading power.
More recently, Ref.~\cite{Rodini:2022wic} established the factorization of quasi-TMDPDFs at next-to-leading power, involving twist-three TMD correlators together with a new lattice-specific nonperturbative function.
Importantly, the  kernel remains a universal quantity and can still be extracted from the twist-three correlators.
In addition, a class of kinematic power corrections associated with the intrinsic transverse momentum has been identified and derived in Ref.~\cite{BrisCuerpo:2026pma}.
These corrections can have a non-negligible impact on lattice calculations performed at finite hadron momentum $P^z$.
At small transverse separation $b_\perp$, the quasi-TMDPDF can be expanded in terms of collinear PDFs convoluted with perturbatively calculable matching coefficients.
In particular, the Sivers quasi-TMD can be matched onto the twist-three Qiu-Sterman function~\cite{Ji:2020jeb}, providing a promising avenue for determining this quantity from lattice-QCD calculations.

Alongside these theoretical developments, lattice calculations of TMD observables within the LaMET framework have also progressed substantially.
As in the case of quasi-PDFs, lattice renormalization remains a major source of systematic uncertainty.
The RI/MOM scheme was first explored for the nonperturbative renormalization of staple-shaped Wilson-line operators~\cite{Ebert:2019tvc,Constantinou:2019vyb,Shanahan:2019zcq,Alexandrou:2023ucc}, where one typically needs to determine the operator-mixing pattern among all sixteen Dirac structures entering quark quasi-TMDs~\cite{Shanahan:2019zcq}.
This approach was later refined into the RI/xMOM scheme~\cite{Green:2017xeu,Green:2020xco}, in which renormalization is carried out at each quark--Wilson-line vertex~\cite{Avkhadiev:2023poz}.
As a result, operator mixing is restricted to structures permitted by the hypercubic symmetry of the lattice~\cite{Ji:2021uvr}.
Meanwhile, an alternative renormalization strategy, analogous to the hybrid scheme, was proposed in Ref.~\cite{Zhang:2022xuw} and subsequently employed in lattice calculations.

Beyond renormalization, perturbative matching has also undergone substantial improvements.
The first implementation of NNLO matching appeared in Refs.~\cite{LatticePartonCollaborationLPC:2022myp,Avkhadiev:2023poz}, together with the resummation of large logarithms~\cite{Ebert:2022fmh}.
Furthermore, Ref.~\cite{Avkhadiev:2023poz} demonstrated the presence of sizable power corrections in the transition region $xP^z b_\perp \sim 1$, where neither collinear nor TMD factorization alone provides a complete description.
These corrections can be quantified perturbatively through an unexpanded matching kernel~\cite{Avkhadiev:2023poz}, motivating the development of a unified factorization framework that smoothly interpolates between the small- and large-$b_\perp$ regimes.
Finally, Ref.~\cite{Liu:2023onm} showed that the imaginary part of the matching kernel for quasi-TMDWFs contains a linear renormalon associated with the same underlying physics as the Wilson-line mass renormalon.
Applying the LRR scheme to the imaginary part substantially improves the convergence of the perturbative expansion and yields better agreement between theory and lattice data in the  kernel calculations~\cite{Liu:2023onm,Avkhadiev:2024mgd}.
Consequently, the resummation of large logarithms, the use of unexpanded matching kernels, and the LRR scheme have become standard components of modern lattice TMD calculations.

\subsubsection{The Coulomb-gauge approach} 
\label{subsec:cg}

The LaMET framework for quark TMDPDFs has now reached a stage where the leading-power factorization, the soft-factor subtraction, and the perturbative matching are theoretically established, and a lattice-computable definition of the soft function is available. While the systematic understanding of subleading-power corrections remains an active research topic, TMDPDFs can be determined within the leading-power LaMET framework through appropriate large-momentum extrapolations, without explicitly calculating these corrections.

At the same time, the field has progressed from exploratory studies toward the systematic control of uncertainties.
In lattice calculations, the staple-shaped Wilson line with a large spatial extent $\eta$ constitutes a major source of statistical noise in quasi-TMD matrix elements.
Furthermore, owing to the linear power divergence in the Wilson-line self-energy, the matrix elements are exponentially suppressed as
$\exp\!\big[-\delta m(a)\,(2\eta+b_\perp)\big],$
where $\delta m(a)\propto 1/a$,
while the noise level typically does not decrease at a comparable rate.
Consequently, the signal-to-noise ratio (SNR) deteriorates rapidly with increasing $b_\perp$, significantly restricting the range of $b_\perp$ accessible with current computational resources.
In addition, the staple-shaped operator exhibits a complicated operator-mixing pattern on the lattice~\cite{Constantinou:2019vyb,Shanahan:2019zcq,Green:2020xco,Ji:2021uvr,Alexandrou:2023ucc} and is subject to additional power corrections arising from the presence of multiple length scales~\cite{Ebert:2019okf}.
Although gauge-link smearing and Wilson-flow techniques~\cite{Luscher:2010iy} can mitigate the power divergence and improve the SNR to some extent, they also introduce additional systematic effects that must be carefully quantified.

To improve the precision of TMD calculations, a new approach based on Coulomb-gauge (CG) correlators has recently been proposed within the LaMET framework~\cite{Gao:2023lny,Zhao:2023ptv}.
The quasi-TMDs defined from CG correlators are given by
\begin{align}\label{eq:cgqtmd}
    \tilde B(x,b_\perp,\mu,P^z) &\equiv {P^z\over 2P^t}\int_{-\infty}^\infty \frac{db^z}{2\pi} e^{i x P^z b^z} \langle P| \bar{\psi}(b)\gamma^t \psi(0)\Big|_{\vec{\nabla}\cdot \vec{A}=0}|P\rangle\,.
\end{align}
These correlators belong to the same universality class as the conventional gauge-invariant (GI) definition~\cite{Hatta:2013gta,Ji:2020ect}, since both reduce to the same lightcone TMD correlator in the infinite-momentum limit~\cite{Zhao:2023ptv}.
The most distinctive feature of the CG correlators is the absence of Wilson lines, which has important consequences for their renormalization properties and the SNR in lattice calculations.

In the literature, substantial progress has been made toward a complete proof of the renormalizability of QCD in the CG~\cite{Zwanziger:1998ez,Baulieu:1998kx,Niegawa:2006ey,Niegawa:2006hg}.
These studies demonstrate that the quark field remains multiplicatively renormalizable despite the explicit breaking of Lorentz symmetry.
In parallel, lattice calculations have shown that equal-time quark correlators are multiplicatively renormalizable and free of linear divergences~\cite{Burgio:2012ph,Gao:2023lny,Zhang:2024omt}.
Consequently, the renormalization of these nonlocal operators reduces to quark wave-function renormalization, which can be determined precisely within a static regularization-independent momentum-subtraction scheme~\cite{Gao:2026hix}.
Owing to the absence of Wilson lines and the associated linear divergences, the SNR of CG quasi-TMD matrix elements improves exponentially with increasing $b_\perp$ relative to the GI case.
This feature enables lattice calculations to probe significantly deeper into the nonperturbative region and has been demonstrated to remain robust for $b_\perp \gtrsim 1.0\text{ fm}$~\cite{Bollweg:2024zet,Bollweg:2025iol,Bollweg:2025ecn}.
Furthermore, the renormalization procedure on the lattice is considerably simplified, thereby reducing the corresponding systematic uncertainties.
Finally, the three-dimensional rotational symmetry preserved in the CG suppresses operator mixing and facilitates access to larger off-axis hadron momenta~\cite{Gao:2023lny,Gao:2026hix}.

It should be noted that the CG condition does not completely fix the gauge freedom in QCD.
On the lattice, gauge fixing is performed independently on each time slice. 
Consequently, time-dependent correlation functions are expected to vanish after averaging over gauge configurations, whereas equal-time correlation functions remain fully gauge-fixed and are nonzero.
Besides, CG fixing is known to admit multiple solutions to the gauge-fixing condition, known as the Gribov copies~\cite{Gribov:1977wm,Singer:1978dk}. In addition, the finite numerical precision of gauge fixing in lattice simulations can also introduce small variations in the resulting gauge-fixed configurations. These imply potential systematic uncertainties that are specific to gauge-dependent correlators; the conventional gauge-invariant quasi-TMDs defined with Wilson lines require no gauge fixing and are free of this ambiguity at the level of the correlator. We note, however, that gauge-fixing is also commonly used for the nonperturbative renormalization of gauge-invariant operators (e.g., RI/MOM-type schemes), where Gribov copies can likewise play a role.

Currently, a theoretical framework for quantifying the effects of Gribov copies on hadronic matrix elements of CG correlators is not yet available, but numerical studies provide strong evidence that these effects are well controlled and negligible in existing practical calculations. Reference~\cite{Zhang:2024omt} demonstrated that lattice artifacts associated with imperfect gauge fixing can be minimized by achieving sufficiently high gauge-fixing precision. Meanwhile, Ref.~\cite{Gao:2024fbh} systematically investigated the effects of Gribov copies on CG quark correlators in the vacuum and pion states using high-precision gauge fixing. The study found that Gribov copies introduce a statistical uncertainty that is inversely proportional to the square root of the number of gauge configurations, with a magnitude negligible compared to the statistical uncertainties of the correlators. Although these findings are empirical and may depend on the specific observable, similar tests can be readily performed for other observables of interest with a small computational overhead.

At large hadron momentum $P^z$, the CG quasi-TMD is related to the TMDPDF through a LaMET expansion~\cite{Zhao:2023ptv},
\begin{equation}\label{eq:fact}
  {\tilde B(x, b_\perp, \mu, P^z) \over \tilde S_C(b_\perp,\mu)} = |C(xP^z/\mu)|^2  \exp\left[{1\over 2}K(b_\perp,\mu) \ln{4(xP^z)^2\over \zeta}\right]f(x, b_\perp, \mu, \zeta)\,,
\end{equation}
where $C(xP^z/\mu)$ is a perturbative matching coefficient, and $\tilde S_C(b_\perp,\mu)$ is a soft factor that compensates for the difference in the $b_\perp$ dependence between the quasi-TMD and the TMDPDF.
The above relation has the same structure as that of the GI quasi-TMDs~\cite{Ebert:2019okf,Ji:2019sxk,Ji:2019ewn,Ebert:2022fmh}. It can be derived within SCET~\cite{Bauer:2001yt,Bauer:2002uv,Beneke:2002ph,Bauer:2003mga} and has been verified at one-loop order~\cite{Zhao:2023ptv}.

The matching kernel is currently known at NLO~\cite{Zhao:2023ptv}. 
Reference~\cite{Liu:2023onm} found, within the large-$\beta_0$ approximation, that it contains a linear renormalon, implying a linear power correction in the TMD expansion. In practical lattice calculations, both the  kernel and TMDPDFs exhibit good convergence with increasing hadron momentum $P^z$, reaching a level comparable to state-of-the-art calculations based on gauge-invariant quasi-TMDs. This observation suggests that the corresponding linear power correction may be numerically small, a possibility that can be further tested by examining the perturbative convergence of the matching kernel beyond NLO.

The factorization formula in Eq.~\eqref{eq:fact} also sheds clear insight into the role of Gribov copies in the CG quasi-TMDs. According to the SCET derivation of the factorization formula for the CG quasi-TMD~\cite{Zhao:2023ptv}, the leading-power contributions come from the collinear and soft sectors in the large-momentum expansion. In the collinear sector, the effective Lagrangian reduces to that of lightcone QCD, where the Gribov copies are absent. The soft sector is still described by a CG Lagrangian in which the Gribov ambiguity persists, but it is expected to cancel against the same ambiguity in the soft factor $\tilde S_C$, so that residual effects appear only in the power-suppressed corrections in Eq.~\eqref{eq:fact}. 
Therefore, even if the Gribov copies introduce a nonzero systematic effect in the infinite-statistics and physical limits, their contribution is power suppressed in the large-momentum expansion of the CG quasi-TMDs.

The soft factor $\tilde S_C$ can be determined from a meson form factor introduced in Ref.~\cite{Ji:2019sxk},
\begin{align}
    F(b_\perp,P^z) &= \langle \pi(-\vec{P}) | j_1(b_\perp) j_2(0) | \pi(\vec{P})\rangle\,,
\end{align}
where $\pi(\vec{P})$ denotes a pion state with momentum $\vec{P}=(0,0,P^z)$, and $j_{1,2}=\bar{\psi}\Gamma_{1,2}\psi$ are light-quark currents.
At large $P^z$, $F$ satisfies a Drell-Yan--like factorization formula~\cite{Ji:2019sxk,Ji:2021znw,Deng:2022gzi},
\begin{align}\label{eq:fact_ff}
    F(b_\perp,P^z) &= \int dx_1 dx_2{H_F(x_1, x_2, P^z,\mu)\over C({x_1P^z\over\mu}) C({x_2P^z\over\mu}) C({\bar{x}_1P^z\over\mu}) C({\bar{x}_2P^z\over\mu})}  {\tilde \phi(x_1, b_\perp, \mu, P^z) \over \tilde S_C(b_\perp,\mu)} {\tilde \phi(x_2, b_\perp, \mu, P^z) \over \tilde S_C(b_\perp,\mu)}\,,
\end{align}
where $\bar{x}=1-x$, and the hard coefficient $H_F$ is known at one loop~\cite{Ji:2019sxk,Deng:2022gzi}.
The quasi-TMD wave function $\tilde \phi$ is defined as
\begin{align}
    if_\pi P^\mu \tilde \phi(x,P^z,b_\perp) = \int_{-\infty}^\infty \frac{db^z}{2\pi} e^{i x P^z b^z} \langle 0| \bar{\psi}(b)\gamma^\mu \psi(0)\Big|_{\vec{\nabla}\cdot \vec{A}=0}|\pi(P)\rangle\,,
\end{align}
which can be computed on the lattice.
By inserting the expressions for $F$ and $\tilde \phi$ into Eq.~\eqref{eq:fact_ff}, one can extract $\tilde S_C$, thereby completing the CG framework for TMD calculations.

\section{TMD results from LaMET Method}\label{sec:Results}

In this section, we review lattice QCD results for  TMD distributions following a physically motivated and methodologically consistent order, prioritizing foundational quantities that underpin TMD factorization and evolution before proceeding to the full TMDPDFs and wave functions. We begin with the Collins-Soper (CS) kernel, the universal nonperturbative input governing TMD rapidity evolution, covering its early conceptual validation, successive methodological refinements, and recent systematic determinations. 
Next, we discuss the progress on the reduced soft function, the rapidity-independent component of the TMD soft function that encodes nonperturbative soft-gluon radiation and is required to connect lattice quasi-TMDs to physical lightcone TMDs. Finally, we summarize  first-principles calculations of the primary phenomenological targets: TMDPDFs and TMDWFs, including both unpolarized and spin-dependent distributions (such as the Boer-Mulders function) for mesons and nucleons. Within each subsection, results are arranged chronologically to highlight the progressive improvement in precision, lattice setup (toward physical quark masses and continuum limit), methodology, and uncertainty quantification across recent years.

\subsection{Collins-Soper kernel}

The foundational methodology for extracting the CS kernel from lattice QCD is based on the key theoretical insight that a suitable ratio of quasi-TMD correlators, evaluated at two different large hadron momenta, cancels the nonperturbative soft contributions and directly isolates the kernel governing rapidity evolution~\cite{Collins:1981uk}.
At the core of this approach is a formal relation connecting the CS kernel $K(b_\perp)$ to the logarithm of the ratio of renormalized quasi-TMD correlators in the large-momentum limit.

As proposed in Refs.~\cite{Ji:2014hxa,Ebert:2018gzl} and systematically formulated within the LaMET factorization framework~\cite{Ebert:2019okf,Ji:2019sxk,Ji:2019ewn,Ebert:2022fmh}, the dependence of quasi-TMD observables on the hadron momentum reproduces the CS evolution.
In this picture, the CS kernel emerges as the quantity governing the momentum evolution of quasi-TMDs, thereby providing a direct and lattice-accessible definition.
An illustration of the corresponding LaMET factorization for meson quasi-TMDWFs is shown in Fig.~\ref{fig:factorization_quasi_TMD}.
In the infinite-momentum limit, the kernel can be extracted through the relation
\begin{align}
K(b_\perp, \mu, x, P_1^z, P_2^z)
&= \frac{1}{\ln(P_1^z/P_2^z)}
\left[
\ln\frac{H(\zeta_{z,2}, \bar{\zeta}_{z,2}, b_\perp, \mu)\,
\tilde f(x, b_\perp, \mu, \zeta_{z,1})}
{H(\zeta_{z,1}, \bar{\zeta}_{z,1}, b_\perp, \mu)\,
\tilde f(x, b_\perp, \mu, \zeta_{z,2})}
\right]
+ \mathcal{O}\!\left(
\frac{\Lambda_{\text{QCD}}^2}{\zeta_z},
\frac{1}{b_\perp^2 \zeta_z}
\right),
\label{eq:cs kernel of ratio}
\end{align}
with the condition $P_1^z \neq P_2^z \gg 1/b_\perp$.
In practical lattice simulations, the hadron momenta are large but finite, and this relation provides the basis for extracting the CS kernel by analyzing the momentum dependence of the computed ratio.
This ratio method has been widely adopted and systematically refined in numerous lattice studies of the CS kernel and related soft functions~\cite{Shanahan:2020zxr,LatticeParton:2020uhz,Schlemmer:2021aij,Shanahan:2021tst,Li:2021wvl,LatticePartonLPC:2022eev,LatticePartonLPC:2023pdv,Shu:2023cot,Avkhadiev:2023poz,Liu:2024sqj,Avkhadiev:2024mgd,Bollweg:2024zet,Bollweg:2025iol,Alexandrou:2025xci,Tan:2025ofx}.
While this approach provides a theoretically well-controlled pathway to determine the CS kernel, early implementations demonstrated that achieving high precision requires careful control of systematic effects associated with the finite-momentum expansion inherent to the LaMET framework.

Lattice QCD calculations of the CS kernel have advanced from early feasibility studies to precision determinations, providing first-principles results that now actively constrain and refine phenomenological models extracted from experimental data.
This evolution, from exploratory calculations limited by substantial systematic uncertainties to modern precision studies of parton distributions and hadron structure, marks a major milestone in nonperturbative QCD study of TMD physics.
It has enabled direct and quantitative comparisons between lattice QCD results and phenomenological analyses, establishing lattice QCD as an important tool for TMD physics.
A summary of lattice-QCD--extracted CS kernel can be found in Table~\ref{tab:CS}.

It is instructive to ask how closely existing calculations realize the condition $P^z_1 \neq P^z_2 \gg 1/b_\perp$ underlying Eq.~\eqref{eq:cs kernel of ratio}, since this hierarchy controls the size of the power corrections neglected in the extraction.
Concretely, the leading power corrections are of order $\Lambda_\text{QCD}^2/\zeta_z$ and $1/(b_\perp^2\zeta_z)$ with $\zeta_z=(2xP^z)^2$, so the relevant dimensionless quantity is $2xP^z b_\perp/\hbar c$, with $\hbar c\simeq 0.197\text{ GeV}\cdot\text{fm}$; for $x\sim 0.5$ this reduces to $P^z b_\perp/\hbar c$.
The maximum momenta in Table~\ref{tab:CS} lie between $1.4$ and $4.3$~GeV, well above $\Lambda_\text{QCD}$, and the corresponding scale $1/b_\perp$ ranges from about $0.2$~GeV at $b_\perp=1$~fm to about $0.7$~GeV at $b_\perp=0.3$~fm.
Combining the $P^z_{\max}$ and $b_\perp^{\max}$ columns, the product $P^z_{\max}b_\perp^{\max}/\hbar c$ is about $4$--$15$ for most calculations, reaching $\approx 22$ for BNL/ANL'25, so the hierarchy $P^z\gg 1/b_\perp$ is satisfied by roughly one order of magnitude at the largest separations.
The RQCD/LHPC'23 ($\approx 2.8$) and MIT/FNAL/ANL'21 ($\approx 3.8$) calculations satisfy the condition marginally, so their results may receive considerable power corrections.
Because the condition becomes easier to meet as $b_\perp$ grows, it is the \emph{small}-$b_\perp$ end that is limiting: the usable window is roughly $b_\perp\gtrsim \text{few}/P^z_{\max}$, i.e.\ $b_\perp \gtrsim 0.2$--$0.3$~fm, and $b_\perp$ must in any case exceed the lattice spacing $a=0.06$--$0.12$~fm.
The upper end of the window, $b_\perp^{\max}$, is instead set by other systematics, such as signal degradation and finite-volume effects at large separations.
The same reasoning restricts the momentum fraction: since $\zeta_z\propto x^2$, the power corrections are controlled for $x$ away from the endpoints, and, for example, ETMC'25 finds an approximately $x$-independent kernel for $0.3<x<0.7$.
Finally, the two momenta must be different but need not be far apart. Each is individually large compared with $1/b_\perp$, while $\ln(P^z_1/P^z_2)$ provides the lever arm that isolates $K$; typical choices are pairs such as $2.22/3.33$~GeV and $1.63/2.45$~GeV (ETMC'25), or ratios $4/3$, $5/4$, $6/5$ (PKU/ETMC'22).
Closer momenta suppress residual momentum dependence but amplify statistical errors through the small denominator.
The residual finite-$P^z$ effects are then controlled through the perturbative matching kernel, whose order, among leading order (LO), next-to-leading order (NLO), next-to-leading logarithm (NLL) and unexpanded next-to-next-to-leading logarithm (uNNLL) in the ``Matching Kernel'' column, determines how well the ratio of hard kernels at $\zeta_{z,1}$ and $\zeta_{z,2}$ is known, and through the ``Syst.~Uncert.'' column, which indicates whether a systematic uncertainty from such effects, for example from varying the momentum pair, is included.

It should be pointed out that the above table-level checks are necessary but not sufficient to demonstrate that the systematic uncertainties are fully under control. Here we discuss four categories of benchmarks of lattice calculations that have been reported in the literature.

\begin{enumerate}[label=(\roman*)]
\item Control of lattice systematics. In addition to the LaMET systematics associated with renormalization, perturbative matching, and finite-momentum power corrections, lattice systematic uncertainties arising from unphysical quark masses, finite lattice spacing and volume, excited-state contamination, and other sources should also be demonstrated to be under control. This is essential for establishing the quantitative reliability of lattice calculations.
\item Agreement with perturbation theory at small $b_\perp$. In the small-$b_\perp$ regime, the kernel is perturbatively calculable, with predictions independent of phenomenological inputs or matching conventions adopted at large $b_\perp$. At finite $P_z$, the window satisfying both the TMD limit $P_z b_\perp \gg 1$ and the perturbative condition $b_\perp \ll \Lambda_\text{QCD}^{-1}$ is typically limited. This limitation can be overcome through a unified, unexpanded matching scheme~\cite{Avkhadiev:2023poz} that smoothly interpolates between the collinear and TMD regions, provided that $xP_z \gg \Lambda_\text{QCD}$. However, the complete framework is still under development. In addition, discretization artifacts may become significant at $b_\perp \sim a$, requiring comparisons between lattice results and perturbative predictions to be performed in the window $a \ll b_\perp \ll \Lambda_\text{QCD}^{-1}$. Once the unified matching scheme is fully developed, this comparison will provide a well-defined benchmark for lattice results.
\item Agreement among independent lattice determinations. Lattice-QCD calculations employ a variety of gauge actions, fermion discretizations, renormalization prescriptions, operator constructions, including staple-shaped Wilson lines and Coulomb-gauge formulations, and matching procedures. Nevertheless, all approaches are expected to yield consistent results in the physical limits, making cross-validation among independent calculations necessary and a powerful benchmark. For such comparisons to be fully reliable, however, the table-level checks discussed above, as well as benchmarks (i) and (ii), should be satisfied.
\item Comparison with global fits. Global fits typically employ phenomenological parametrizations based on certain assumptions and analyze experimental data within a given factorization framework. Since lattice QCD calculations and global fits rely on entirely different sources of information, their comparison is better viewed as a confrontation between theoretical predictions and experimental constraints, rather than a benchmark for validating either approach. In cases where TMD observables are subject to substantial systematic uncertainties, a global analysis combining experimental and lattice data may offer a practical approach at the present stage, as discussed below.
\end{enumerate}
We structure the subsequent discussion along these four lines, emphasizing that criteria (ii) and (iii) offer mutually reinforcing corroboration, while (i) constitutes a direct, quantitative error budget, which cannot be replaced by (ii) or (iii).
In the following section, we discuss recent developments and review the corresponding results presented in the remaining sections.

\subsubsection{Foundational Exploration  and Conceptual Validation}
The CS kernel, which governs the $\zeta$-scale evolution of TMDPDFs, can be isolated through ratios of quasi-TMDPDFs evaluated at different hadron momenta, thereby canceling nonperturbative soft contributions and lattice artifacts.
This provides a first-principles pathway to determine the kernel in the nonperturbative regime $b_\perp \sim \Lambda_\text{QCD}^{-1}$~\footnote{
In the literature, the  transverse  impact parameter $b_\perp$ is frequently denoted as $b_T$ (or simply $b$), whereas the transverse momentum $k_\perp$ is occasionally written as $q_T$ or $q_\perp$.
}, thereby bridging lattice-accessible scales to those relevant for high-energy processes.
Within this framework, MIT'20~\cite{Shanahan:2020zxr} presents the first lattice implementation in quenched QCD, extracting the CS kernel for $250\text{ MeV} < k_\perp < 2\text{ GeV}$.
Using staple-shaped Wilson-line operators and renormalization in the RI/MOM scheme, the study fits quasi-beam functions using Hermite and Bernstein polynomial parameterizations to mitigate Fourier-transform truncation effects.
The resulting CS kernel is consistent with perturbative calculations at small $b_\perp$ and supports phenomenological expectations of constant or linear behavior in the nonperturbative region.
This proof-of-principle study demonstrates the feasibility of the theoretical framework while identifying key sources of systematic uncertainty, including finite-volume effects and power corrections, that require further improvement in future calculations.

Complementing the use of quasi-TMDPDFs, Ref.~\cite{Ji:2019sxk} formulates the theoretical framework by constructing the TMD soft function, including both the CS kernel and the reduced soft function, within a Euclidean lattice framework.
It shows that the soft function, which captures soft gluon radiation, can be extracted either as a form factor of boosted heavy-quark pairs in heavy quark effective theory or through ratios of large-momentum light-meson form factors and quasi-TMDWFs.
In this formulation, the CS kernel governing TMD evolution is obtained from ratios of quasi-TMDWFs evaluated at different hadron momenta.
LPC'20~\cite{LatticeParton:2020uhz} advances this program to a full lattice implementation using a $2+1$ flavor dynamical ensemble, which is more physically realistic than quenched QCD.
It computes both the reduced soft function $S_I$ and the CS kernel.
By simulating quasi-TMDWFs and light-meson form factors, it demonstrates the stability of the reduced soft function across different hadron momenta and shows that the extracted CS kernel is consistent with perturbative results (up to three loops) at small $b_\perp$, as well as with previous nonperturbative studies.
Together, these works complete a more comprehensive nonperturbative picture of TMD factorization by providing the theoretical connection between lattice-accessible observables and the soft function, while also demonstrating the feasibility of such extractions in dynamical QCD.

\subsubsection{Methodological Refinement and  Iterative Advancement}

Initial lattice QCD calculations successfully demonstrated the viability of extracting the CS kernel via LaMET, but they also revealed significant systematic uncertainties that limited their immediate phenomenological impact.
As a result, several studies were conducted to investigate these systematic uncertainties.

RQCD/LHPC'21~\cite{Schlemmer:2021aij} presents a lattice-QCD determination of the nonperturbative CS kernel using CLS $2+1$-flavor dynamical clover fermions on an ensemble with lattice spacing $a = 0.0854\text{ fm}$ and pion mass $m_\pi = 422\text{ MeV}$.
To avoid the Fourier transform, the authors extracted the CS kernel from ratios of the first Mellin moments of quasi-TMDs at different nucleon momenta. The convolution between the matching kernel and physical TMDPDFs in the $z$-space is supplemented by the phenomenological input of the latter~\cite{Vladimirov:2020ofp}, which is found to be numerically insensitive to the transverse separation $b_\perp$. As a result, the kernel is extracted from the ratios of $z=0$ matrix elements, which simplifies the calculation and reduces the statistical errors.
This semi-phenomenological approach led to results consistent across three leading-twist TMD channels ($f_1$, $g_{1T}$, and $h_1$), which numerically demonstrates the universality of the CS kernel in the range $0.8 \le b_\perp \le 2.5\text{ GeV}^{-1}$.
Compared with previous lattice and phenomenological extractions, these lattice results disfavor the commonly used $K(b_\perp)\sim b_\perp^2$ ansatz at large $b_\perp$ and instead support linear or constant asymptotic behavior.

PKU/ETMC'22~\cite{Li:2021wvl} advances calculations of the CS kernel by replacing the conventional RI/MOM scheme with a ratio renormalization approach. 
Critically, the study addresses substantial HT contamination (which was found to flip the signs of matrix elements) by constructing improved pion matrix elements that explicitly cancel dominant HT contributions.
Technical improvements include five-step stout smearing to reduce operator fluctuations, two-state fits to suppress excited-state contamination, and extrapolation of staple-shaped Wilson lines to the asymptotic limit ($L \gtrsim 0.8\text{ fm}$), where the relevant correlations plateau.
The CS kernel is extracted from multiple momentum ratios ($4/3, 5/4$, and $6/5$) using quasi-TMDWF correlators, with results in qualitative agreement with previous lattice determinations.
These improvements refine the nonperturbative extraction of the CS kernel and provide more robust lattice inputs for TMDPDF rapidity evolution.

Notably, both LPC'20~\cite{LatticeParton:2020uhz} and PKU/ETMC'22~\cite{Li:2021wvl} were performed at leading order in perturbation theory, where the CS kernel can be extracted directly from ratios of the $z=0$ matrix elements. Beyond LO, however, the matching kernel acquires a nontrivial $x$ dependence, requiring a Fourier transformation of the quasi-TMD matrix elements, which has become a standard procedure in subsequent calculations.

A dynamical fermion ensemble with $m_\pi = 538\text{ MeV}$ and a target transverse-momentum range $400\text{ MeV} < k_\perp < 1.7\text{ GeV}$ is employed in MIT/FNAL/ANL'21~\cite{Shanahan:2021tst}.
Compared to previous analyses, key improvements in this work include NLO perturbative matching between quasi-TMDPDFs and lightcone TMDPDFs, a full treatment of operator mixing under renormalization, and refined control of power corrections (including higher-twist effects) as well as finite-volume systematics.
The authors compare multiple extraction methods using the same lattice matrix elements, revealing non-negligible effects from both power corrections and matching, thereby demonstrating that their neglect can lead to significant biases.
The extracted CS kernel, obtained by combining results from $\gamma_4$ and $\gamma_3$ Dirac structures, shows qualitative agreement with previous lattice and phenomenological studies, while emphasizing the necessity of NLO matching and careful systematic control for  nonperturbative constraints on TMDPDF rapidity evolution.

LPC'22~\cite{LatticePartonLPC:2022eev} employs an $N_f = 2+1+1$ MILC HISQ ensemble with lattice spacing $a = 0.12\text{ fm}$ and pion mass $m_\pi = 670\text{ MeV}$, using clover valence fermions and three meson momenta in the range $1.72$--$2.58\text{ GeV}$ to extract the CS kernel from ratios of quasi-TMDWFs and perturbative matching kernels.
Key advances include Wilson-loop renormalization to cancel the linear divergences associated with staple-shaped gauge links, averaging over two Dirac structures ($\gamma^t\gamma^5$ and $\gamma^z\gamma^5$) to reduce power corrections, and systematic quantification of uncertainties arising from operator mixing and scale dependence.
The extracted CS kernel spans transverse separations up to $b_\perp \simeq 0.6\text{ fm}$, agrees with perturbative QCD predictions at small $b_\perp$, and is consistent with the SV19~\cite{Scimemi:2019cmh} phenomenological parametrization at larger $b_\perp$.

RQCD/LHPC'23~\cite{Shu:2023cot} focuses on verifying the universality of the CS kernel via lattice QCD, using a CLS ensemble with $N_f = 2+1$ clover-improved Wilson fermions, lattice spacing $a = 0.0854\text{ fm}$, and sea pion mass $m_{\pi,\text{sea}} = 422\text{ MeV}$.
The CS kernel is extracted from ratios of the first Mellin moments of quasi-TMDPDFs with the semi-phenomenological approach in Ref.~\cite{Vladimirov:2020ofp}, for four independent cases: pion and proton targets, each with twist-2 and twist-3 operators.
Key technical ingredients include HYP gauge-link smearing and momentum smearing to improve signal quality, as well as the application of leading-power and next-to-leading-power factorization for different Dirac structures.
The results exhibit consistent CS kernel behavior across all four cases in the intermediate transverse-separation range
$0.8\text{ GeV}^{-1} \le b_\perp \le 2.6\text{ GeV}^{-1},$
providing strong evidence for its universality.
The extracted kernel is qualitatively consistent with previous lattice determinations, phenomenological parametrizations such as SV19~\cite{Scimemi:2019cmh} and MAP22~\cite{Bacchetta:2022awv}, as well as three-loop perturbative calculations.
At the same time, the study highlights the need for improved quantification of the power corrections at small $b_\perp$ and of the uncontrolled systematic uncertainties at much larger $b_\perp$ to achieve higher precision.

LPC'23~\cite{LatticePartonLPC:2023pdv} presents a lattice QCD determination of the CS kernel across multiple lattice ensembles, using both a $2+1$-flavor CLS ensemble with lattice spacing $a = 0.098\text{ fm}$ and pion mass $m_\pi = 333\text{ MeV}$, and a $2+1+1$-flavor MILC ensemble with $a = 0.121\text{ fm}$ and $m_\pi = 310\text{ MeV}$.
The CS kernel is extracted primarily from ratios of quasi-TMDWFs combined with one-loop perturbative matching. Main methodological ingredients include Wilson-loop renormalization to cancel linear and logarithmic divergences, $\lambda$-extrapolation to control large longitudinal-separation effects, and Fierz rearrangement to suppress higher-twist contamination.
The study compares CS kernel extractions based on quasi-TMDWFs and quasi-TMDPDFs, finding that quasi-TMDWFs yield more reliable results with weaker $x$-dependence, thereby avoiding the early breakdown of the LaMET expansion observed in quasi-TMDPDF-based approaches.
The extracted CS kernel agrees with three-loop perturbative QCD ($\text{N}^3\text{LO}$) predictions at small $b_\perp$ and with phenomenological parametrizations at moderate $b_\perp$, while showing good consistency across different lattice ensembles and compatibility with previous lattice determinations within uncertainties.
Systematic uncertainties are quantified through analyses of imaginary-part contributions and cross-ensemble comparisons, reinforcing the CS kernel's role as a reliable nonperturbative input for TMD rapidity evolution.

BNL/ANL'24~\cite{Bollweg:2024zet} performed the first lattice-QCD determination of the
CS kernel using the CG method (see Sec.~\ref{subsec:cg}) on an
ensemble of domain-wall--fermion configurations with lattice spacing
$a=0.086\text{ fm}$ and physical quark masses.
The CS kernel was extracted from
ratios of pion quasi-TMDWFs at next-to-leading logarithmic
accuracy. 
The improved signal at large quark separation enabled the
determination of the CS kernel up to transverse separations of $b_\perp \sim 1.0\text{ fm}$, which is consistent with previous lattice
calculations based on gauge-invariant quasi-TMDWFs and agrees with recent
global analyses.

BNL/ANL'25~\cite{Bollweg:2025iol} presents a new lattice-QCD determination of the CS kernel, leveraging the CG approach to avoid Wilson-line-induced linear divergences and to improve the signal-to-noise ratio.
The CS kernel is extracted from ratios of quasi-TMDWFs  computed at different pion momenta (e.g., $n_z \in \{8,9,10\}$, corresponding to $P^z$ values up to $4.30\text{ GeV}$) on a $2+1$-flavor HotQCD ensemble with fine lattice spacing $a = 0.06\text{ fm}$. The calculations use valence pion masses of $300\text{ MeV}$ for the main analysis and $670\text{ MeV}$ for the quasi-TMDWF calculations.
The extraction incorporates NLL renormalization-group resummation (RGR) to account for large logarithmic contributions in the hard kernel, thereby improving the accuracy of perturbative matching.
At $\mu = 2\text{ GeV}$, the extracted CS kernel shows good agreement with three-loop perturbative results in the small-$b_\perp$ region, while remaining consistent with recent lattice determinations and global fits at larger $b_\perp$, where perturbation theory is expected to break down.
Systematic uncertainties, dominated by power corrections associated with the heavier pion mass ($670\text{ MeV}$) and mild residual $x$-dependence in the momentum ratios, are quantified through jackknife resampling and variations of momentum-pair choices.

ETMC'25~\cite{Alexandrou:2025xci} employs two $N_f = 2+1+1$ twisted-mass fermion ensembles with lattice spacing $a = 0.093\text{ fm}$, lattice volumes $24^3 \times 48$ and $32^3 \times 64$, and valence pion masses of $640\text{ MeV}$ and $830\text{ MeV}$, respectively.
The CS kernel is extracted from ratios of momentum-space quasi-TMDWFs evaluated at different pion momenta, such as $2.22/3.33\text{ GeV}$ for the smaller-volume ensemble and $1.63/2.45\text{ GeV}$ for the larger-volume ensemble.
The analysis employs asymmetric staple-shaped Wilson-line operators and neglects operator mixing, which is verified to be numerically negligible through an RI$'$-short scheme analysis.
Renormalization is carried out nonperturbatively using the short-distance ratio (SDR) scheme, with the resulting quantities converted to the $\overline{\text{MS}}$ scheme for direct comparison with phenomenological studies.
Both LO and NNLO matching corrections are incorporated, with NLO effects found to be more significant for the ensemble with smaller boost momenta.
The extracted CS kernel exhibits approximate $x$-independence in the range
$0.3 < x < 0.7$,
where power corrections are expected to be suppressed, and shows negligible imaginary parts.
The results are consistent across the two pion masses and indicate minimal finite-volume effects.

\subsubsection{Precision Determination and Systematic Controls}
Subsequent lattice calculations, by achieving control over discretization effects, quark-mass dependence, and perturbative matching uncertainties, now provide first-principles determinations of the CS kernel with sufficient precision to directly compare with and distinguish among phenomenological models.
In this subsection, we focus on these updated studies, which incorporate improved control of lattice systematics and aim toward precision determinations of the CS kernel.

\begin{figure}[t]
\centering\includegraphics[width=0.48\textwidth]{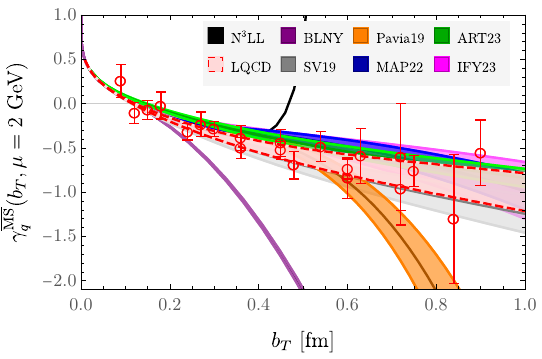}
\caption{\label{fig:phenocomparison_Avkhadiev:2024mgd} 
Comparison of lattice calculation of the CS kernel with phenomenological parameterizations~\cite{Landry:2002ix,Scimemi:2019cmh,Bacchetta:2019sam,Bacchetta:2022awv,Moos:2023yfa,Isaacson:2023iui} of experimental data (BLNY, SV19, Pavia19, MAP22, ART23, IFY23), and perturbative results from Refs.~\cite{Collins:2014jpa,Li:2016ctv,Vladimirov:2016dll} ($\mathrm{N}^3\mathrm{LL}$).}
\end{figure}

A new generation of lattice studies, spearheaded by MIT/FNAL/ANL'24~\cite{Avkhadiev:2024mgd}, marks a turning point by achieving the first systematic determination of the CS kernel with controlled quark-mass dependence, operator renormalization, and discretization effects.
Using three $2+1+1$-flavor MILC ensembles with lattice spacings $a = 0.09$, $0.12$, and $0.15\text{ fm}$ at near-physical pion mass, the calculation employs NNLL $b_\perp$-unexpanded matching and RI/xMOM renormalization.
The continuum-extrapolated results shown in Fig.~\ref{fig:phenocomparison_Avkhadiev:2024mgd} agree with recent phenomenological parameterizations (e.g., SV19 and ART23) at large $b_\perp \gtrsim 0.6\text{ fm}$, while disfavoring older models. These results provide systematic first-principles constraints for TMD evolution.

\begin{figure*}[http]
\centering
\includegraphics[width=0.7\textwidth]{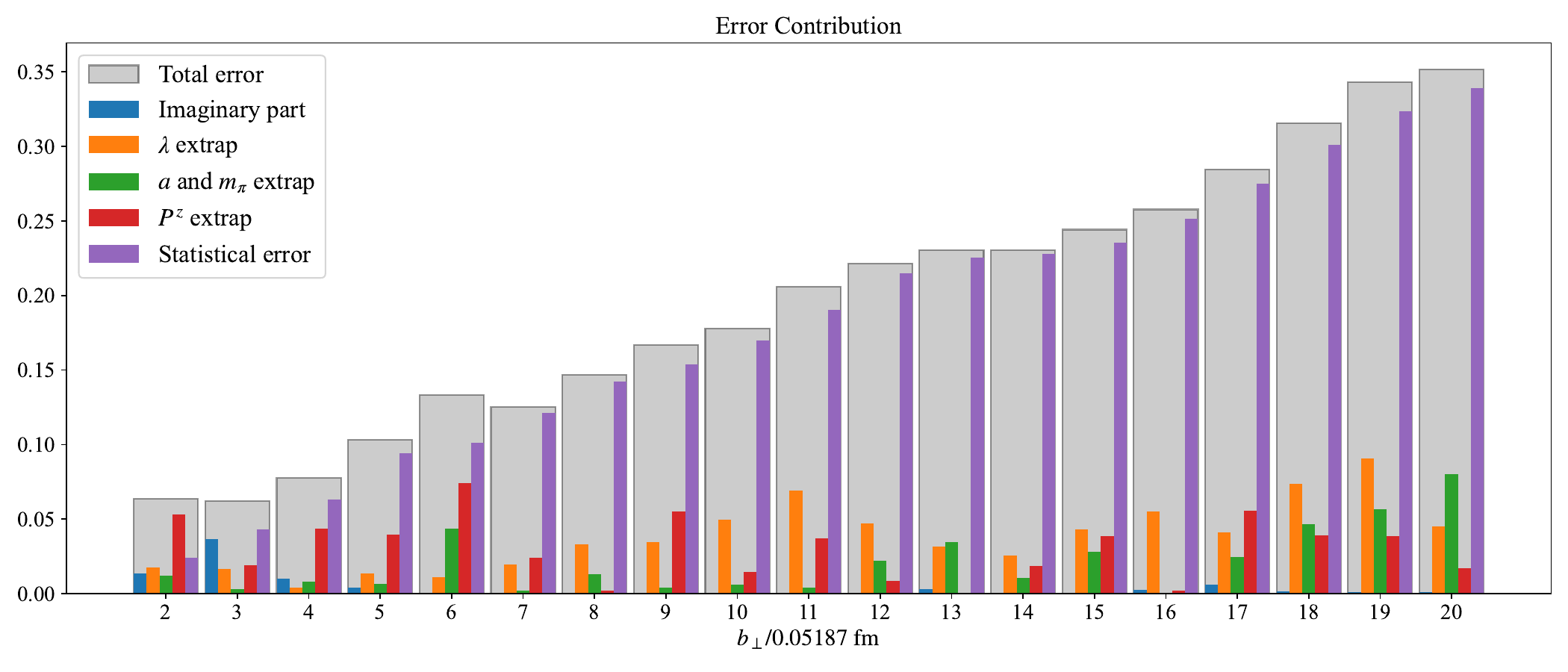}
\caption{LPC'25's~\cite{Tan:2025ofx} error budget for the CS kernel, showing statistical and systematic uncertainties from different sources. The total uncertainties are obtained by adding these contributions in quadrature.
}
\label{fig:System_Error_Contribution}
\end{figure*}

\begin{figure*}[http]
\centering
{\includegraphics[width=0.45\textwidth]{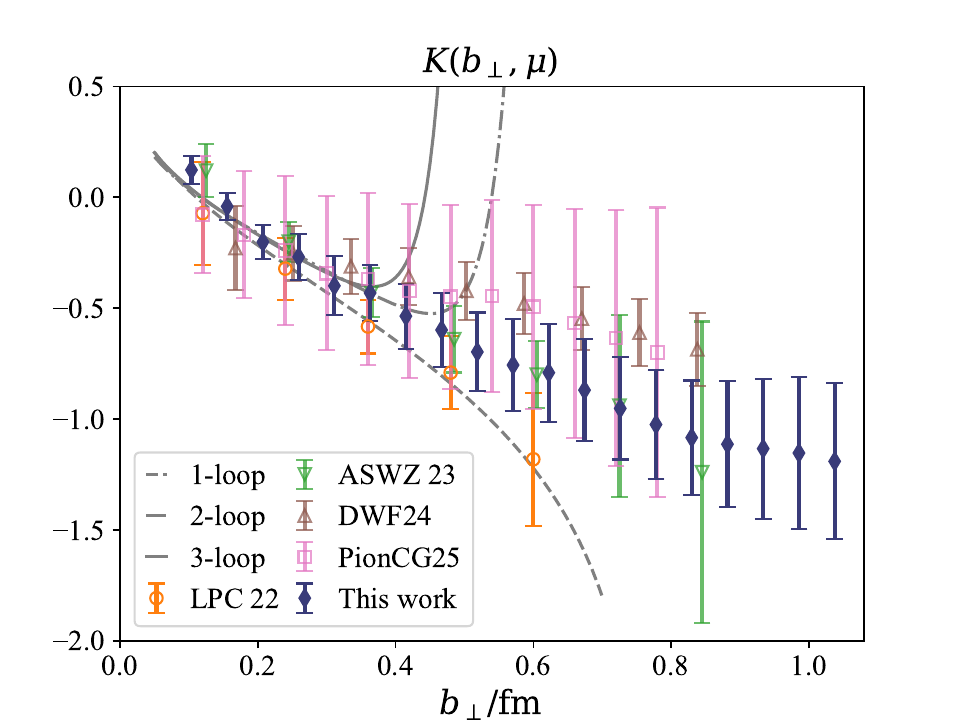}}
{\includegraphics[width=0.45\textwidth]{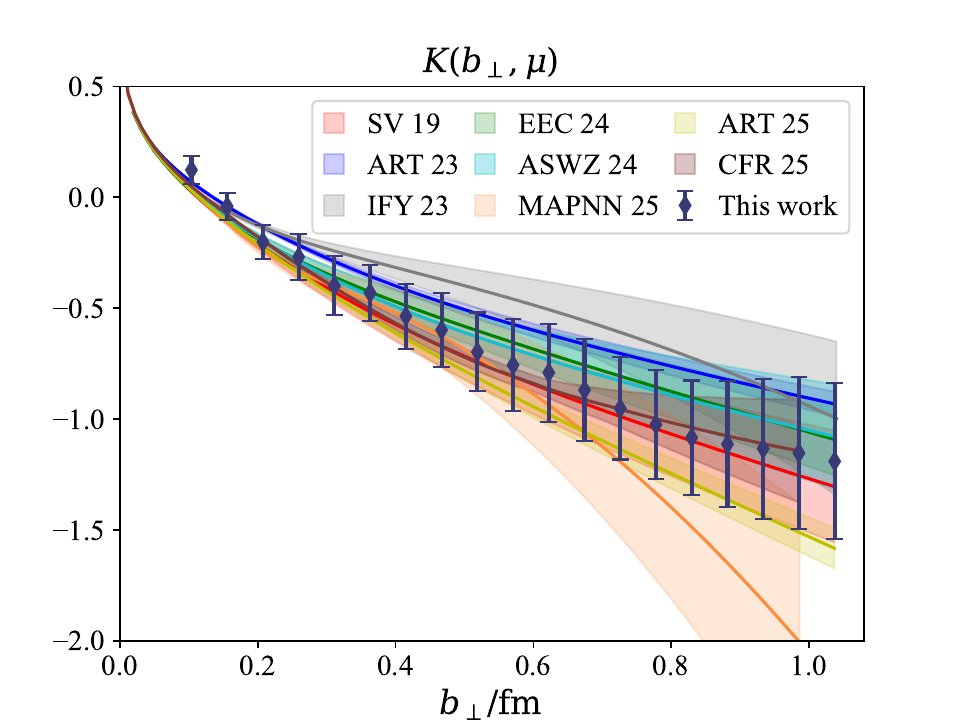}}
  \caption{Left panel: comparison of  the $K(b_\perp, \mu)$ results from LPC'25~\cite{Tan:2025ofx} denoted as ``This work'' with others lattice calculations by LPC'22~\cite{LatticePartonLPC:2022eev}, MIT/FNAL/ANL'24~\cite{Avkhadiev:2023poz} (labeled ``ASWZ24''), BNL/ANL'24~\cite{Bollweg:2024zet} (labeled ``DWF24''), BNL/ANL'25~\cite{Bollweg:2025iol} (labeled ``PionCG25''), as well as the perturbative calculations up to three loops~\cite{Li:2016ctv, Moch:2017uml}.
  Right panel: comparison of LPC'25 with phenomenological extractions from SV19~\cite{Scimemi:2019cmh}, ART23~\cite{Moos:2023yfa}, IFY23~\cite{Isaacson:2023iui}, EEC24~\cite{Kang:2024dja}, MIT/FNAL/ANL'24~\cite{Avkhadiev:2024mgd} (labeled as ``ASWZ24'' in the plot), MAPNN25~\cite{Bacchetta:2025ara}, ART25~\cite{Moos:2025sal}, CFR25~\cite{Camarda:2025lbt}.
}
  \label{fig:CS_kernel_result_LPC25}
\end{figure*}

LPC'25~\cite{Tan:2025ofx} stands as a landmark achievement in modern lattice-QCD calculation of TMD physics, representing the culmination of years of methodological advances toward high precision in the nonperturbative regime of nucleon structure. 
These include improved control over the continuum, physical quark mass, and infinite-momentum limits, which remain important sources of systematic uncertainty in existing lattice calculations of TMD-related observables.
The calculation employs hypercubic (HYP) smearing to improve signal-to-noise, nonperturbative renormalization, and $b_\perp$-unexpanded matching, enabling a reliable determination up to $b_\perp \sim 1\text{ fm}$. 
The results with their systematic uncertainties are illustrated in Fig.~\ref{fig:System_Error_Contribution}. As shown in Fig.~\ref{fig:CS_kernel_result_LPC25}, the LPC'25~\cite{Tan:2025ofx} results agree with three-loop perturbative QCD at small $b_\perp$ and are consistent with the recent phenomenological extractions.

The progress outlined above in determining the CS kernel reflects the evolution of lattice calculations from qualitative validation toward quantitative predictions that can distinguish among phenomenological parameterizations at the current level of precision.

\begin{landscape}
\begin{longtable}{|l|c|c|c|c|c|c|c|p{1.5cm}|}
\hline
\textbf{Collaboration(s)}
& \boldmath{$N_f$}
& \boldmath{$M_\pi^\textbf{val}$} \textbf{(MeV)}
& \boldmath{$a$} \textbf{(fm)}
& \textbf{Method}
& \textbf{\makecell{Matching\\ Kernel}}
&
\makecell{\boldmath{$P_z^\text{max}$}\\
\textbf{(GeV)} }
& \boldmath{$b_\perp^\text{max}$}~\textbf{(fm)}
& \textbf{\makecell{Syst. \\Uncert.}}
\\
\hline\hline
MIT'20~\cite{Shanahan:2020zxr} & 0 & 1200 &  0.06 & Quasi beam function & LO & 2.58 & 0.8
& No  \\ \hline
LPC'20~\cite{LatticeParton:2020uhz} & 2+1  & 547  & 0.098 & Quasi TMDWF & LO & 2.11 & 0.4
& No
\\ \hline
RQCD/LHPC'21~\cite{Schlemmer:2021aij} & 2+1  & 422  & 0.0854 & \makecell{Mellin moments \\of quasi TMDPDF} & -- & 2.27  & 0.5
& Estimated 
\\ \hline
MIT/FNAL/ANL'21~\cite{Shanahan:2021tst} & 2+1+1  & 538  & 0.12 & Quasi beam function & NLO & 1.5  & 0.5
& No
\\ \hline
PKU/ETMC'22~\cite{Li:2021wvl} & 2+1+1 & $ 827$  & 0.093 & Quasi TMDWF & LO & 3.3  & 0.3
& No
\\ \hline
LPC'22~\cite{LatticePartonLPC:2022eev} & 2+1+1  & 670  & 0.12 & Quasi TMDWF & NLO & 2.58  & 0.6
& Yes
\\ \hline
RQCD/LHPC'23~\cite{Shu:2023cot} & 2+1 & 686  &  0.0854 & \makecell{Mellin moments \\of quasi TMDPDF} & -- & 1.36  & 0.4
& No
\\ \hline
\hline
LPC'23~\cite{LatticePartonLPC:2023pdv} & 2+1+1& [220, 670]  & [0.098,0.121] & Quasi TMDWF & -- & 3.16  & 0.6
& Yes 
\\ \hline
MIT/FNAL/ANL'23~\cite{Avkhadiev:2023poz} & 2+1+1 & 148.8  & 0.12 & Quasi TMDWF & uNNLL & 2.16  & 0.8
& Yes
\\ \hline
MIT/FNAL/ANL'24~\cite{Avkhadiev:2024mgd}
 & 2+1+1
 & 148.8  & \makecell{$[0.09, 0.15]$ \\ $a\rightarrow 0 $} & Quasi TMDWF & uNNLL & 2.36  & 0.8
 & Yes
 \\ \hline
BNL/ANL'24~\cite{Bollweg:2024zet} & 2+1 &  140  & 0.084 & Quasi TMDWF & NLL & 1.85  & 1.0 
& Yes
 \\ \hline
BNL/ANL'25~\cite{Bollweg:2025iol} & 2+1 &  300  & 0.06 & Quasi TMDWF & NLL & 4.3  & 1.0 
& No
\\ \hline
ETMC'25~\cite{Alexandrou:2025xci}& 2+1+1 & [640, 830]  & 0.093 & Quasi TMDWF & -- & 3.336  & 0.5
& No
\\ \hline
LPC'25~\cite{Tan:2025ofx} & 2+1 & [136, 320] & \makecell{ [0.052, 0.105]
 \\ $a\rightarrow 0 $} &
Quasi TMDWF & uNLO & 3.0 & 1.0 & Yes    \\
\hline \hline
\caption{Summary of lattice calculations of the CS kernel to date.
MIT'20~\cite{Shanahan:2020zxr} is the first CS kernel extraction, while
LPC'20~\cite{Zhang:2020dbb} is the first extraction of CS kernel via quasi-TMDWFs. 
\label{tab:CS}
}
\end{longtable}
\end{landscape}

\subsection{Reduced soft function}

In TMD factorization, the soft function describes soft gluon radiation that is not collinear to either the incoming or outgoing partons.
Compared to the CS kernel, which governs the rapidity evolution of the soft function, the reduced soft function has been less explored in lattice QCD due to the presence of rapidity divergences and related renormalization challenges.
In this subsection, we collect various results for the reduced soft function in Table~\ref{tab:intrinsic_soft_function}.
Note that neither the continuum limit nor the limit of physical quark mass has been achieved in these calculations.
As a result, the current lattice extractions should be regarded as exploratory, providing qualitative insight rather than precision inputs for phenomenology.

LPC'20~\cite{LatticeParton:2020uhz} presents a pioneering work that carries out the first-ever lattice QCD computation of the rapidity-independent reduced soft function $S_I$ using a 2+1-flavor CLS dynamical ensemble with spacing $a=0.098$ fm and valence pion mass of 547 MeV. The $S_I$ is determined through the ratio between the large-momentum pseudoscalar meson form factor $F(b_\perp,P^z)$ and the squared magnitude of the zero-longitudinal-momentum quasi-TMDWF $|\phi(0,b_\perp,P^z)|^2$ under leading-order perturbative factorization, after removing ultraviolet divergences from quasi-TMDWFs using rectangular spacelike Wilson loop vacuum expectation values to eliminate Wilson-line self-energy singularities. Through joint two-state fits of the form factor correlators and two-point quasi-TMDWF correlators across three distinct boosted pion momenta $P^z=1.05,1.58,2.11$~GeV, corresponding to boost factors $\gamma=2.17,3.06,3.98$, LPC'20~\cite{LatticeParton:2020uhz} obtains the renormalization-scale-independent ratio of reduced soft function referenced to a  transverse separation $b_{\perp,0}=a$, then converts these bare lattice outcomes into the  $\overline{\text{MS}}$ renormalized reduced soft function by combining measured lattice ratios with perturbatively computed short-distance one-loop soft function values at the reference scale. The final extracted $S_{I,\overline{\text{MS}}}$ results at varying transverse distance $b_\perp$ show excellent consistency across all three meson boost momenta and reasonable agreement with one-loop perturbative QCD predictions at small \(b_\perp\), with quantified systematic uncertainties stemming from operator mixing effects (reaching roughly $5\%$ near $b_\perp\approx0.6$ fm) and residual finite-boost power corrections neglected in the leading-order LaMET approximation.

PKU/ETMC'22~\cite{Li:2021wvl} uses an $N_f = 2+1+1$ clover-improved twisted-mass fermion ensemble with lattice spacing $a = 0.093\text{ fm}$ and pion masses $m_\pi \in [350, 827]\text{ MeV}$ to address key challenges in reliable extractions of TMD-related matrix elements.
It identifies substantial higher-twist contamination arising from operators such as $\gamma_5$, $\sigma^{02}$, $\sigma^{12}$, and $\sigma^{23}$, which can even flip the sign of matrix elements. To mitigate this, improved pion matrix elements are constructed, e.g.
$\frac{1}{2}\left(F_{\gamma_5 \gamma_1} + F_{\gamma_1}\right),$
which explicitly cancel these higher-twist contributions.
The renormalization is performed using a ratio scheme (instead of RI/MOM), enabling cleaner cancellation of ultraviolet and logarithmic divergences. Systematic effects are further controlled through extrapolation of staple-shaped Wilson line lengths to infinity ($L \gtrsim 0.8\text{ fm}$) and two-state fits to suppress excited-state contamination.
The extracted reduced soft function agrees qualitatively with one-loop perturbative results at small transverse separations ($b_\perp \sim 0.2\text{ fm}$) and large external momenta ($\sim 3.3\text{ GeV}$), while showing clear nonperturbative behavior (slower decay) at larger $b_\perp$. It also exhibits weak pion-mass dependence, consistent with factorization expectations.

LPC'23~\cite{LatticePartonLPC:2023pdv} addresses the pivotal need for precise, nonperturbative determinations of the soft function, specifically its rapidity-independent component $S_I$ (also referred to as the reduced soft function). Using a $2+1$-flavor CLS ensemble with lattice spacing $a = 0.098\text{ fm}$ and pion mass $m_\pi = 547\text{ MeV}$, the reduced soft function $S_I$ is extracted from ratios of large-momentum-transfer pseudoscalar meson form factors and quasi-transverse-momentum-dependent wave functions, within a leading-order factorization framework.
Key technical steps include constructing quasi-TMDWFs with staple-shaped Wilson lines, canceling linear divergences using vacuum expectation values of rectangular Wilson loops, and performing joint fits of three-point and two-point correlators to isolate ground-state contributions. The analysis also accounts for systematic uncertainties from operator mixing (up to $\sim 5\%$ at $b_\perp \sim 0.6\text{ fm}$) and scale dependence of the strong coupling, with $\mu \in [1/2,2]\times 1/b_\perp$.
The results show consistency across pion momenta ($1.05$, $1.58$, and $2.11\text{ GeV}$), with $S_I$ agreeing with one-loop perturbative predictions at small transverse separations and exhibiting clear nonperturbative behavior at larger $b_\perp$. 

To further verify the universality of the reduced soft function across different lattice ensembles, LPC'23~\cite{LatticePartonLPC:2023pdv} employs two additional ensembles: a CLS $2+1$-flavor clover ensemble with lattice spacing $a = 0.098\text{ fm}$ and valence pion mass $m_\pi^{\text{val}} = 662\text{ MeV}$, and a MILC $2+1+1$-flavor HISQ ensemble with $a = 0.121\text{ fm}$ and $M_\pi^{\text{val}} = 670\text{ MeV}$.
Building on earlier studies, this work advances the methodology by incorporating one-loop perturbative matching in place of tree-level matching and by applying Fierz rearrangement to suppress substantial higher-twist contamination.
Key technical ingredients include the renormalization of quasi-TMDWFs using Wilson loops to cancel linear and pinch-pole divergences, together with a nonperturbative $Z_O$ factor to remove logarithmic divergences; two-state fits of three-point correlation functions to isolate ground-state contributions in the meson form factors; and the combination of Dirac structures ($\gamma_\perp + \gamma_\perp \gamma_5$) through Fierz rearrangement to eliminate dominant higher-twist effects arising from operators such as $\gamma_5$ and $\sigma_{02}$.
The results show that one-loop matching brings the extracted reduced soft function into closer agreement with one-loop perturbative predictions at small transverse separations ($b_\perp \sim 0.1$--$0.2\text{ fm}$), while larger $b_\perp$ values exhibit clear nonperturbative behavior, characterized by a slower decay than predicted by perturbation theory.
The two ensembles yield consistent results except at small $b_\perp$, where discretization effects become significant. The study also demonstrates the universality of the reduced soft function through comparison with previous lattice determinations and demonstrates its utility for extracting physical TMD wave functions and TMDPDFs.
These advances strengthen the reliability of nonperturbative reduced soft-function constraints relevant for TMD factorization in processes such as Drell-Yan production.

BNL/ANL'25~\cite{Bollweg:2025iol} employs the CG approach to avoid explicit Wilson lines, thereby simplifying renormalization and improving the signal-to-noise ratio. Using a $2+1$-flavor HotQCD ensemble with lattice spacing $a = 0.06\text{ fm}$, pion mass $m_\pi = 300\text{ MeV}$, and pion momenta up to $4.30\text{ GeV}$, the reduced soft function is extracted from ratios of large--momentum-transfer pion form factors and renormalized quasi-TMDWFs.
Key methodological advances include NLL RGR of the Sudakov kernel and the use of Fierz rearrangement to suppress higher-twist contamination.
The extracted reduced soft function agrees with one-loop perturbative predictions at small transverse separations $b_\perp$ and exhibits clear nonperturbative behavior at larger $b_\perp$, consistent with recent lattice and phenomenological studies.
The CG approach also enables reliable constraints at larger transverse separations, extending the accessible range to $b_\perp \gtrsim 1\text{ fm}$.

Last but not least, ETMC'25~\cite{Alexandrou:2025xci} presents a lattice-QCD study of the reduced soft function (or reduced soft function) within the LaMET framework, using two $N_f = 2+1+1$ twisted-mass fermion ensembles with $a = 0.093\text{ fm}$, lattice volumes $24^3 \times 48$ and $32^3 \times 64$, and pion masses of $640$ and $830\text{ MeV}$, respectively.
The reduced soft function is extracted through factorization of large--momentum-transfer meson form factors, obtained from four-point correlation functions, together with renormalized quasi-TMDWFs constructed using asymmetric staple-shaped Wilson lines.
Key methodological advances include verification that operator mixing is numerically negligible, enabling the use of the nonperturbative short-distance ratio (SDR) renormalization scheme; suppression of higher-twist contamination through linear combinations of Dirac structures, such as
$\frac{1}{2}\left(\gamma_\perp + \gamma_5 \gamma_\perp\right)$;
and incorporation of NLO matching corrections.
The extracted reduced soft function agrees with perturbative predictions at small transverse separations $b_\perp$ and exhibits clear nonperturbative behavior at larger $b_\perp$. The results show good consistency across the two ensembles, indicating minimal finite-volume effects, and are in agreement with previous lattice determinations.

\begin{longtable}{|l|c|c|c|c|c|c|p{1.5cm}|}
\hline
\textbf{Collaboration(s)}
& \boldmath{$N_f$}
& \boldmath{$M_\pi^\textbf{val}$} \textbf{(MeV)}
&  \boldmath{$a$} \textbf{(fm)}
& \textbf{\makecell{Matching\\ Kernel}}
&
\makecell{\boldmath{$P_z^\text{max}$}\\
\textbf{(GeV)} }
& \boldmath{$b_\perp^\text{max}$}~\textbf{(fm)}
& \textbf{\makecell{Syst. \\Uncert.}}
\\
\hline\hline
LPC'20~\cite{Zhang:2020dbb} & 2+1  & 547  & 0.098 & LO & 2.11~GeV & 0.7
& No
\\ \hline
PKU/ETMC'22~\cite{Li:2021wvl} & 2+1+1 & $\{350, 827\}$  & 0.093 & LO & 3.3  & 0.7
& No
 \\ \hline
LPC'23~\cite{LatticePartonLPC:2023pdv} & 2+1+1&  670  & [0.098,0.121] & NLO & 2.58  & 0.6
& No
 \\ \hline
BNL/ANL'25~\cite{Bollweg:2025iol} & 2+1 &  300  & 0.06 & NLO & 2.58/4.3  & 1.1
& No
\\ \hline
ETMC'25~\cite{Alexandrou:2025xci}& 2+1+1 & 640  & 0.093 &  NLO & 3.336  & 0.5
& No
\\
\hline
\hline
\caption{Summary of lattice calculations of the reduced soft function.
LPC'20~\cite{Zhang:2020dbb} is the first lattice QCD calculation. PKU/ETMC'22~\cite{Li:2021wvl} and ETMC'25~\cite{Alexandrou:2025xci} reduced the high-twist effects in form factors.
BNL/ANL'25~\cite{Bollweg:2025iol} used the CG method.}\label{tab:intrinsic_soft_function}
\end{longtable}

\subsection{Transverse-momentum--dependent parton distribution functions  and wave functions}

Due to significant progress in determining the CS kernel and reduced soft function, together with advances in the quasi-TMD formalism, lattice QCD has begun to provide \textit{ab initio} determinations of the proton and pion TMDPDFs.
These studies reveal a nearly identical transverse structure for the two hadrons and provide a important benchmark for global analyses.
Similarly, the TMDWFs, which are fundamental theoretical objects describing the 3D hadron structure, can now also be computed on the lattice.
In this section, we review these recent calculations of TMDPDFs and TMDWFs.
A summary of the current status of lattice determinations of TMDPDFs and TMDWFs is provided in Table~\ref{tab:TMDPDF_TMDWFs}.

\begin{figure*}[!th]
\centering
\includegraphics[scale=0.9]{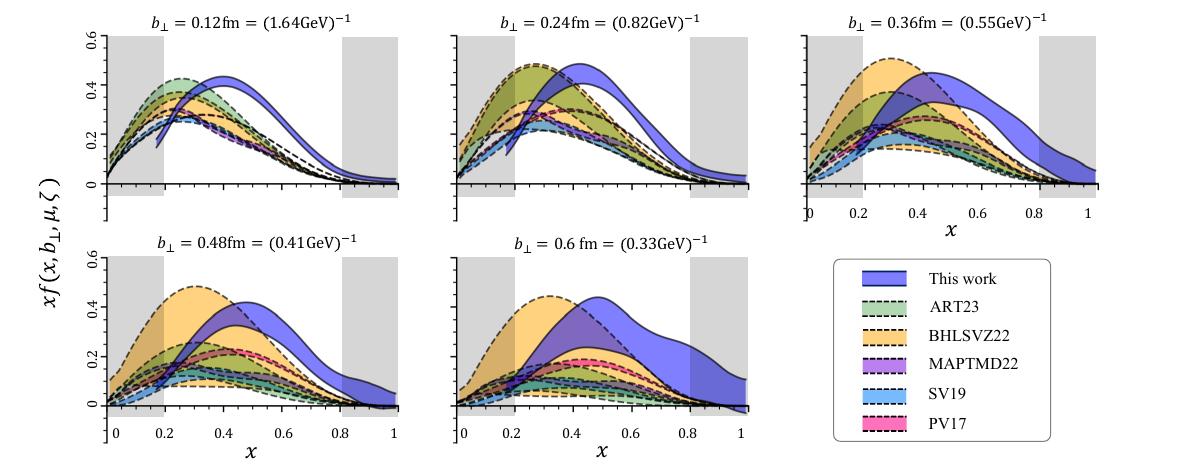}
\caption{LPC'22's~\cite{LatticePartonCollaborationLPC:2022myp} unpolarized isovector nucleon TMDPDFs, $x f(x, b_{\perp}, \mu, \zeta)$, are presented at renormalization and rapidity scales $\mu = \sqrt{\zeta} = 2\text{ GeV}$, extrapolated to the physical pion mass of 135~MeV and the infinite-momentum limit $P^z \to \infty$. The results are compared with global fits from ART23~\cite{Moos:2023yfa}, BHLSVZ22~\cite{Bury:2022czx}, MAPTMD22~\cite{Bacchetta:2022awv}, SV19~\cite{Scimemi:2019cmh}, and PV17~\cite{Bacchetta:2017gcc}.
The colored bands include both statistical and systematic uncertainties, while the shaded grey regions indicate the endpoint regions where LaMET-based predictions are not reliable.
}
\label{fig:Fig_finalTMDPDFs}
\end{figure*}

\begin{figure*}[!th]
\centering
\includegraphics[scale=0.95]{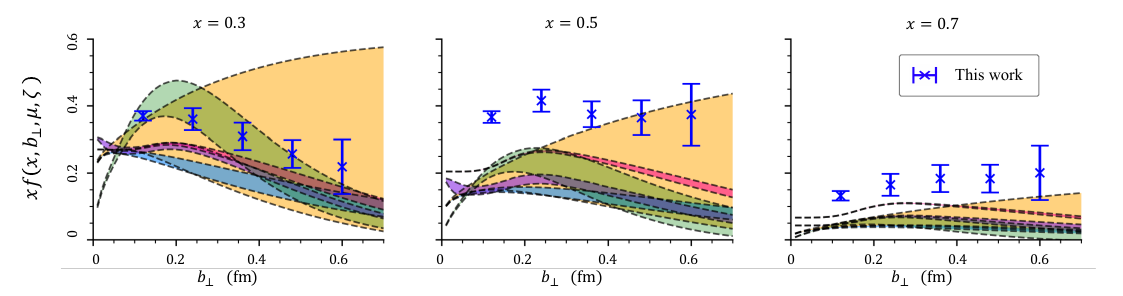}
\caption{LPC'22's~\cite{LatticePartonCollaborationLPC:2022myp} results for the TMDPDFs $x f(x, b_{\perp}, \mu = 2\text{ GeV}, \zeta = 4\text{ GeV}^2)$ at longitudinal momentum fractions $x = 0.3$, $0.5$, and $0.7$, together with corresponding phenomenological results.
The labels for the latter are the same as those in Fig.~\ref{fig:Fig_finalTMDPDFs}.
}
\label{fig:Fig_finalTMDPDFs_with_b}
\end{figure*}

LPC'22~\cite{LatticePartonCollaborationLPC:2022myp} performed the first lattice calculation of isovector unpolarized TMDPDFs, representing an important milestone. This work addresses long-standing challenges in the nonperturbative determination of TMDPDFs by adopting state-of-the-art techniques. In particular, it employs a $2+1+1$ MILC ensemble with HISQ sea quarks, implements renormalization via Wilson loops and short-distance matrix element matching, and includes next-to-next-to-leading order (NNLO) perturbative kernels with RGR.
Results are shown in Fig.~\ref{fig:Fig_finalTMDPDFs} and Fig.~\ref{fig:Fig_finalTMDPDFs_with_b}. 
By extrapolating to the physical pion mass and infinite momentum limit, 
and by carefully quantifying statistical and systematic uncertainties (including those arising from $\lambda$-extrapolation, physical quark mass and momentum extrapolations, and differences in Lorentz structure), the study demonstrates qualitative consistency between lattice results and phenomenological TMDPDF fits.

\begin{figure}
\centering
\includegraphics[scale=0.6]{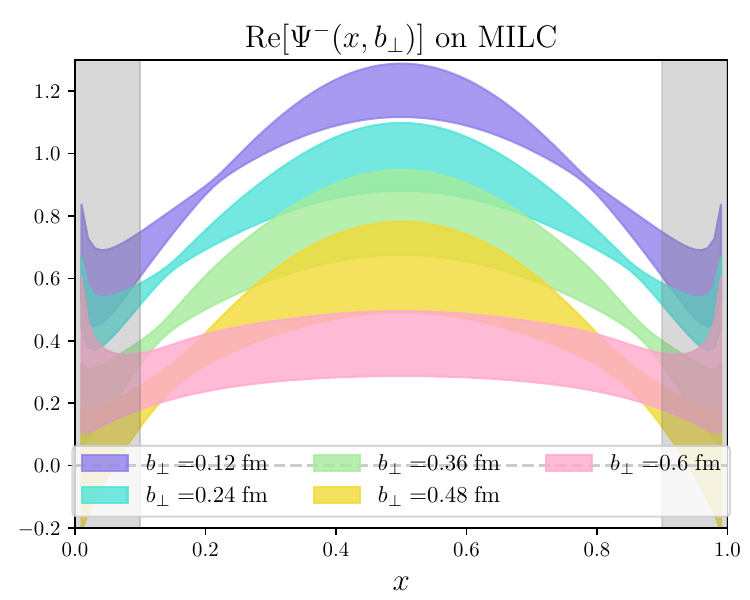}
\includegraphics[scale=0.6]{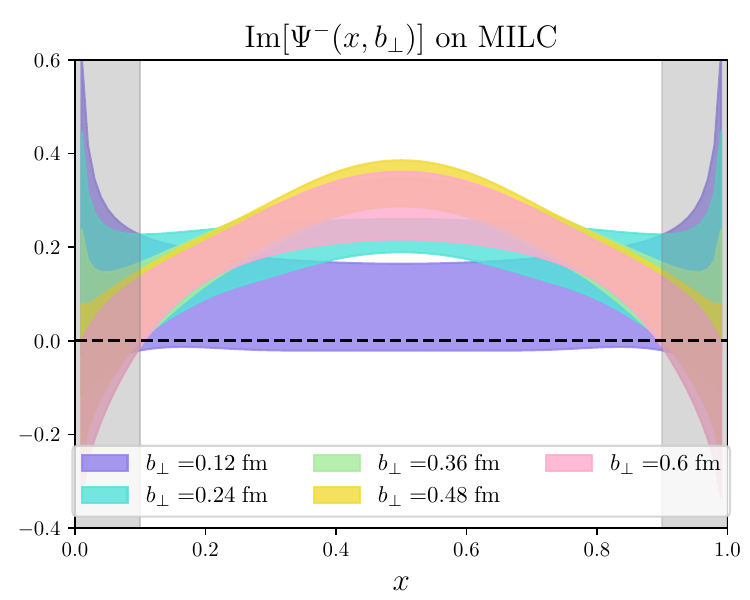}
\caption{LPC'23's~\cite{LatticeParton:2023xdl} results for the real (left) and imaginary (right) parts of the pion TMDWF obtained from the MILC ensemble.
The TMDWFs approach the infinite-$P^z$ limit with rapidity scale $\zeta = (6\text{ GeV})^2$ and renormalization scale $\mu = 2$~GeV.
}
\label{fig:lcwf}
\end{figure}

\begin{figure}
\centering
\includegraphics[scale=0.65]{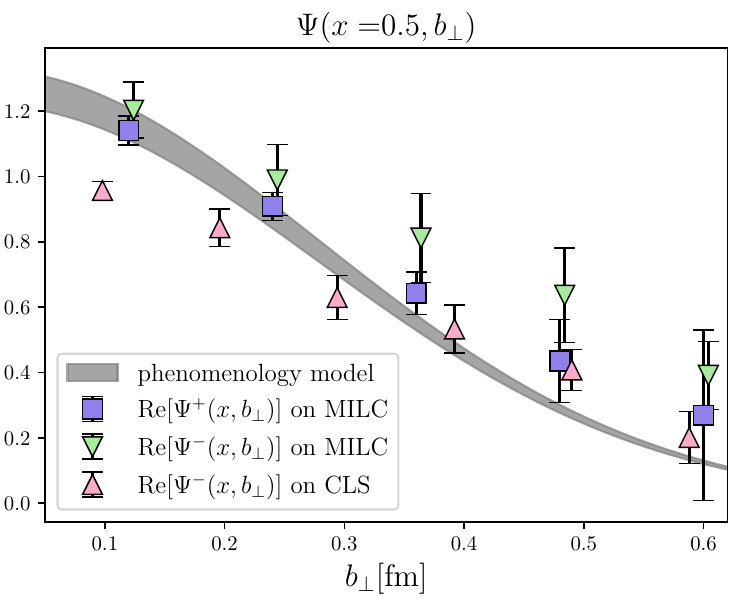}
\caption{Comparison of the TMDWFs from LPC'23~\cite{LatticeParton:2023xdl} with a phenomenological model~\cite{Lu:2007hr} at $x = 0.5$, evaluated at rapidity and renormalization scales $\{\zeta,\mu\} = \{(6\text{ GeV})^2, 2\text{ GeV}\}$.}
\label{fig:pheno}
\end{figure}

To fill a longstanding gap in \textit{ab initio} studies of hadronic 3D structure in exclusive processes, LPC'23~\cite{LatticeParton:2023xdl} presented the first lattice calculation of the pion TMDWF. 
The calculation employed MILC's $2+1+1$-flavor HISQ ensemble and CLS's $2+1$-flavor clover-fermion ensemble to enable a nontrivial cross-validation of the results.
The study removed linear and logarithmic divergences through Wilson-loop and quark--Wilson-line vertex renormalization, systematically controlled excited-state contamination, and utilized analytical extrapolation techniques to avoid unphysical oscillations in the Fourier transform.
The renormalized quasi-TMDWFs were then combined with one-loop perturbative matching, CS evolution kernels, and reduced soft function to obtain the corresponding lightcone TMDWFs.
The resulting distributions, shown in Figs.~\ref{fig:lcwf} and~\ref{fig:pheno}, exhibit consistent dependence on the transverse separation $b_\perp$ across the two ensembles and show qualitative agreement with phenomenological results. 
In addition, the calculation revealed non-negligible imaginary components, which are often neglected in phenomenological analyses, as well as distinct behaviors of $\psi^+$ and $\psi^-$ arising from the orientation of Wilson lines. 
These results challenge several simplifying assumptions commonly adopted in phenomenological models and provide reliable first-principles input for the study of exclusive processes. 
More broadly, they represent an important step toward a deeper understanding of nonperturbative QCD dynamics, including confinement.

\begin{figure}[h!]
\centering
\includegraphics[width=0.45\textwidth]{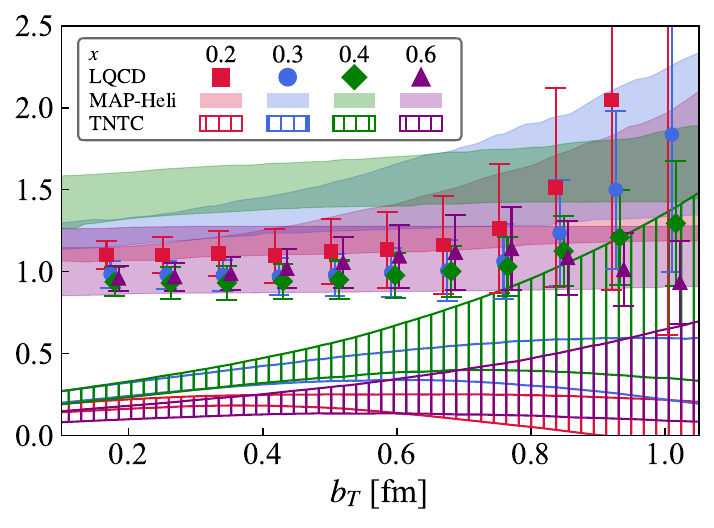}
\includegraphics[width=0.45\textwidth]{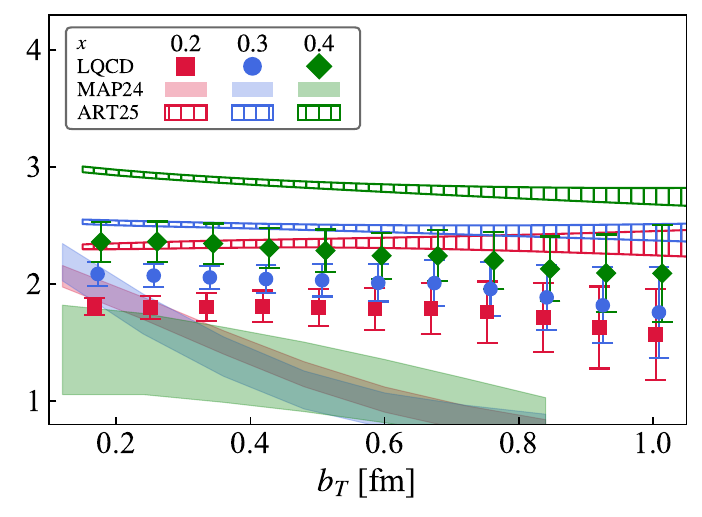}
\caption{
Left: BNL/ANL'25b~\cite{Bollweg:2025ecn} results (filled symbols labeled ``LQCD") 
of the isovector helicity to isovector unpolarized TMDPDFs, 
shown as a function of $b_T$ for several values of $x$.
The hatched bands indicate the corresponding global-fit results from MAP-Heli~\cite{Bacchetta:2024yzl} and TNTC~\cite{Yang:2024drd}.
Right: BNL/ANL'25b~\cite{Bollweg:2025ecn} results (filled symbols labeled ``LQCD") 
of up- to down-quark unpolarized TMDPDFs, 
shown as a function of $b_T$ for several values of $x$.
The hatched bands denote the corresponding global-fit results from MAP24~\cite{Bacchetta:2024qre} and ART25~\cite{Moos:2025sal}.
}
\label{fig:HeliUnpol-UnpolUD}
\end{figure}

BNL/ANL'25b~\cite{Bollweg:2025ecn} presents a lattice calculation of the proton's isovector helicity TMDPDF and flavor-dependent unpolarized TMDPDFs for up and down quarks at physical quark masses. 
This work employs the CG method (see Sec.~\ref{subsec:cg}) within the LaMET
framework. 
Using domain-wall fermion discretization and $2+1$-flavor gauge configurations, the study probes nonperturbative transverse separations up to $b_\perp \sim 1$~fm, corresponding to transverse momenta $k_\perp \sim 200$~MeV, and constructs renormalization-group--invariant ratios for direct comparison with phenomenological fits.
The results reveal nearly identical transverse structures for the isovector helicity and unpolarized TMDPDFs at moderate $x$, along with only mild flavor dependence in the unpolarized distributions. Lattice results, global fits from MAP-Heli~\cite{Bacchetta:2024yzl} and TNTC~\cite{Yang:2024drd}  are collected in Fig.~\ref{fig:HeliUnpol-UnpolUD}. 
These results can distinguish among different global fits, establish an important first-principles benchmark for future TMDPDF analyses, and support upcoming experimental efforts.

\begin{figure}[thbp]
\centering
\includegraphics[width=.58\textwidth]{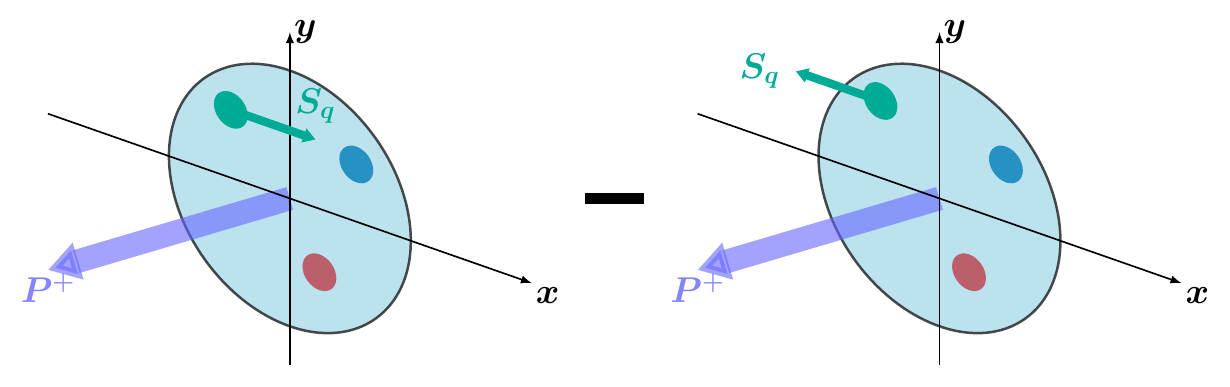}
\centering
\includegraphics[width=0.4\textwidth]{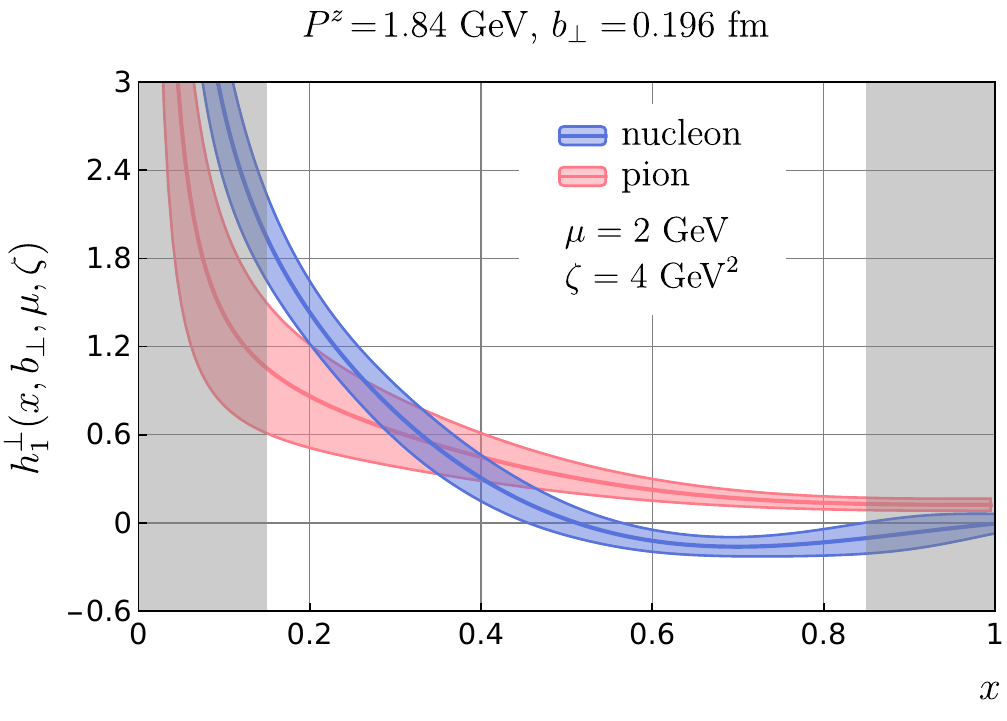}
\caption{
Left:
Illustration of the physical interpretation of the Boer-Mulders TMDPDF, where $S_q$ denotes the quark spin.
Right:
Comparison of the Boer-Mulders functions for the nucleon and pion, shown for $P^z = 1.84$~GeV and $b_\perp = 0.196$~fm as an example,from LPC'24~\cite{LatticeParton:2024mxp} and LPC'25~\cite{LatticeParton:2025eui}.
The shaded regions indicate estimated unreliable regions, where power corrections are also expected to become significant.
}
\label{fig:BM:intrp}
\end{figure}

LPC'24~\cite{LatticeParton:2024mxp} and  LPC'25~\cite{LatticeParton:2025eui},
focus on the time-reversal-odd Boer-Mulders function, describing the correlation between quark transverse polarization and momentum, for the pion and nucleon, respectively.
An illustration of this function and lattice results can be found in Fig.~\ref{fig:BM:intrp}.
The pion-focused work employs three lattice ensembles with spacings $a=0.098,0.085,0.064$~fm, pion masses close to 350~MeV, and momenta up to 1.8~GeV, implementing NNLO perturbative matching and RGR, extrapolating to the continuum and infinite-momentum limits, and finding the function decays with increasing transverse separation $b_\perp$ and approaches zero at $b_\perp\approx 0.5$~fm.
The nucleon-focused study uses an ensemble with $a=0.098$~fm and pion mass 338~MeV, proton momenta up to 2.11~GeV, and applies the same advanced perturbative matching and RGR; its results exhibit a similar $b_\perp$-decay trend, share the same sign as the pion's Boer-Mulders function, but show a more rapid decrease with momentum fraction $x$ compared to the pion.
Together, these first-principles calculations fill the gap in nonperturbative constraints for $T$-odd TMDPDFs, validate consistency across hadron species, and provide useful inputs for phenomenological analyses of single-spin asymmetries.

\begin{figure}[th!]
    \centering
    \includegraphics[width=0.4\linewidth]{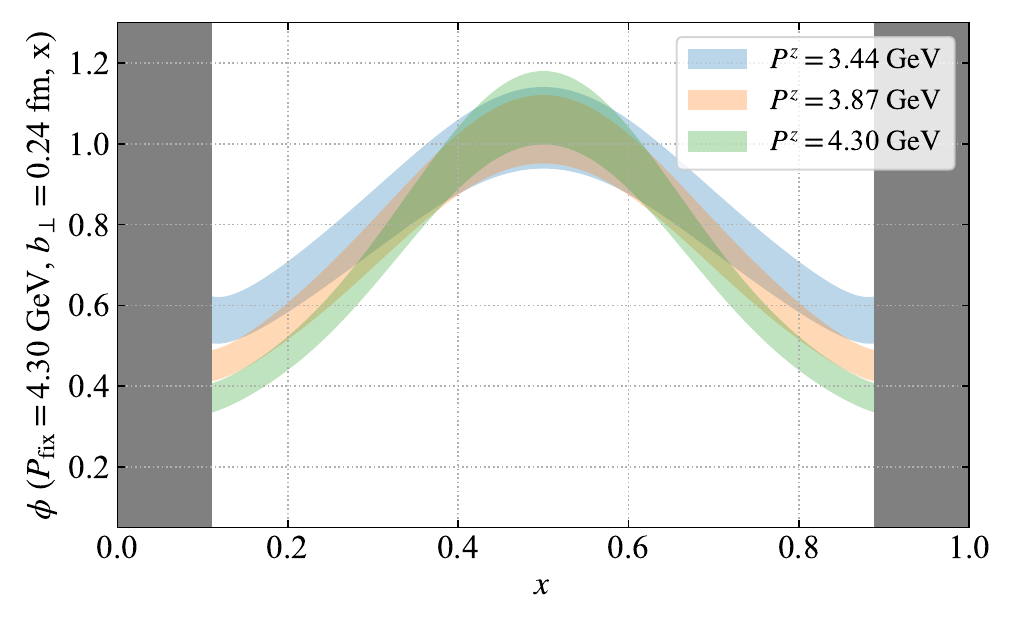}
    \includegraphics[width=0.4\linewidth]{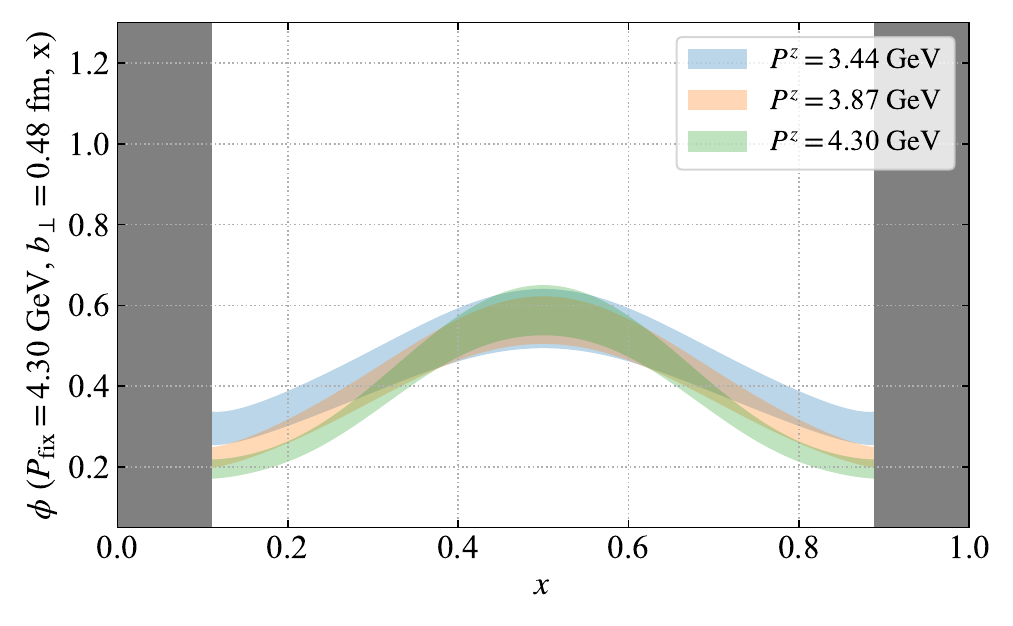}
    \caption{BNL/ANL'25a's~\cite{Bollweg:2025iol} lightcone pion TMDWF at $y_n = 0$, $P_{\text{fix}} = 4.30$~GeV, and $\mu = 2$~GeV, shown as a function of momentum fraction $x$ for different hadron momenta $P^z$, and for two transverse separations $b_\perp = 4a$ (upper panel) and $b_\perp = 8a$ (lower panel).
    The shaded gray bands ($x < 0.11$ and $x > 0.89$) indicate endpoint regions where the estimated combined systematic uncertainties exceed $30\%$.
    }
    \label{fig:tmdwf_xdep}
\end{figure}

\begin{figure}[th!]
    \centering
    \includegraphics[width=0.45\linewidth]{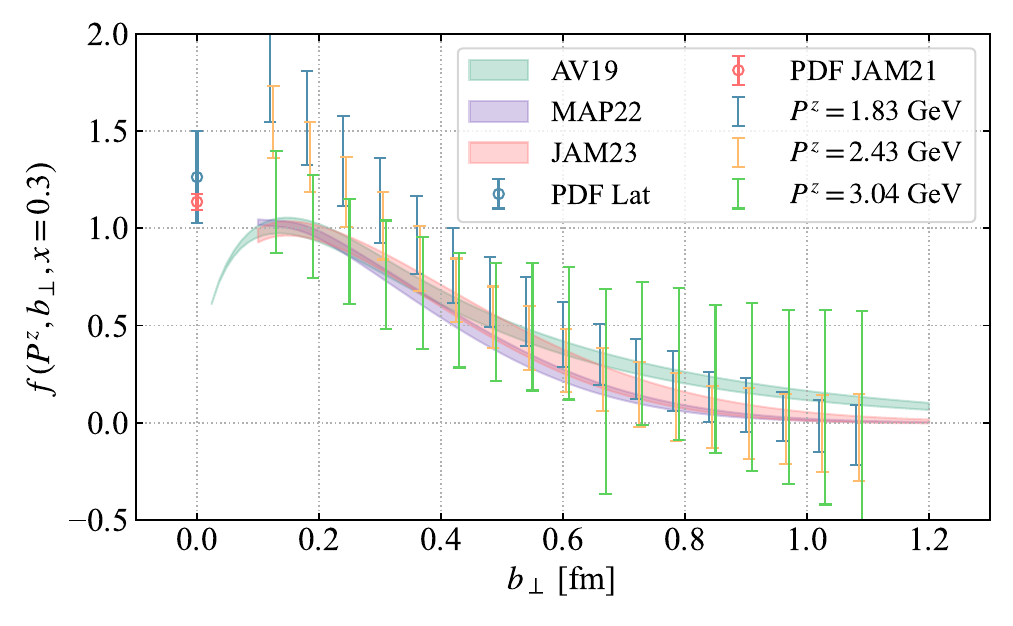}
    \includegraphics[width=0.45\linewidth]{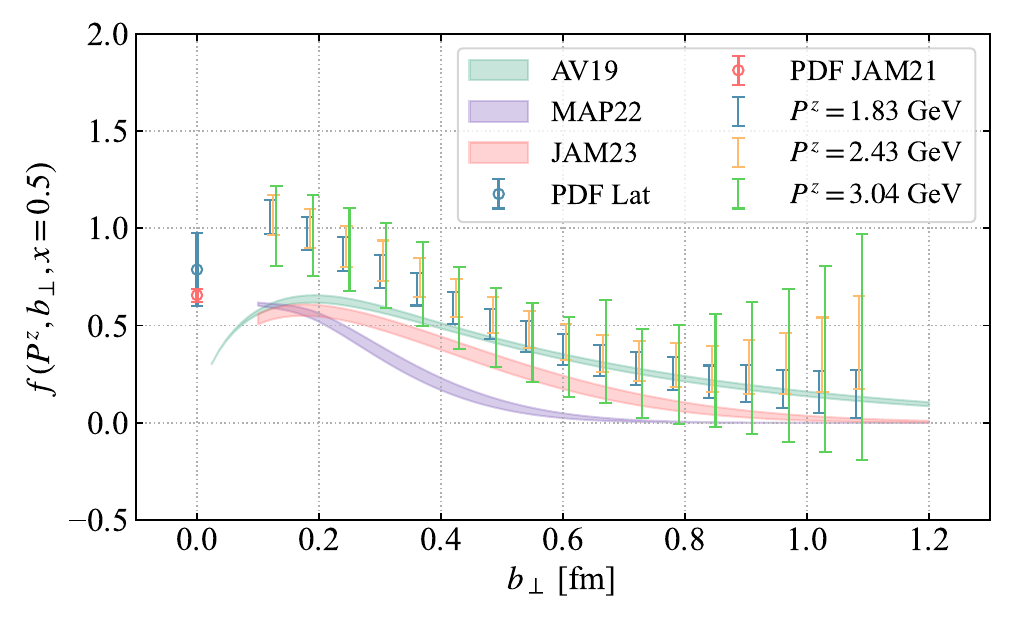}
    \caption{BNL/ANL'25a's~\cite{Bollweg:2025iol} unpolarized lightcone pion TMDPDF at $x = 0.3$ and $0.5$ in $b_\perp$ space, extending up to $b_\perp \gtrsim 1$~fm and shown for different hadron momenta $P^z$.
    The lattice results are compared with phenomenological determinations from AV19~\cite{Vladimirov:2019bfa}, MAP22~\cite{Cerutti:2022lmb}, and JAM23~\cite{Barry:2023qqh}, as well as with the collinear PDF obtained from lattice QCD (``PDF Lat'')~\cite{Gao:2023lny} and global fit (``PDF JAM2''1)~\cite{Barry:2021osv}.}
    \label{fig:tmdpdf_bdep}
\end{figure}

BNL/ANL'25a~\cite{Bollweg:2025iol}  computes pion unpolarized valence-quark TMDPDFs using the CG approach, based on a $2+1$-flavor HotQCD ensemble with $a = 0.06$~fm and pion masses of 300 and 670~MeV; results are shown in Fig.~\ref{fig:tmdpdf_bdep}.
Combining NLL matching, reduced soft function, and CS evolution kernels, the study extracts $x$-dependent TMDPDFs that show reasonable agreement with phenomenological parametrizations at moderate $x$.
BNL/ANL'25a~\cite{Bollweg:2025iol} also computes pion TMDWFs within the same framework; results are shown in Fig.~\ref{fig:tmdwf_xdep}.
Quasi-TMDWFs are extracted at hadron momenta up to 4.30~GeV and combined with reduced soft function and CS kernel to obtain lightcone TMDWFs, which show qualitative agreement with phenomenological fits and extend coverage to transverse separations $b_\perp \gtrsim 1$~fm.

\begin{landscape}
\begin{longtable}{|l|c|c|c|c|c|c|c|p{1.5cm}|}
\hline
\textbf{Quantity}
&
\textbf{Collaboration(s)}
& \boldmath{$N_f$}
& \boldmath{$M_\pi^\textbf{val}$} \textbf{MeV)}
&  \boldmath{$a$} \textbf{(fm)}
& \textbf{\makecell{Matching\\ Kernel}}
&
\makecell{\boldmath{$P_z^\text{max}$}\\
\textbf{GeV)} }
& \textbf{\makecell{Syst. \\Uncert.}}
\\
\hline\hline
\hline
Nucleon unpolarized TMDPDF & LPC'22~\cite{LatticePartonCollaborationLPC:2022myp} & 2+1+1 & 310, 220~MeV & 0.12 & NNLO & 2.58~GeV &  Yes
\\
\hline
Pion TMD wave function & LPC'23~\cite{LatticeParton:2023xdl} & 2+1+1 & 670~MeV & 0.12 & NLO & 2.64~GeV &  No
\\
\hline
Pion Boer-Mulders function & LPC'24~\cite{LatticeParton:2024mxp} & 2+1 & 350~MeV &  \{0.098, 0.085, 0.064\}&  NNLO & 1.8~GeV   & No
\\
\hline
Nucleon Boer-Mulders function & LPC'25~\cite{LatticeParton:2025eui} & 2+1& 338~MeV &  0.098 & NNLO & 2.21~GeV &   No 
\\
\hline
Pion unpolarized TMDPDF/TMDWFs & BNL/ANL'25a~\cite{Bollweg:2025iol} & 2+1 & 300~MeV & 0.06 & NLO & 3.04~GeV &  No 
\\
\hline
Proton  Unpolarized and Helicity  TMDPDFs  & BNL/ANL'25b~\cite{Bollweg:2025ecn} & 2+1 & 135~MeV & 0.084 & NLO & 1.62~GeV &  No 
\\
\hline
\hline
\caption{Summary of other lattice-QCD TMD-related calculations.} \label{tab:TMDPDF_TMDWFs}
\end{longtable}
\end{landscape}

\section{Toward Phenomenology and Next-Generation Developments}\label{sec:Developments}

\subsection{Impact on phenomenology}
\label{sec:pheno}

The TMDPDFs are key to 3D nucleon imaging in momentum space.
In global analyses based on the Collins-Soper-Sterman formalism~\cite{Collins:1981uk,Collins:1984kg,Collins:2011zzd} for a factorization formula like Eq.~\eqref{eq:fact_thm_generic}, the TMDPDF is usually parameterized at initial scales $(\mu_i,\zeta_i)\sim (1,1)$~GeV and evolved to arbitrary scales $(\mu,\zeta)\sim (Q,Q)$ as
\begin{equation}
    f(x, b_\perp, \mu, \zeta) = f(x, b_\perp, \mu_i, \zeta_i) \exp\Big(\int_{\mu_i}^\mu {d\mu'\over \mu'}\gamma_\mu(\mu',\zeta)\Big)\left({\zeta\over \zeta_i}\right)^{K(b_\perp,\mu)/2}\,,\label{eq:tmdpdf_evolution}
\end{equation}
where $\gamma_\mu$ is the UV anomalous dimension of the TMDPDF.
At large $b_\perp$, both the nonperturbative TMDPDF $f(x, b_\perp, \mu_i, \zeta_i)$ and
the CS kernel $K(b_\perp,\mu)$ must be modeled or priored, posing a major challenge
for uncertainty quantification. Lattice calculations of both quantities can provide
useful constraints on the global analyses, because the lattice quantities are the
same objects that enter the fits. The Collins-Soper (CS) kernel is universal and enters
through the evolution factor of Eq.~\eqref{eq:tmdpdf_evolution}, while the lattice
TMDPDFs are converted from the lattice renormalization scheme to $\overline{\text{MS}}$
and evolved with the same kernel to common values of $(\mu,\zeta)$, after which they
are compared with the fitted $f(x,b_\perp,\mu,\zeta)$ on the same footing as the
experimental data. Renormalization-group--invariant ratios, such as those used in the
Coulomb-gauge calculations~\cite{Bollweg:2025ecn}, can be compared with global fits
even before an absolute normalization is available.

With improved systematic control, especially with continuum extrapolation and physical quark masses~\cite{Avkhadiev:2024mgd,Tan:2025ofx}, lattice calculations have already reached a precision sufficient to differentiate the phenomenological models for the CS kernel and nonperturbative TMDPDFs. 
A joint extraction of the CS kernel from integrating experimental Drell-Yan data and lattice-QCD results was conducted in Ref.~\cite{Avkhadiev:2025wps}.
Based on a neural-network parametrization, the study adopted two complementary strategies, Bayesian reweighting of existing TMD fits and simultaneous fitting of experimental and lattice data, to quantify the impact of lattice inputs.
A key discovery was that the inclusion of lattice data (from continuum-extrapolated and finite lattice-spacing ensembles) shifts the central value of the nonperturbative parameter $g_2$ (governing large-$b_\perp$ CS-kernel behavior) by about $10\%$ and reduces its uncertainty by 40--50\%, while maintaining consistency with experimental data (no tension observed in the Drell-Yan description).
This new direction addresses the longstanding problem that the CS kernel has no direct experimental observables, using lattice QCD's direct nonperturbative constraints to mitigate model dependence in TMD evolution.
It provides a promising framework for 
integrating multiple data types, offering a more precise and robust path to constrain TMD-related quantities and deepen the understanding of nucleon 3D structure.

Meanwhile, it should be noted that lattice calculations and global fits overlap in a well-defined kinematic region. Present lattice
calculations cover roughly $x\in[0.2,0.8]$, with the endpoints excluded by power
corrections of order $\Lambda_\text{QCD}/(xP^z)$, and $b_\perp\lesssim 1$~fm,
corresponding to transverse momenta $k_\perp\gtrsim 0.2$~GeV, whereas the experimental
data extend down to $x\sim 10^{-3}$ and up to $q_\perp\sim 10$~GeV. The overlap is
therefore at moderate $x$ and low transverse momentum, which is precisely where the
global fits are dominated by their nonperturbative $b_\perp$-dependent priors; this is
where the lattice input has the largest impact. Reaching smaller
$x$ requires larger hadron momenta, which is one of the limitations discussed in
Sec.~\ref{sec:conclusion}.

\begin{figure}
    \centering
    \includegraphics[width=0.5\linewidth]{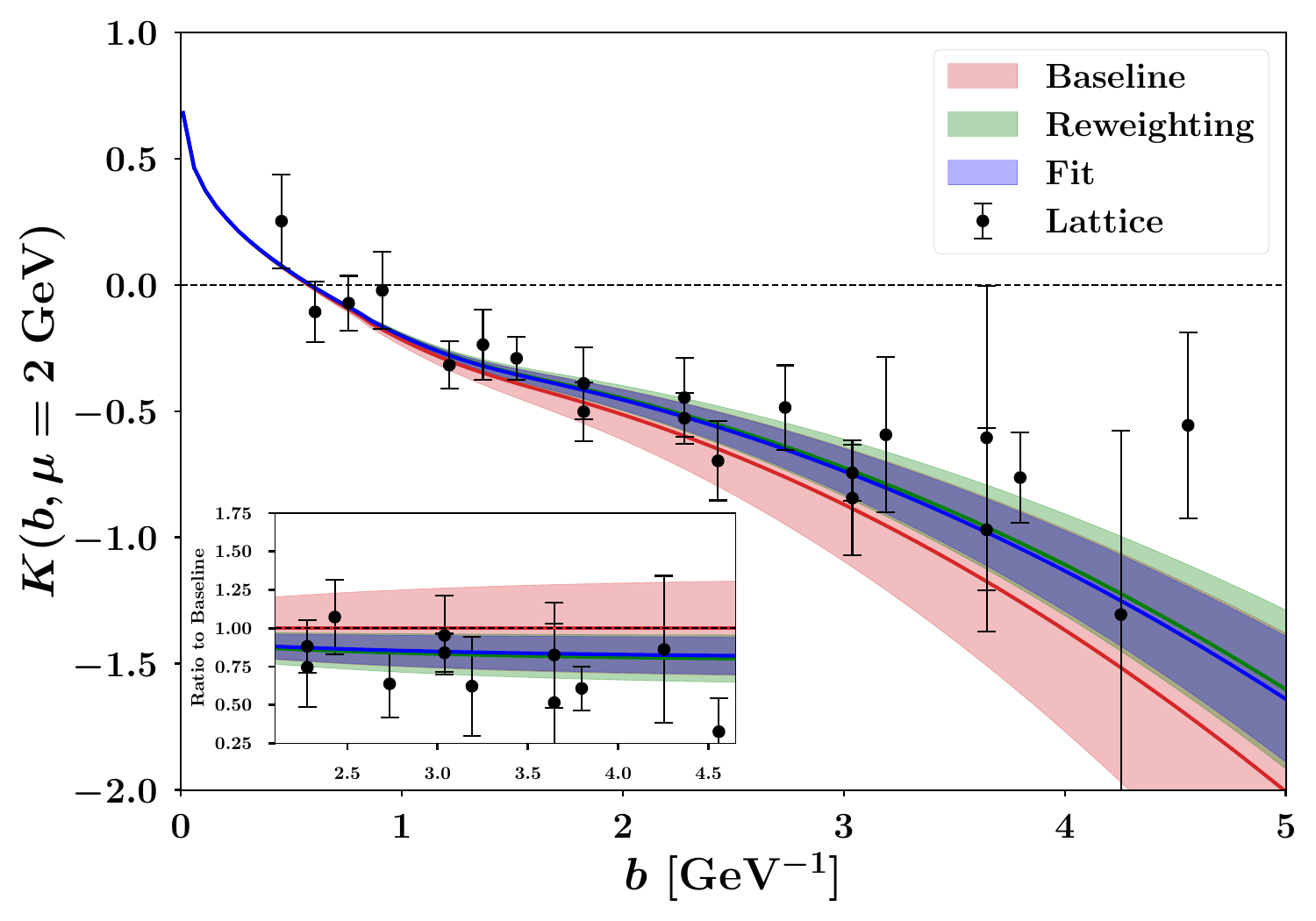}
    \caption{The CS kernel as a function of $b$ at $\mu= 2$~GeV as obtained from the baseline of fit of Ref.~\cite{Bacchetta:2025ara} (red band), from  reweighting (green band), and from the simultaneous fit of experimental and lattice data (blue band). Bands correspond to one-$\sigma$ uncertainties. Data points correspond to the extraction of Ref.~\cite{Avkhadiev:2024mgd}. The bottom inset shows the ratio to the baseline predictions at large values of $b$.}
    \label{fig:CS_comparison}
\end{figure}

Recently, an analysis of the CS kernel leveraging the artificial intelligence agent PhysMaster has been presented in Refs.~\cite{Miao:2025sms,Tan:2026ier}. Notably, using bare lattice data, PhysMaster yields results for the CS kernel that are consistent with those reported in Ref.~\cite{Tan:2025ofx}, featuring similar central values and comparable uncertainty bounds. This consistency is particularly encouraging, as it validates the potential of AI-driven tools to streamline and reinforce nonperturbative QCD calculations. Currently, efforts to enhance PhysMaster by integrating automatic lattice QCD calculation workflows are underway, which is expected to further boost efficiency and reliability in future TMD physics studies.

Unlike the CS kernel, a continuum-extrapolated pion or nucleon TMDPDF is not available yet. Nevertheless, recent calculations based on single lattice ensembles~\cite{LatticePartonCollaborationLPC:2022myp,Bollweg:2025iol,Bollweg:2025ecn} demonstrate that it is possible to reach the precision to differentiate contemporary global analyses. For example, as shown in Fig.~\ref{fig:HeliUnpol-UnpolUD}, the ratio of the isovector helicity TMDPDF to the unpolarized one predicted by lattice QCD~\cite{Bollweg:2025ecn} shows better agreement with MAP-Heli~\cite{Bacchetta:2024yzl} over TNTC~\cite{Yang:2024drd}. Although the lattice-spacing dependence has not been quantified yet, based on previous calculations of other parton observables, it is expected that they will not significantly change the behavior of these renormalization-group--invariant ratios.

As measurements of Drell-Yan cross sections at the LHC have reached percent-level precision, their transverse-momentum spectra have become increasingly sensitive to nonperturbative inputs, including the CS kernel and TMDPDFs, up to $q_\perp\sim 10\text{ GeV}$~\cite{Billis:2024dqq}. This sensitivity can directly affect precision determinations of observables such as the $W/Z$ boson masses~\cite{CDF:2022hxs,CMS:2024lrd} and the strong coupling constant~\cite{Cridge:2025wwo}. In 2022, the CDF Collaboration reported a precision measurement of the $W$ boson mass~\cite{CDF:2022hxs} that exceeded the contemporary Standard-Model prediction by more than $7\sigma$, although this result was not confirmed by subsequent measurements, including the recent CMS analysis~\cite{CMS:2024lrd}. It has been argued in Ref.~\cite{Bacchetta:2018lna} that flavor-dependent nonperturbative TMDPDF parameters can induce a non-negligible shift in the extracted $W$ mass.

Another example is the energy-energy correlator (EEC)~\cite{Moult:2025nhu} in the $e^+e^-$ collision process~\cite{Electron-PositronAlliance:2025fhk}. In the back-to-back region, the observable admits a TMD factorization description, where the CS kernel enters as a key nonperturbative input for predicting the EEC observables. In particular, the track-based EEC has achieved substantially improved angular resolution experimentally. The theoretical prediction~\cite{Jaarsma:2025tck} shows remarkable agreement with data over a broad angular range, except in the region corresponding to transverse momenta of $1$--$10\text{ GeV}$. This discrepancy is primarily attributed to nonperturbative inputs, among which the CS kernel was taken from the lattice calculation in Ref.~\cite{Avkhadiev:2024mgd}. Better constraint on the CS kernel and the remaining nonperturbative input will be crucial for resolving this inconsistency, thus enabling more accurate and precise Standard Model predictions for the EEC observables.

In summary, precision lattice calculations of TMD physics are playing an increasingly important role in both high-energy particle and nuclear experiments.

\subsection{Other TMD observables}
\label{sec:other}

With significant progress made in the lattice calculation of quark CS kernel and TMDPDFs, the natural next step is to extend to the gluonic sector. The gluon quasi-TMDs can be defined in a similarly to the quark case, where the adjoint Wilson lines are constructed from gauge links in the fundamental representation~\cite{Fan:2018dxu}. The factorization formula that relates the gluon quasi-TMD to the lightcone TMDPDF has been derived in Ref.~\cite{Ebert:2022fmh}, which follows the same form as Eq.~\eqref{eq:quasiTMDmatch}. Since the UV divergences of the gluon quasi-TMD correlators depend on the choice of Lorentz indices in the gluon field strength tensor, the matching coefficients need to be modified accordingly, which have been computed at one-loop order~\cite{Schindler:2022eva,Zhu:2022bja,Zhao:2022qym,Xie:2025rrw}.

However, the method to calculate the soft factor in the adjoint representation is not available yet. Although the operator definition of the soft factor can be obtained from the derivation in Ref.~\cite{Ebert:2022fmh}, it cannot be directly simulated on the lattice. One may construct a similar meson form factor as Eq.~\eqref{eq:meson_FF}, where the quark currents are replaced by gluonic bilinear operators, and check if it satisfies a factorization formula like Eq.~\eqref{eq:form_fac_bare}. But the other challenge is the computational cost of such matrix elements, as the leading-power contribution is expected to come from the disconnected diagrams at large momentum transfer~\cite{LatticeParton:2020uhz}.

Part of the gluon TMD program is already within reach. Without the soft factor, it is still possible to calculate the gluon CS kernel and ratios of spin-dependent TMDPDFs. Thanks to the absence of mixing between the quark and gluon channels in the factorization formula~\cite{Ebert:2022fmh}, such a calculation is also greatly simplified, since the disconnected quark contributions are computationally expensive. The first lattice calculation of the gluon CS kernel is carried out in Ref.~\cite{Avkhadiev:2026xyf}, where nonvanishing results are obtained at nonperturbative $b_\perp$. Simulations with CG-fixed correlators have also demonstrated that a gluon separation
of about $1.0$~fm is accessible~\cite{Good:2024iur}.

The full gluon TMDPDF is not yet accessible. The soft factor in the adjoint
representation has no lattice-simulable definition, although its operator form can be
obtained from the derivation of Ref.~\cite{Ebert:2022fmh}. Without it, the absolute
normalization of the gluon TMDPDF cannot be fixed. Whether a meson form factor built
from gluonic bilinears in analogy with Eq.~\eqref{eq:meson_FF} satisfies a factorization
formula analogous to Eq.~\eqref{eq:form_fac_bare} remains to be examined. Furthermore, the leading-power contribution is expected to come from
disconnected diagrams, whose cost is a key obstacle~\cite{LatticeParton:2020uhz}.
The corresponding CG matching formalism is still under development.

In any case, the simulation of gluonic observables, such as the gauge-invariant gluon
quasi-PDF matrix elements~\cite{Fan:2018dxu,Fan:2020cpa,HadStruc:2021wmh,Salas-Chavira:2021wui,HadStruc:2022yaw,Fan:2022kcb,Delmar:2023agv,Good:2023ecp,Good:2024iur,ChenChen:2025amm,NieMiera:2025inn}, is usually very noisy and requires high
statistics. When it comes to the gluon quasi-TMD case, the large staple-shaped Wilson
line makes it even more challenging to obtain a practical signal without gauge-link
smearing~\cite{NieMiera:2025inn} or Wilson flow~\cite{Luscher:2010iy}. This makes the
CG method rather appealing, since the absence of Wilson lines and the exponential
improvement of SNR at large $b_\perp$ will benefit the computation significantly.

The Wigner distributions~\cite{Belitsky:2003nz}, defined in a quantum phase space of the impact parameter and transverse momentum, provide the complete information for hadron imaging. In addition to giving the GPDs and TMDs, the Wigner distributions also hold the keys to describing the canonical orbital angular momentum and spin-orbit correlations in the nucleon~\cite{Ji:2020ena}. Since the Wigner distributions are defined from the same TMD correlators, the LaMET formalism for the latter can be readily applied to them with modification of the matching coefficient to account for the off-forward matrix elements.

The double parton distributions (DPDs) describe the possibility of finding two partons in a hadron, as compared to the PDFs that describe the probability density of a single parton. The effect of double-parton scattering can be significant for precision measurements due to the additional radiation from secondary partonic collisions, which makes the DPDs a desirable nonperturbative target for lattice QCD calculations. So far, only the moments of the DPDs have been computed on the lattice~\cite{Bali:2020mij,Bali:2021gel,Reitinger:2024ulw,Diehl:2025kcr}, while methods to calculate their $x$-dependence within the LaMET framework have been proposed~\cite{Zhang:2023wea,Jaarsma:2023woo}. In particular, Ref.~\cite{Jaarsma:2023woo} has carried out a thorough one-loop study of the DPDs at nonperturbative transverse separations, which also follow rapidity evolutions and contain soft contributions like the TMDs. Although the method to compute the corresponding soft factor is still under investigation, lattice QCD calculations can already provide useful inputs for constraining their $x$-dependence.

\subsection{Current and future development} 
\label{sec:future}

The CS kernel is a universal quantity that describes not only the rapidity evolution of TMDs but also their vacuum properties. 
To be more specific, it is a cusp anomalous dimension related to Wilson lines with a large hyperbolic angle or rapidity difference, which also defines the soft functions. Apart from using the meson form factor in Eq.~\eqref{eq:meson_FF}, Ref.~\cite{Ji:2019sxk} also proposes that one may construct the Wilson-loop operator for the soft function with the moving heavy-quark effective theory formulation. Motivated by this idea, Refs.~\cite{Francis:2023gik,Francis:2024odu,Francis:2026gxd} constructed close-to-lightcone Wilson lines on the lattice through an auxiliary field with a complex direction, which can be solved iteratively through its Green-function equation~\cite{Mandula:1990fit}. 
These Wilson lines are then used to construct a vacuum soft function with the expected rapidity dependence through analytical continuation~\cite{Liu:2022nnk}, which can be used to extract the CS kernel.
In an exploratory lattice calculation on multiple pure gauge ensembles~\cite{Francis:2026gxd,Francis:2026czb}, it is demonstrated that this method can achieve extremely high statistical precision compared to existing methods. The rapidity dependence of such a soft function can be nicely described by an \textit{ansatz} based on CS evolution. Although the lattice renormalization and perturbative matching constitute a significant and the dominant source of systematic uncertainty, this method is still capable of achieving precision comparable to state-of-the-art calculations~\cite{Francis:2026czb}. It is worth further studying the renormalization and matching of these Wilson-line operators, the extraction of the TMD soft function, as well as the broader applications of the auxiliary-field formalism.
In addition, the finite length of the auxiliary Wilson line, the discretization used in solving its Green-function equation, and the truncation of the iterative solver must be controlled; the calculation presented so far uses pure-gauge ensembles, so the inclusion of sea quarks remains to be demonstrated. 

Thanks to the absence of linear divergences in the CG correlators (see Sec.~\ref{subsec:cg})~\cite{Gao:2023lny,Zhang:2024omt}, renormalization and Fourier transformation commute with each other. As a result, their discrete Fourier transform defines a unique momentum-density operator on a finite periodic lattice, which can be directly simulated. 
Therefore, parameterization of the long-distance matrix elements in coordinate-space for Fourier transform~\cite{Dutrieux:2025jed,Chen:2025cxr,Dutrieux:2025axb} is avoided
and replaced by exponentially suppressed finite-volume effects. 
This idea is proposed in Ref.~\cite{Grebe:2026qmt}, where direct lattice simulations of momentum-space quasi-PDF operators show good agreement with the conventional Fourier-transform approach supplemented by an asymptotic extrapolation of the long-distance tail~\cite{Ji:2026vir}. This provides a nontrivial cross-validation of both approaches in controlling uncertainties in the $x$ dependence of quasi-distributions. Moreover, the momentum-space method can also be extended to quasi-TMDs in both $b_\perp$ and $k_\perp$ spaces, as demonstrated in Ref.~\cite{Grebe:2026qmt}. These developments provide a simple and promising avenue for further improving the precision of LaMET calculations.

Recently, a new class of kinematically enhanced interpolation operators for highly boosted hadrons has been proposed to improve the SNR in lattice simulations~\cite{Zhang:2025hyo}. The key idea is to construct the interpolation operators which have maximal overlap with the leading lightcone Fock states of the hadron~\cite{Ji:2003yj}, which lead to an enhancement factor proportional to the square of the Lorentz boost factor. This corresponds to ${\cal O}(100)$ and ${\cal O}(10)$ improvement in statistical precision for physical pion and proton moving at $P^z\sim 3$~GeV~\cite{Zhang:2025hyo,Reitinger:2026hta}, which will greatly improve the precision of parton physics calculations with LaMET and the other approaches.

At such high momentum, the gap between the excited and ground states of the hadron becomes narrow, which makes it harder to extract the desired quasi-distribution matrix elements.
The recently developed Lanczos method for analyzing lattice correlation functions can provide two-sided bounds for the extraction of energy levels~\cite{Wagman:2024rid,Hackett:2024xnx}, thus reducing the model biases that contribute to the systematic uncertainties. This method has already been applied to the extraction of quasi-TMD matrix elements in LaMET calculations~\cite{Avkhadiev:2026xyf,Grebe:2026qmt}. 
Moreover, the operators need to be renormalized and matched exactly as the conventional ones, so the same renormalization and perturbative-matching systematics apply, and the discretization artifacts and the continuum extrapolation require explicit study in the future.

\section{Summary and Outlook}\label{sec:conclusion}

In this review, we have summarized the rapid progress in understanding TMDs from lattice QCD, highlighting their essential role in revealing the 3D momentum and spin structure of hadrons.
Recent developments have established systematic frameworks, most notably LaMET, that enable direct access to TMD observables from Euclidean lattice calculations, significantly advancing beyond earlier efforts based on the Lorentz-invariant method.
Within this framework, further methodological improvements such as the Coulomb-gauge formulation have enhanced numerical efficiency and control over lattice systematics.
These approaches have made it possible to compute a broad class of TMD-related quantities from first principles, including the CS kernel, reduced soft function, as well as TMDPDFs and TMDWFs.
Together, these quantities provide a unified picture of both perturbative evolution and nonperturbative structure underlying transverse-momentum dynamics in QCD.

Lattice-QCD studies have evolved from early exploratory investigations into quantitatively controlled calculations with increasing precision.
In particular, the extraction of the CS kernel has become a mature and robust methodology, while ongoing progress in computing TMDPDFs and TMDWFs demonstrates the feasibility of accessing the full transverse and longitudinal structure of hadrons directly from QCD.
These advances establish lattice QCD as a powerful complement to phenomenological analyses, offering independent nonperturbative input for global fits and the interpretation of experimental data.

Looking ahead, continued improvements in lattice methodology and computational resources are expected to significantly enhance the precision and scope of TMD calculations.
A key direction is the systematic control of all relevant limits, including the continuum, physical quark-mass, and large-momentum extrapolations.
Future developments in renormalization schemes, higher-order perturbative matching, and power corrections will be essential for further reducing the theoretical uncertainties in the determinations of TMD observables across a wide kinematic range.
These advances will also strengthen the connection between lattice results and experimental measurements, particularly in the regime where nonperturbative effects dominate.
Beyond methodological improvements, expanding the range of accessible TMD observables represents an important future frontier.
Current lattice studies already demonstrate the feasibility of computing both unpolarized and spin-dependent TMDs, and future efforts are expected to include a broader class of quark and gluon distributions, as well as connections to small-$x$ physics and 3D hadron tomography.
With upcoming experimental facilities, such as the Electron-Ion Collider, poised to deliver high-precision data sensitive to transverse-momentum dynamics, lattice QCD will play an increasingly crucial role in providing first-principles inputs for global analyses.
Ultimately, the combined progress in theory, lattice simulation, and experiment will enable a comprehensive and quantitatively reliable picture of the 3D structure of hadrons.

\section*{Acknowledgments}
We thank Yizhuang Liu for helpful comments on this manuscript and Jin-Xin Tan for making the graphics in Fig.~\ref{fig:factorization_quasi_TMD}.
The work of HL is partially supported by the US National Science Foundation under grant PHY~2514533 and
by the U.S.~Department of Energy under contract DE-SC0024644.
WW is supported in part by the National Natural Science Foundation of China under grant Nos.~12125503, and 12305103.
YZ is supported by the U.S. Department of Energy, Office of Science, Office of Nuclear Physics through Contract No.~DE-AC02-06CH11357, the Early Career Award through Contract No.~DE-SCL0000017, and the Quark-Gluon Tomography (QGT) Topical Collaboration under award No.~DE-SC0023646. 

\bibliography{refs}
\end{document}